\documentclass[pra,twocolumn,amsmath,amssymb,superscriptaddress]{revtex4-1}
\pdfoutput=1  
\usepackage{epsfig,amsmath}
\usepackage{subfigure}
\usepackage{graphicx}
\usepackage{dcolumn}
\usepackage{stmaryrd}
\usepackage{mathrsfs}
\usepackage{pifont}
\usepackage{amsthm}
\usepackage{amssymb}
\usepackage{bm}
\usepackage{latexsym}
\usepackage[colorlinks=true,linkcolor=blue,citecolor=blue]{hyperref}
\usepackage{color}
\usepackage{epstopdf}
\usepackage{booktabs}
\usepackage{amsmath}
\usepackage{threeparttable}
\usepackage{times} 
\usepackage{hyperref}
\usepackage{orcidlink}
\begin{document}
	\title{Magnon Blockade  with Skyrmion Qubit-Magnon Coupling in a Hybrid Quantum System}
	
	\author{Si-Tong Jin\orcidlink{0009-0007-0399-4193}}
	\affiliation{School of Physics, Dalian University of Technology, Dalian 116024, China}
	\author{Shi-wen He}
	\affiliation{School of Physics, Dalian University of Technology, Dalian 116024, China}
	\author{Zi-long Yang}
	\affiliation{School of Physics, Dalian University of Technology, Dalian 116024, China}
	\author{Xuanxuan Xin}
	\affiliation{College of Arts and Science, Qingdao Binhai University, Qingdao 266555, China.}
	\author{Chong Li}
	\email{Email address: lichong@dlut.edu.cn}
	\affiliation{School of Physics, Dalian University of Technology, Dalian 116024, China}

	\date{\today}
	
	\begin{abstract}

Magnon blockade is a fundamental quantum phenomenon for generating single-magnon state, which gradually becomes one of the candidates for quantum information processing. In this paper, we propose a theoretical scheme to generate the magnon blockade in a hybrid system consisting of a YIG micromagnet and a skyrmion. Considering weak probing of the magnon and driving of the skyrmion qubit, the second-order correlation function is analytically derived, and the optimal condition for realizing the magnon blockade is identified. Under the optimal condition, we systematically analyze the behavior of the second-order correlation function \( g^{(2)}(0) \) under different parameter regimes. Our analysis shows that with appropriate driving and probing field intensities, the magnon blockade effect can be significantly enhanced, effectively suppressing multi-magnon states and facilitating the generation of high-purity single-magnon states exhibiting pronounced antibunching. Furthermore, we explore the physical mechanisms underlying the magnon blockade, revealing the coexistence and interplay of conventional and unconventional magnon blockade. This scheme provides a versatile all-magnetic platform for generating high purity single-magnon sources.
\end{abstract}

\maketitle
	
\section{Introduction}
Hybrid quantum systems combine the advantages of different subsystems, playing a key role in quantum information processing, networking, and sensing \cite{doi:10.1088/0031-8949/2009/t137/014001, doi:10.1038/nature07127, RevModPhys.85.623, doi:10.1073/pnas.1419326112}. Examples of hybrid systems are magnonic systems \cite{PhysRevB.108.024105,PhysRevA.107.033516, doi:10.1088/1367-2630/ad327c, PhysRevLett.116.223601, PhysRevLett.125.117701,PhysRevX.3.041023,PRXQuantum.2.040314}, optomechanical systems \cite{10.1126/science.1156032, 10.1002/andp.201200226, 10.1103/RevModPhys.86.1391, Bowen2015, PhysRevLett.129.063602}, and photonic-qubit systems \cite{10.1103/PhysRevLett.131.050601}, and so on. Magnons excited in YIG materials, with high spin density, frequency tunability, long lifetime, and enhanced coherence, are promising quantum information carriers \cite{doi:10.1088/0022-3727/43/26/264002, doi:10.1038/nphys3347, doi:10.1063/5.0152543, YUAN20221,ZARERAMESHTI20221,PhysRevLett.111.127003,PhysRevLett.113.083603,PhysRevLett.123.107701,PhysRevB.94.224410,10.1103/PhysRevA.106.012609}. The unique properties of magnons have facilitated hybrid magnonic systems like cavity optomagnonic systems \cite{PhysRevB.108.024105,PhysRevA.107.033516,doi:10.1088/1367-2630/ad327c}, cavity magnonic systems \cite{PhysRevLett.116.223601,10.1063/5.0138391} and qubit-magnon systems \cite{PhysRevLett.125.117701,PhysRevX.3.041023,PRXQuantum.2.040314}. In these systems, magnon modes effectively couple with quantum information carriers, including phonon modes, microwave photons, optical photons, and superconducting qubits \cite{PhysRevLett.120.133602,doi:10.1126/sciadv.1501286,PhysRevLett.129.243601, PhysRevLett.119.147701,PhysRevResearch.2.022027,PhysRevLett.129.037205,PhysRevB.100.134421, PhysRevA.101.042331, PhysRevLett.130.193603}. Nontrivial quantum phenomena in these configurations highlight the potential of hybrid magnonic systems as platforms for exploring and controlling complex quantum interactions \cite{PhysRevA.107.023709, 10.1103/PhysRevA.110.053710, 10.1103/PhysRevA.109.022442, PhysRevLett.130.193603,PhysRevB.100.134421,PhysRevA.101.042331,PhysRevA.103.063708,XIONG20238,PhysRevB.106.184426,doi:10.1088/1367-2630/ad0b20,PhysRevB.105.094422,PhysRevLett.123.127202,PhysRevApplied.13.014053,PhysRevA.103.043706,PhysRevLett.124.053602,PhysRevA.105.022624,PhysRevB.93.174427,PhysRevA.101.063838,Xu:21,PhysRevA.103.052411,PhysRevA.106.013705,PhysRevA.108.053702,PhysRevB.102.100402,PhysRevA.109.023710,PhysRevA.110.012459,Wang:20,Zhu:21,PhysRevA.108.053703,PhysRevA.109.033709,doi:10.1038/s42254-023-00583-2,PhysRevA.110.033702, doi:10.1103/PhysRevA.107.023709}.

Skyrmions, as magnetic nanostructures with particle-like characteristics, exhibit novel magnetic effects  \cite{RevModPhys.94.035005,doi:10.1063/1.5048972,PhysRevB.104.L100415,PRXQuantum.3.040321}. Their most notable attribute is their topological protection \cite{doi:10.1142/s0217979219300056,doi:10.1088/1361-6463/ab8418}, which enables skyrmions to withstand external disturbances. This topological protection enables state manipulation with ultra-low currents and modulation of electric and magnetic fields  \cite{PhysRevLett.127.067201, PhysRevLett.132.193601, doi:10.1038/nnano.2016.234, doi:10.1038/s41586-024-07859-2}, facilitating logical transitions within a richly anharmonic operational regime and making them robust information carriers \cite{doi:10.1063/5.0027042,RevModPhys.94.035005}. Consequently, skyrmions hold significant promise for applications in spintronics and quantum computing \cite{BOGDANOV1994255,doi:10.1038/nature05056,doi:10.1126/science.1166767,doi:10.1038/nature09124,doi:10.1126/science.1195709,doi:10.1038/nphys2231,PhysRevLett.108.267201,doi:10.1038/nnano.2013.243,doi:10.1038/nnano.2013.29,doi:10.1038/natrevmats.2017.31,PhysRevLett.128.227204}. Experimentally, skyrmions can be generated in centrosymmetric crystals, such as Gd$_{2}$PdSi$_{3}$ \cite{doi:10.1126/science.aau0968}. In frustrated magnets, skyrmions exhibit internal degrees of freedom related to helicity \cite{doi:10.1038/ncomms9275,PhysRevB.93.064430,doi:10.1126/science.aau0968,pmid:29167418,doi:10.1088/1367-2630/aba1b3,PhysRevLett.130.106701}. By quantizing collective helicity coordinates, two types of qubits can be constructed \cite{PhysRevLett.127.067201}, with skyrmion qubits achieving microsecond-scale coherence times, making them suitable for quantum state manipulation. Tunable coherent and dissipative coupling has been achieved between YIG magnon modes and the quantized helicity of skyrmions in skyrmion qubit-magnon systems \cite{PhysRevLett.132.193601}. This robust all-magnetic coupling mediates long-range coherent interactions between a magnon and skyrmion qubits, thereby realizing the scalability of skyrmion qubits and providing a promising platform for exploring quantum effects.
	
In particular, magnon blockade (MB) has attracted special attention as a purely quantum effect \cite{PhysRevA.110.033702,PhysRevLett.130.193603,PhysRevB.100.134421,PhysRevA.101.042331,PhysRevA.101.063838,Xu:21,PhysRevA.103.052411,PhysRevA.106.013705,PhysRevA.108.053702,PhysRevB.102.100402,PhysRevA.109.023710, PhysRevA.110.012459}. The MB effect arises from two mechanisms: energy-level anharmonicity, which inhibits the excitation of a second magnon after the first magnon is excited to the single-magnon state, termed conventional magnon blockade (CMB)~\cite{PhysRevB.100.134421, PhysRevA.101.042331, PhysRevA.108.053702, PhysRevA.109.023710}, and destructive interference between transition pathways to the two-magnon state, known as as unconventional magnon blockade (UMB)~\cite{PhysRevA.101.042331}. The MB effect has been theoretically studied in hybrid ferromagnet-superconductor systems \cite{PhysRevB.100.134421}. Subsequent studies proposed schemes to achieve antibunched light via MB, including quantum-interference-enhanced MB in YIG-superconducting circuits \cite{PhysRevA.101.042331}, and tunable MB in ferromagnet-superconductor systems with dual qubits \cite{PhysRevB.104.224434}, and enhanced MB in optomechanical-magnetic systems using quantum destructive interference among three indirect transition paths \cite{PhysRevA.109.043712}. The standard indicator for verifying a stable single-magnon state is \( g^{(2)}(0) \rightarrow 0 \) \cite{PhysRevLett.107.063601,PhysRevA.82.032101,PhysRevA.99.013804}. Analogous to photon blockade \cite{Wang:20,Zhu:21,PhysRevA.108.053703,PhysRevA.109.033709,doi:10.1038/s42254-023-00583-2, PhysRevA.109.033709,doi:10.1088/1367-2630/ac6a46,PhysRevA.96.053810,PhysRevLett.121.153601} and phonon blockade \cite{PhysRevA.96.013861,PhysRevA.94.063853,APL2021}, MB enables the preparation of single-magnon states, offering the possibility of quantum control at the single-magnon level.

In this work, we theoretically study the realization of MB in YIG ferromagnetic insulators. The system comprises a YIG micromagnet strongly coupled to a resonant skyrmion qubit \cite{PhysRevLett.132.193601}. Probing and driving fields are applied to the YIG and the skyrmion qubit, with the latter modulated by electric fields. Building on this setup, we analyze the steady-state properties of the coupled system and derive the optimal MB condition, $\Delta/\gamma \approx \pm \frac{1}{2} g_{ms}/\gamma$, which is consistent with numerical simulations \cite{PhysRevB.100.134421, Carmichael1999}. Using the optimal condition, we investigate the effect of the probing-to-driving field strength ratio on the MB effect. At a ratio of approximately 6, the second-order correlation function $g^{(2)}(0) \sim 10^{-7}$, indicating the strongest MB effect. This result underscores the critical role of the probing-to-driving field strength ratio in optimizing the MB effect. Furthermore, we investigate the mechanisms of MB, revealing the coexistence of CMB and UMB. CMB arises from the intrinsic anharmonicity and energy level splitting in the magnon-qubit system, where the non-equidistant energy spectrum inhibits multi-magnon excitations. UMB is induced by quantum interference effects between transition pathways under the influence of both driving and probing fields. Together, these mechanisms ensure the suppression of multi-magnon states and enhance the MB effect. In another scenario, the skyrmion qubit is exclusively modulated by magnetic fields, and the tprobing is restricted to magnons. Here, MB persists with $g^{(2)}(0) \sim 10^{-2}$, and the optimal condition \( \Delta/\gamma \approx \pm \frac{1}{2} \tilde{g}_{ms}/\gamma \) remains valid. Therefore, our study provides theoretical support for the generation of a single-magnon source, which is essential for investigating fundamental aspects of quantum magnons and for potential applications in magnon-based quantum simulation and hybrid quantum devices \cite{PhysRevLett.123.107701,PhysRevLett.123.107702,PhysRevLett.132.193601}. 
\section{Model and HAMILTONIAN}\label{modelandHam}
\begin{figure}
	\centering
	\includegraphics[width=0.9\linewidth, height=0.3\textheight]{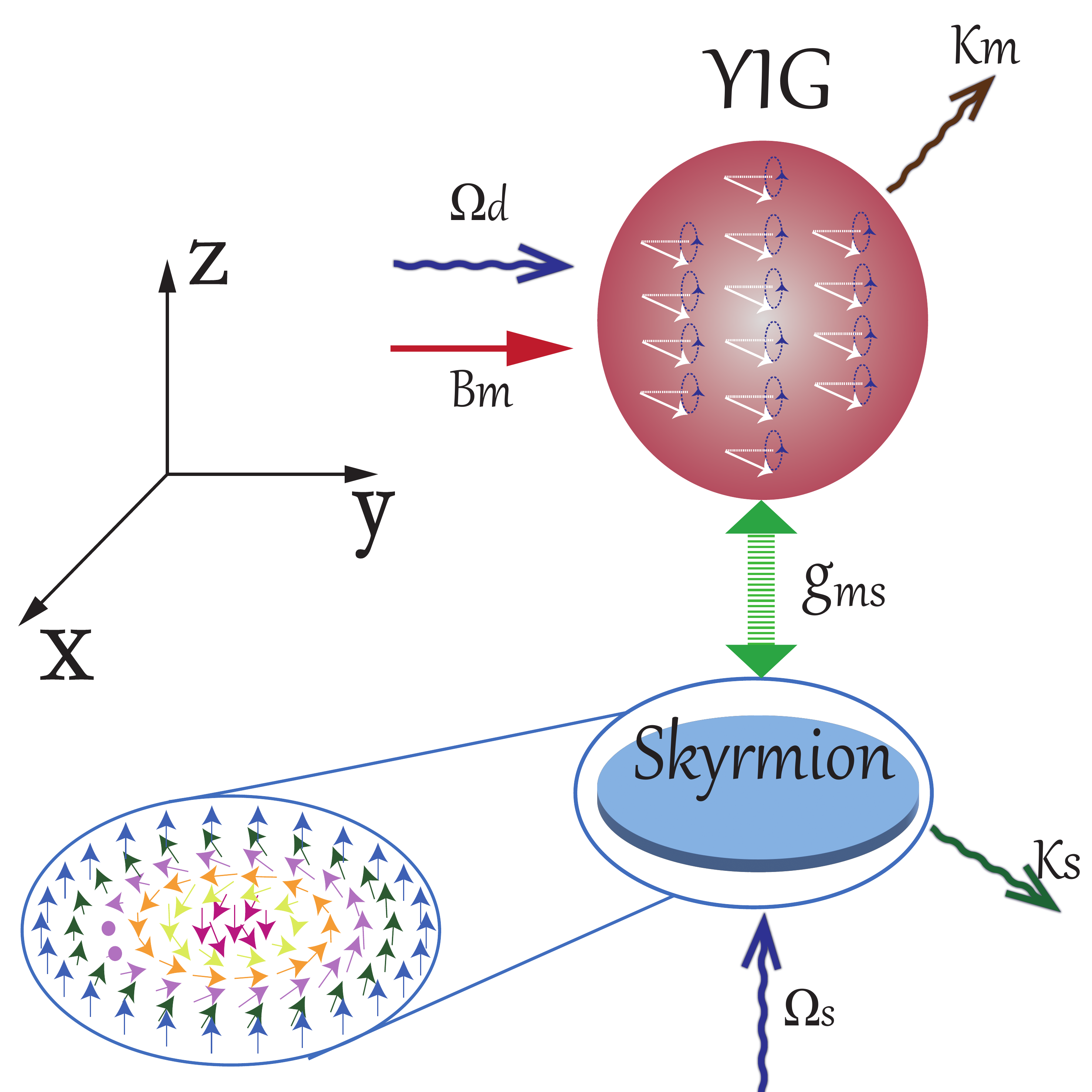}
	\caption{Schematic representation of the hybrid magnon-Skyrmion system.The magnon is probed by an external microwave field \( \Omega_m \), while the skyrmion is driven by a microwave field \( \Omega_s \). The magnon mode interacts with the skyrmion mode through a coupling strength \( g_{ms} \). The dissipative processes for both the magnon and skyrmion modes are represented by the damping rates \( \kappa_m \) and \( \kappa_s \), respectively.}
	\label{fig:fig1}
\end{figure}

As shown in Fig. ~\ref{fig:fig1}, we study a hybrid quantum system,  integrating magnon modes in a YIG micromagnet and skyrmion modes beneath it. Strong coupling between the skyrmion qubit and magnon can be tuned by varying their vertical separation and the radius of the YIG sphere, as the coupling strength depends on both factors \cite{PhysRevLett.132.193601}.
Within the YIG sphere, spin wave modes with long coherence times can be induced through the application of a uniform bias magnetic field $B_m$ \cite{PhysRevLett.132.193601,PhysRevB.101.125404,PhysRevLett.124.093602}. The magnon represents spin waves, collective excitations of spin ensembles in magnetic materials, while the skyrmion is a topological excitation characterized by rotations or twists in the spin structure. The coupling between the magnon and the skyrmion arises from the interaction between the magnon's collective spin wave oscillation and the skyrmion's localized spin motion within its topological structure. 

Here, we focus on the Kittel mode, which features the coherent precession of all spins in the micromagnet, with spins maintaining identical phase and amplitude. The free Hamiltonian of the magnon mode is \( \hat{H}_m = \omega_m \hat{m}^\dagger \hat{m} \) (with \( \hbar = 1 \)), where \( \hat{m}^\dagger \) and \( \hat{m} \) are the creation and annihilation operators of the magnon mode, respectively, and \( \omega_m = \gamma_e B_m \) is the resonance frequency, with \( \gamma_e \) being the gyromagnetic ratio \cite{PhysRevLett.132.193601}. By quantizing the collective coordinates, the skyrmion can be treated as a qubit that can be controlled and manipulated. The energy levels of the \( \hat{S}_z \) skyrmion qubit are anharmonic \cite{PhysRevLett.127.067201,PhysRevLett.132.193601,doi:10.1063/5.0177864,PhysRevResearch.6.023067}, the effective Hamiltonian of the \( \hat{S}_z \) qubit within the subspace \( \{|0\rangle, |1\rangle \} \) is \( \hat{H}_{S_z} = \frac{K_0}{2} \hat{\sigma}_z - \frac{E_z}{2} \hat{\sigma}_x \). The parameters \( K_0 \) and \( E_z \) are externally modulated, with \( K_0 \) controlled by magnetic fields and \( E_z \) by electric fields, enabling the manipulation of the qubit's states \cite{PhysRevLett.127.067201}. The interaction Hamiltonian is 
$\hat{H}_{\text{int}} = \frac{\tilde{g}_{ms}}{2} (\hat{m} + \hat{m}^\dagger) \hat{\sigma}_x 
+ \frac{g_{ms}}{2} (\hat{m} + \hat{m}^\dagger) \hat{\sigma}_z$, where \( \tilde{g}_{ms} \) describes the coupling between the X- and Y-direction magnetic field generated by the YIG sphere and the skyrmion qubit, and \( g_{ms} \) describes the coupling between the Z-direction magnetic field from the YIG sphere and the skyrmion qubit. The maximum coupling strength of $\tilde{g}_{ms}$ can reach up to 40\,\text{MHz}, whereas $g_{ms}$ can achieve up to 200\,\text{MHz} \cite{PhysRevLett.132.193601}.

To achieve and dynamically manipulate MB, we introduce a microwave drive applied to the skyrmion qubit with a driving frequency \( \omega_s \). The corresponding Hamiltonian is \( H_s = \Omega_{s} \cos(\omega_s t) \hat{\sigma}_z \) \cite{PhysRevLett.132.193601}. Additionally, to observe the MB effect, a weak probe field is introduced into the system, with Hamiltonian \( H_d = \Omega_d (\hat{m}^\dagger e^{-i\omega_d t} + \hat{m} e^{i\omega_d t}) \) \cite{doi:10.1007/s10773-016-3180-y,PhysRevB.100.134421}, where \( \omega_d \) and \( \Omega_d \) are the probe field's frequency and intensity, respectively. The total Hamiltonian is given by
\begin{align}
	\hat{H}_{\text{total}} = & \, w_m \hat{m}^\dagger \hat{m} + \frac{K_0}{2} \hat{\sigma}_z - \frac{E_z}{2} \hat{\sigma}_x \notag \\
	& + \frac{\tilde{g}_{ms}}{2} (\hat{m} + \hat{m}^\dagger) \hat{\sigma}_x + \frac{g_{ms}}{2} (\hat{m} + \hat{m}^\dagger) \hat{\sigma}_z \notag \\
	& + \Omega_s \cos(\omega_s t) \hat{\sigma}_z + \Omega_d (\hat{m}^\dagger e^{-i\omega_d t} + \hat{m} e^{i\omega_d t}).
\end{align}
When the skyrmion qubit operates near the degeneracy point, its Hamiltonian $\hat{H}_{S_z}$ can be expressed in terms of the eigenstates $\lvert \psi_2 \rangle = \frac{\sqrt{2}}{2} \lvert 1 \rangle - \frac{\sqrt{2}}{2} \lvert 0 \rangle$ and $\lvert \psi_1 \rangle = \frac{\sqrt{2}}{2} \lvert 1 \rangle + \frac{\sqrt{2}}{2} \lvert 0 \rangle$. In the subspace spanned by $\{ \lvert \psi_2 \rangle, \lvert \psi_1 \rangle \}$, the Pauli operators are defined as $\tilde{\sigma}_z = \lvert \psi_2 \rangle \langle \psi_2 \rvert - \lvert \psi_1 \rangle \langle \psi_1 \rvert$ and $\tilde{\sigma}_x = \lvert \psi_2 \rangle \langle \psi_1 \rvert + \lvert \psi_1 \rangle \langle \psi_2 \rvert$. The coupling term \( \tilde{g}_{ms} \) does not induce energy level transitions but causes a small shift (\( \tilde{g}_{ms} \approx 50.1 \, \text{MHz} \)) in the skyrmion qubit energy levels. Because this shift is much smaller than the gigahertz-scale resonant frequencies (\( E_z \approx \omega_m \approx 9.8 \, \text{GHz} \)), it can be neglected \cite{PhysRevLett.132.193601,PhysRevResearch.6.023067}. Using the rotating-wave approximation (RWA) and considering the time-evolution operator in the rotating frame, $U = e^{-i \left( \frac{\omega_s \tilde{\sigma}_z t}{2} + w_d t \hat{m}^\dagger \hat{m} \right)}$, the hybrid quantum system Hamiltonian can be simplified. Thus, the effective Hamiltonian can be written as
\begin{align}
	H_{\text{eff}} = & \, \Delta_m \hat{m}^\dagger \hat{m} + \frac{\Delta_s}{2} \tilde{\sigma}_z + \frac{g_{ms}}{2} (\hat{m} \tilde{\sigma}_+ + \hat{m}^\dagger \tilde{\sigma}_-) \notag \\
	& + \frac{\Omega_s}{2} (\tilde{\sigma}_+ + \tilde{\sigma}_-) + \Omega_d (\hat{m}^\dagger + \hat{m}). \tag{2}
\end{align}
where \( \Delta_m \equiv \omega_m - \omega_d \) (\( \Delta_s \equiv E_z - \omega_s \)) represents the detuning between the magnon mode and the probe field frequency, while \( \Delta_s \) represents the detuning between the skyrmion qubit frequency and the driving frequency.  We choose the parameter conditions as \( \omega_d \approx \omega_s \), \( \omega_d \gg g_{ms} \), and \( \omega_s \gg \Omega_s \), with  \( \Delta = \Delta_s = \Delta_m \), which can be flexibly tuned in the current experimental setup. 

\section{Magnon Blockade and OPTIMAL CONDITIONS}\label{modelandHam}
\subsection{The solution of the second-order correlation function}
In the weak-driving regime, where \( \Omega_d, \Omega_s \ll k_m, k_s \), and the effects of quantum jumps on short time scales are neglected, the wave function of the composite system can be approximated within the framework of an analytical model \cite{PhysRevA.103.052411, PhysRevB.100.134421, PhysRevA.101.042331, PhysRevA.106.013705, PhysRevA.87.013839, PhysRevA.96.053810, PhysRevApplied.12.044065, PhysRevA.110.012459, PhysRevA.108.053702}. In particular, the dynamics of the system can be described within an approximate framework using a non-Hermitian Hamiltonian
	\begin{equation}
		\begin{aligned}
			\hat{H}_{\mathrm{non}} = & \, \hat{H}_{\mathrm{eff}} - i \frac{k_m}{2} m^{\dagger} m 
			& - i \frac{k_s}{2} \tilde{\sigma}_{+} \tilde{\sigma}_{-} 
		\end{aligned}
		\tag{3}
	\end{equation}
	thus, the Hilbert space of the composite system can be truncated to a subspace corresponding to a few excitations, and the system's evolution can be effectively approximated by a pure state, as described below
\begin{align}
	|\psi\rangle = & \, C_{g',0}|g',0\rangle + C_{g',1}|g',1\rangle + C_{e',0}|e',0\rangle \nonumber \\ 
	& + C_{e',1}|e',1\rangle + C_{g',2}|g',2\rangle \tag{4}
\end{align}
	where \( C_{m',s } \) (\( m= g',e'  \; \text{and} \;  s=0,1,2 \)) are the probability amplitudes for the states \( |ms\rangle \), and \( |g(e), n\rangle \) represents the ground (excited) state of the skyrmion qubit and the Fock state of the magnon mode, respectively.
	
	Based on the Schrödinger equation \( i \frac{d}{dt} |\psi(t)=\hat{H}_{\text{non}} |\psi(t)\rangle \), the dynamical evolution of the coefficients \( C_{ms} \) can be derived from the equations.
	The corresponding dynamical equations can then be written as
	\begin{align}
		i \dot{C}_{g', 0} &= \frac{\Omega_s}{2} C_{e', 0} + \Omega_d C_{g', 1} \tag{5a} \\ \vspace{-1em}
		i \dot{C}_{e', 0} &= ( \Delta_s - i \frac{k_s}{2} ) C_{e', 0} + \frac{g_{ms}}{2} C_{g', 1} + \frac{\Omega_s}{2} C_{g', 0} \notag\ \tag{5b} \\ \vspace{-1em}
		i \dot{C}_{g', 1} &= ( \Delta_m - i \frac{k_m}{2} ) C_{g', 1} + \frac{g_{ms}}{2} C_{e', 0} + \Omega_d C_{g', 0} \notag\  \tag{5c} \\ \vspace{-1em}
		i \dot{C}_{e', 1} &= ( \Delta_m + \Delta_s - i \frac{k_m}{2} - i \frac{k_s}{2} ) C_{e', 1} \notag  + \frac{\sqrt{2}}{2} g_{ms} C_{g', 2} \\
		&\quad+ \frac{\Omega_s}{2} C_{g', 1} + \Omega_d C_{e', 0} \notag \tag{5d} \\ \vspace{-1em}
		i \dot{C}_{g', 2} &= ( \Delta_m - i \frac{k_m}{2} ) C_{g', 2} + \sqrt{2} \Omega_d C_{g', 1} \notag + \frac{\sqrt{2}}{2} g_{ms} C_{e', 1} \tag{5e} \vspace{-1em}
	\end{align}
	then its steady-state solution can be obtained from
	\begin{align}
		0 &= \frac{\Omega_s}{2} C_{e', 0} + \Omega_d C_{g', 1}, \tag{6a} \\
		0 &= (\Delta_s - i \frac{k_s}{2}) C_{e', 0} + \frac{g_{ms}}{2} C_{g', 1} + \frac{\Omega_s}{2} C_{g', 0}, \tag{6b} \\
		0 &= (\Delta_m - i \frac{k_m}{2}) C_{g', 1} + \frac{g_{ms}}{2} C_{e', 0} + \Omega_d C_{g', 0}, \tag{6c} \\
		0 &= (\Delta_m + \Delta_s - i \frac{k_m}{2} - i \frac{k_s}{2}) C_{e', 1} + \frac{\sqrt{2}}{2} g_{ms} C_{g', 2} \notag \\
		&\quad + \frac{\Omega_s}{2} C_{g', 1} + \Omega_d C_{e', 0}, \tag{6d} \\
		0 &= (\Delta_m - i \frac{k_m}{2}) C_{g', 2} + \sqrt{2} \Omega_d C_{g', 1} + \frac{\sqrt{2}}{2} g_{ms} C_{e', 1}. \tag{6e}
	\end{align}
In the weak-driving regime (\( \Omega_d, \Omega_s \ll \kappa_m, \kappa_s \)), where \( |C_{g',0}| \gg |C_{g',1}|, |C_{g',2}| \) and \( |C_{e',0}| \gg |C_{g',3}|, |C_{e',1}|, |C_{e',2}| \). By applying perturbation theory, we neglect the higher-order terms that contribute minimally in the equation of motion for the lower-order variables. We can derive the steady-state solutions for the probability amplitudes	
\begin{align}
		C_{g', 0} & \approx 1 \tag{7a} \\
		C_{e', 0} & = \frac{-2 \Omega_{d} g_{ms} + 2 \tilde{\Delta}_{m} \Omega_{s}}{-4 \tilde{\Delta}_{m} \tilde{\Delta}_{s} + g_{ms}^{2}} \tag{7b} \\
		C_{g', 1} & = \frac{4 \tilde{\Delta}_{s} \Omega_{d} - \Omega_{s} g_{ms}}{-4 \tilde{\Delta}_{m} \tilde{\Delta}_{s} + g_{ms}^{2}} \label{eq:7c} \tag{7c} \\
		C_{e', 1} & = \frac{-\Omega_{s} D A - 2 A \Omega_{d} E + 8 \tilde{\Delta}_{s} \Omega_{d} g_{ms} - 2 g_{ms}^{2} \Omega_{d} \Omega_{s}}{(-4 \tilde{\Delta}_{m} \tilde{\Delta}_{s} + g_{ms}^{2})(2 A B - g_{ms}^{2})} \tag{7d} \\
		C_{g', 2} & = \frac{-2 \sqrt{2} B C + \Omega_{s} D g_{ms} + 2 \Omega_{d} E g_{ms}}{(-4 \tilde{\Delta}_{m} \tilde{\Delta}_{s} + g_{ms}^{2})(2 \sqrt{2} A B - \sqrt{2} g_{ms}^{2})} \label{eq:7e}\tag{7e}
	\end{align}
	
where \( 
\tilde{\Delta}_{m} = \Delta_m - i \frac{k_{m}}{2}, \,
\tilde{\Delta}_{s} = \Delta_s - i \frac{k_{s}}{2}, \,
A = 2 \big( \Delta_m - i \frac{k_{m}}{2} \big), \,
B = \big( \Delta_m + \Delta_s \big) - i \frac{k_{m}}{2} - i \frac{k_{s}}{2}, \,
C = 4 \sqrt{2} \tilde{\Delta}_{s} \Omega_{d}^{2} - \sqrt{2} g_{ms} \Omega_{d} \Omega_{s}, \,
D = 4 \tilde{\Delta}_{s} \Omega_{d} - \Omega_{s} g_{ms}, \,
E = -2 \Omega_{d} g_{ms} + 2 \tilde{\Delta}_{m} \Omega_{s}.
\)

With the steady-state amplitudes in  Eqs.~\eqref{eq:7c} and \eqref{eq:7e}, the function \( g^{(2)} \) can then be expressed by
\begin{align}
		g^{(2)}(0) & \approx \frac{2 \left| C_{g', 2} \right|^{2}}{\left| C_{g', 1} \right|^{4}}\notag= \frac{2 \left| \eta_1 \right|^{2} \cdot \left| \chi \right|^{2}}{\left|\eta_2 \right|^{2} \left|D \right|^{4}} \notag\tag{8}\label{eq:8}
\end{align}
in which \( \eta_1 = -2 \sqrt{2} B C + \Omega_{s} D g_{ms} + 2 \Omega_{d} E g_{ms},  \eta_2 = 2 \sqrt{2} A B - \sqrt{2} g_{ms}^2, \chi = -4 \tilde{\Delta}_{m} \tilde{\Delta}_{s} + g_{ms}^2 \).

The population in the state \( |g',1\rangle \) can be optimized by
\begin{align}
	\Delta_s \Delta_m & \approx \tilde{\Delta}_s \tilde{\Delta}_m = \frac{1}{4} g_{ms}^2 \tag{9}
		\label{eq:9}
\end{align}
\subsection{Mechanisms of The Magnon Blockade}

\begin{figure}
	\centering
	\includegraphics[width=1\linewidth]{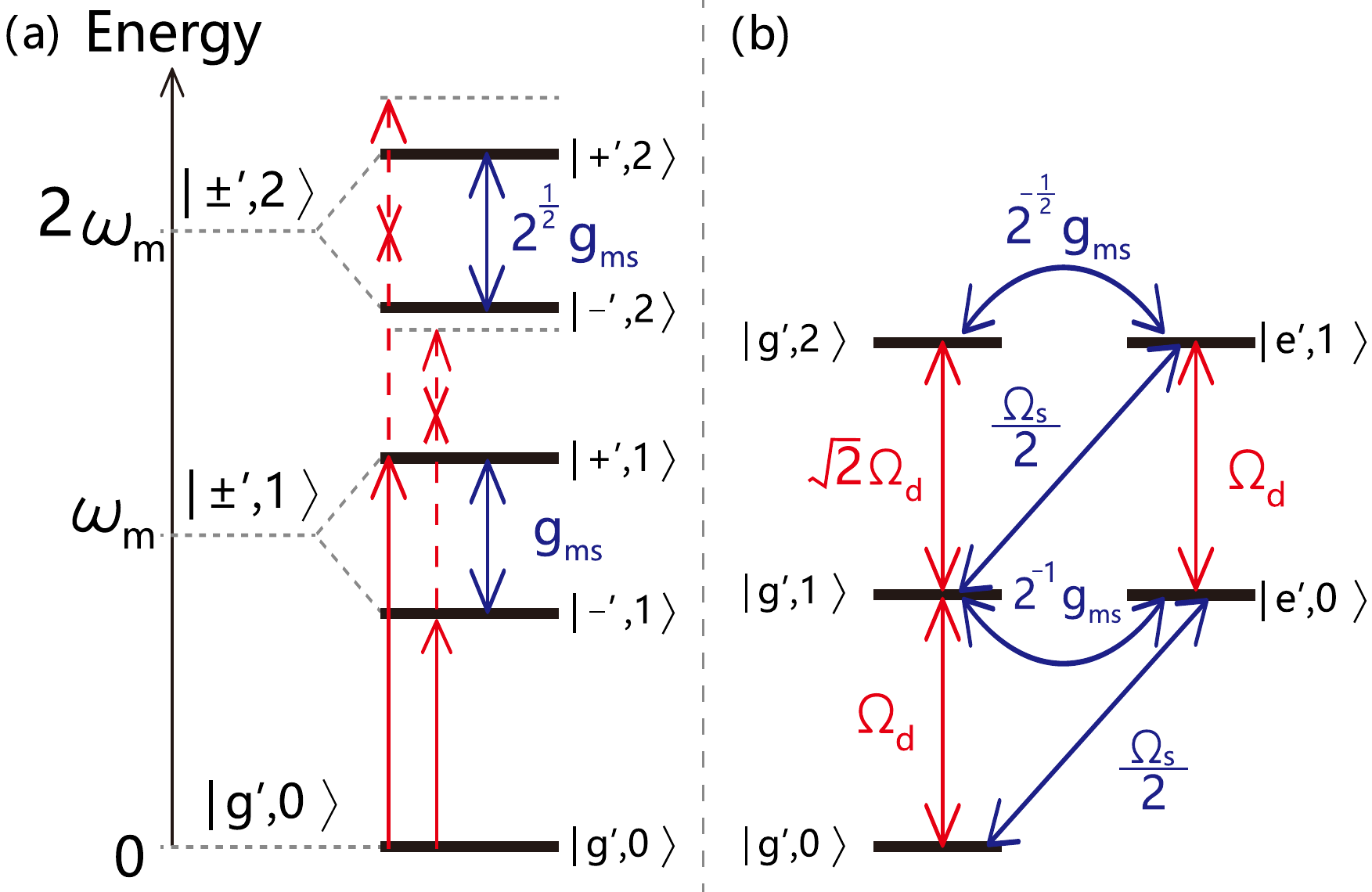}
	\caption{(a) Energy spectrum of the skyrmion qubit-magnon system, showing vacuum Rabi splitting induced by magnon-qubit coupling. (b) Transition pathways in the skyrmion qubit-YIG system.}
	\label{fig:fig2}
\end{figure}

To explore the mechanism of magnon blockade in a skyrmion-qubit-magnon system, we compute the eigenfunctions and examine the energy spectrum. The ground state is \( |g^\prime, 0\rangle \) with eigenvalue \( E_0 = 0 \), while for \( n \geq 1 \), the excited (dressed) states are expressed as
\[
|\psi_{n,\pm}\rangle = C_{g',n}|g',n\rangle + C_{e',n-1}|e',n-1\rangle,\tag{10}
\]
where the normalized coefficients are
\[
C_{g',n} = \pm\frac{g_{ms}\sqrt{n}}{\sqrt{g_{ms}^2 n + 4[n\omega_m - E_{n,\pm}]^2}},\tag{11}
\]
and
\[
C_{e',n-1} = \pm \sqrt{\frac{4[n\omega_m - E_{n,\pm}]^2}{g_{ms}^2 n + 4[n\omega_m - E_{n,\pm}]^2}}.\tag{12}
\]
The associated eigenvalues for \( \psi_{n,\pm} \) are given by
\[
E_{n,\pm} = \frac{(2n-1)\omega_m + E_z}{2} \pm \frac{\sqrt{[\omega_m - E_z]^2 + g_{ms}^2 n}}{2}.\tag{13}
\]

These coefficients and eigenvalues describe the dressed states in the system and ensure their normalization.

Under the strong-coupling regime, the real parts of the eigenvalues provide an accurate description of the energy spectrum, while the imaginary components, associated with dissipation, can be considered negligible \cite{PhysRevB2009}. 

For the case (\(\omega_m = E_z\)), the eigenvalues are \(E_{n,\pm} = n\omega_m \pm \frac{g_{ms}\sqrt{n}}{2}\), and the eigenstates are \(|\psi_{n,\pm}\rangle = \frac{1}{\sqrt{2}}|g',n\rangle \pm \frac{1}{\sqrt{2}}|e',n-1\rangle\). The physical mechanism of the magnon blockade can be explained by the energy-level structure shown in Fig.~\ref{fig:fig2}\hyperref[fig:fig2]{(a)}. Transitions between the states \(|g', 0\rangle\) and \(|\pm', 1\rangle\) occur when the detuning satisfies $\omega_m - \omega_d = \pm \Delta$ ($\Delta = \frac{1}{2}g_{ms}$). Conversely, transitions from \(|\pm', 1\rangle\) to \(|\pm', 2\rangle\) are suppressed due to the qubit-induced vacuum Rabi splitting, which requires the condition \(2\omega_m \pm 2\Delta \neq 2\omega_m \pm \frac{\sqrt{2}}{2}g_{ms}\). This suppression arises from the anharmonicity in the energy spectrum caused by the vacuum Rabi splitting. The anharmonicity prevents the second magnon excitation from resonantly reaching the higher states \(|\pm', 2\rangle\), confining the system to the single-magnon regime and leading to the CMB effect.
In Fig.~\ref{fig:fig2}\hyperref[fig:fig2]{(b)}, the transition pathways under the combined influence of the probe field and the drive field are depicted. The probe field primarily induces transitions between magnon states, such as 
\(|g',0\rangle \xrightarrow{\Omega_d} |g',1\rangle \xrightarrow{\sqrt{2}\Omega_d} |g',2\rangle\). 
Meanwhile, the drive field facilitates transitions between magnon states and qubit states through coupling mechanisms, for example, 
\(|g',0\rangle \xrightarrow{\frac{\Omega_s}{2}} |e',0\rangle \xrightarrow{2^{-1}g_{ms}} |g',1\rangle \xrightarrow{\frac{\Omega_s}{2}} |e',1\rangle \xrightarrow{2^{-1/2}g_{ms}} |g',2\rangle\). 
These pathways may interfere coherently, suppressing the population of the quantum state \(|g',2\rangle\) and leading to the emergence of the UMB effect.
\subsection{Numerical Simulation of the Second-Order Correlation Function}
Considering the damping effects induced by the system-bath coupling, the dissipative dynamics of the hybrid skyrmion qubit-magnon system, with a focus on the magnon mode, can be described by the master equation \cite{Carmichael1999}
	\begin{equation}
		\begin{aligned}
			\frac{\partial \rho}{\partial t} = &-i[H_{\text{eff}}, \rho] + \frac{\kappa_m}{2} (n_{\text{th}} + 1)\mathcal{L}_m[\rho]  \\
			&+ \frac{\kappa_m}{2} n_{\text{th}} \mathcal{L}'_m[\rho] + \frac{\kappa_s}{2} \mathcal{L}_{\tilde{\sigma}}[\rho],
			\label{eq:3}
		\end{aligned} \tag{14}
	\end{equation}

Here, \( \kappa_s \) and \( \kappa_m \) denote the decay rates of the skyrmion qubit and magnon mode, with \( \kappa_s = \kappa_m \) assumed for simplicity. The Pauli operators \( \tilde{\sigma}_+ = | \psi_2 \rangle \langle \psi_1 | \) and \( \tilde{\sigma}_- = | \psi_1 \rangle \langle \psi_2 | \) are defined using the skyrmion qubit eigenstates \( | \psi_1 \rangle \) and \( | \psi_2 \rangle \). The system density matrix is \( \rho \), and the Lindblad superoperators are defined below
\begin{align}
		\mathcal{L}_m[\rho] &= 2m\rho m^{\dagger} - m^{\dagger} m\rho - \rho m^{\dagger} m, \tag{15} \\
		\mathcal{L}_{\tilde{\sigma}}[\rho] &= 2\tilde{\sigma}_- \rho \tilde{\sigma}_+ - \tilde{\sigma}_+ \tilde{\sigma}_- \rho - \rho \tilde{\sigma}_+ \tilde{\sigma}_-. \tag{16}
\end{align}
	
	The phenomenon of magnon blockade can be quantified using the second-order correlation function, given by \cite{Scully1997}
	\begin{equation}
		g^{(2)}(0) = \frac{\langle m^\dagger m^\dagger mm \rangle}{\langle m^\dagger m \rangle^2}. \tag{17}\label{eq:17}
	\end{equation}

If \( g^{(2)}(0) > 1 \), this indicates that the steady-state magnon number distribution follows super-Poissonian statistics, exhibiting magnon bunching. Conversely, if \( 0 < g^{(2)}(0) < 1 \), the magnon number distribution satisfies sub-Poissonian statistics, known as magnon antibunching \cite{PhysRevLett.107.063601,PhysRevA.82.032101,PhysRevA.99.013804}. Specifically, when \( g^{(2)}(0) \rightarrow 0 \), magnon blockade occurs.
\begin{figure}
	\centering
	\includegraphics[width=0.9\linewidth]{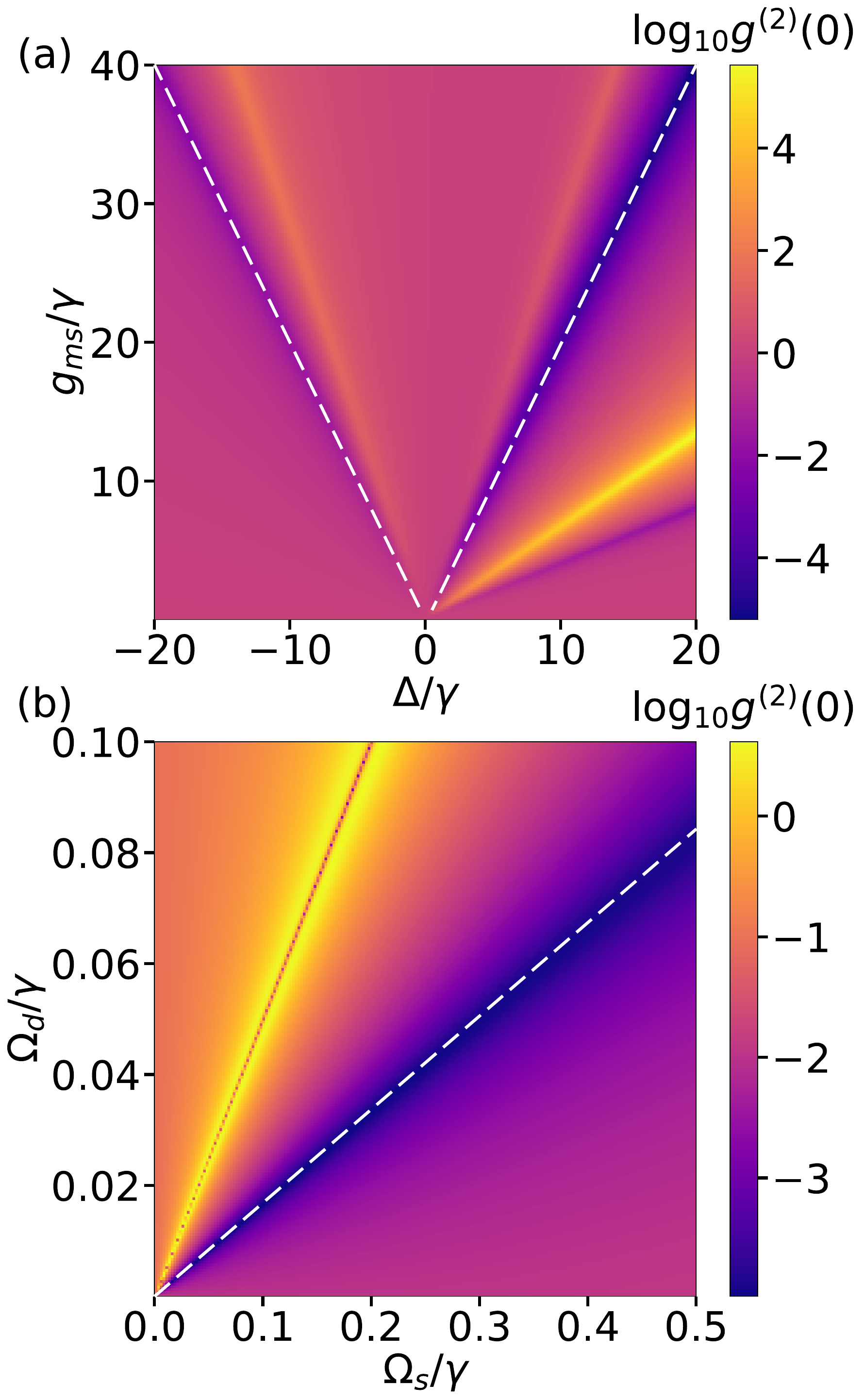}
	\caption{The steady-state equal-time second-order correlation function $\log_{10} g^{(2)}(0)$ varies with the driving detuning $\Delta/\gamma$ and the coupling strength $g_{ms}/\gamma$ in (a), and with the driving field $\Omega_s/\gamma$ and the probe field $\Omega_d/\gamma$ in (b). Frequencies are normalized to $\gamma = 2\pi \times 1$ MHz for consistency. In (a), the parameters are $\Omega_s/\gamma = 0.06$ and $\Omega_d/\gamma = 0.01$, while in (b), $\Delta/\gamma = 9.8$ and $g_{ms}/\gamma = 19.6$. The other parameters are $\kappa_m/\gamma = \kappa_s/\gamma = 1$ and $n_{th} = 0$.}
	\label{fig:fig3}
\end{figure}

To verify the optimal condition in Eq.~\eqref{eq:9}, we numerically solve the master equation \eqref{eq:3} by the Python package QuTiP \cite{JOHANSSON20131234,JOHANSSON20121760}. In Fig.~\ref{fig:fig3}\hyperref[fig:fig3]{(a)}, the steady-state second-order correlation function $\log_{10} g^{(2)}(0)$ depends on the drive detuning $\Delta / \gamma$ and the coupling strength $g_{ms} / \gamma$ between the magnon and the skyrmion qubit. The transition of \( g^{(2)}(0) \) from strong quantum correlations to classical correlations is illustrated by the color gradient, which ranges from deep purple to yellow. The system shows pronounced antibunching near the white dashed lines, where $\Delta / \gamma \approx \pm \frac{1}{2} g_{ms} / \gamma$, and \( g^{(2)}(0) \) reaches its minimum. This result highlights the emergence of the MB effect, aligning with the analytically derived the optimal condition. Fig.~\ref{fig:fig3}\hyperref[fig:fig3]{(b)} provides a detailed analysis of \( g^{(2)}(0) \) as a function of the driving field intensity \( \Omega_s/\gamma \) and the probe field intensity \( \Omega_d/\gamma \). As \( \Omega_s/\gamma \) and \( \Omega_d/\gamma \) increase, \( g^{(2)}(0) \) exhibits significant changes and reaches its minimum near the white dashed lines. The results indicate that the ratio \( \Omega_s/\Omega_d \approx 6 \) corresponds to the optimal condition for realizing the MB effect.

\begin{figure}
	\centering
	\includegraphics[width=0.9\linewidth]{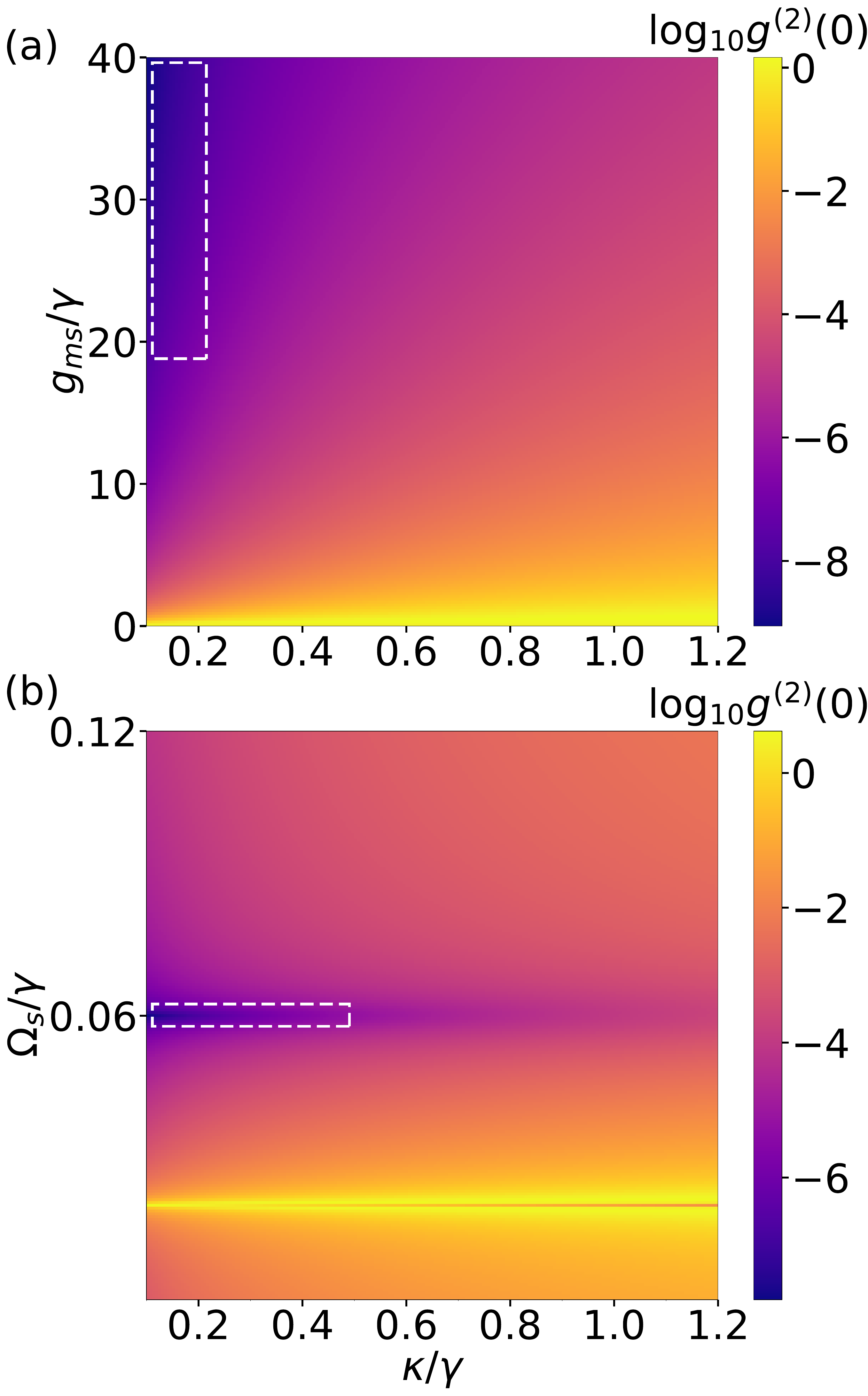}
	\caption{The steady-state $\log_{10} g^{(2)}(0)$ is plotted as a function of (a) the coupling strength $g_{ms}$ and (b) the driving field intensity $\Omega_s/\gamma$, as well as the dissipation rate $\kappa$, under the condition $\Delta/\gamma \approx \frac{1}{2} g_{ms}/\gamma$. In (a), $\Omega_s/\gamma = 0.06$ and $\Omega_d/\gamma = 0.01$, while in (b) $g_{ms}/\gamma = 19.6$, and $n_{th} = 0$.}
	\label{fig:fig4}
\end{figure}
Both the probe field and the driving field play a crucial role in enhancing the MB effect. In Fig.~\ref{fig:fig4}, we investigate the second-order correlation function \( g^{(2)}(0) \) as a function of the coupling strength \( g_{ms} \), driving field intensity \( \Omega_s \), and dissipation rate \( \kappa \). In Fig.~\ref{fig:fig4}\hyperref[fig:fig4]{(a)}, \( \log_{10} g^{(2)}(0) \) remains extremely low (deep purple) under conditions of high \( g_{ms} \) and low \( \kappa \), demonstrating a pronounced antibunching effect. With \( g_{ms}/\gamma = 19.6 \) fixed, Fig.~\ref{fig:fig4}\hyperref[fig:fig4]{(b)} explores the influence of \( \Omega_s \) and \( \kappa \) on \( g^{(2)}(0) \). For \( \Omega_d/\gamma = 0.01 \) and \( \Omega_s/\gamma \approx 0.06 \), \( \log_{10} g^{(2)}(0) \) achieves its minimum within the dissipation range \( \kappa/\gamma \) between 0.1 and 0.6, corresponding to the strongest MB effect.
This confirms that $\Omega_s / \Omega_d \approx 6$ is optimal for \( \kappa/\gamma \) in the range 0.1 to 1.2. These results provide the key parameter values for \( \Omega_s \), \( \kappa \), and \( \Omega_d \) that maximize the MB effect.
\begin{figure}
	\centering
	\includegraphics[width=0.9\linewidth, height=0.5\textheight]{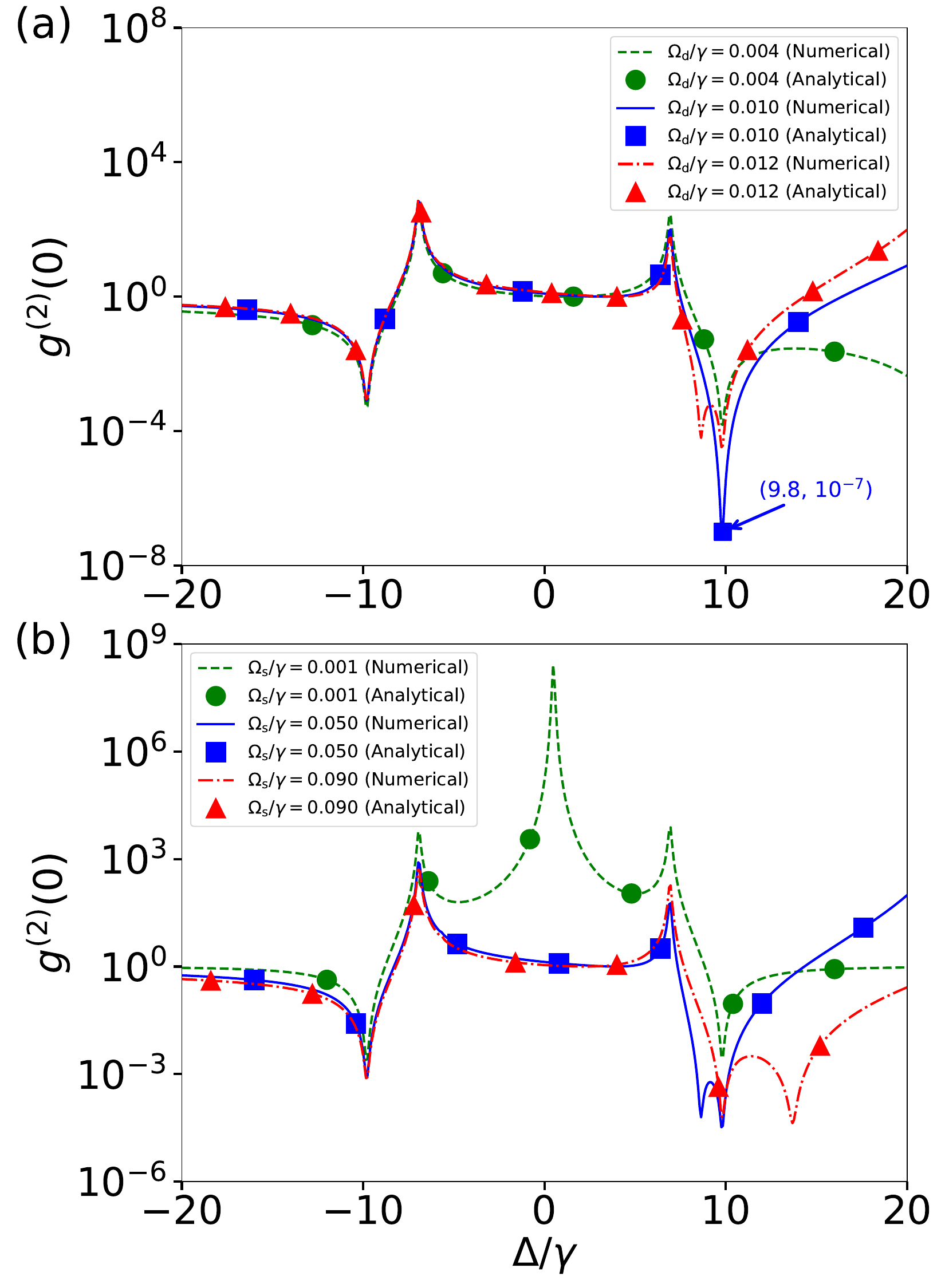}
	\caption{The equal-time second-order correlation function \( g^{(2)}(0) \) is shown as a function of the normalized detuning \( \Delta/\gamma \). As illustrated in (a), for the magnon drive field intensity \( \Omega_d/\gamma = 0.004 \), 0.01, and 0.012, the Skyrmion drive field \( \Omega_s/\gamma \) is kept constant at 0.06. In (b), for the skyrmion drive field intensity \( \Omega_s/\gamma = 0.001 \), 0.05, and 0.09, the magnon drive field \( \Omega_d/\gamma \) is fixed at 0.01. The coupling strength between the magnon and skyrmion modes is \( g_{\text{ms}}/\gamma = 19.6 \). The other system parameters are $\kappa_m/\gamma = \kappa_q/\gamma = 0.15$, and $n_{th} = 0$.}
	\label{fig:fig5}
\end{figure}

Next, we observe a significant dependence of the second-order correlation function \( g^{(2)}(0) \) on the driving detuning \( \Delta/\gamma \), as shown in in Fig.~\ref{fig:fig5}. In Fig.~\ref{fig:fig5}\hyperref[fig:fig5]{(a)}, the variation of \( g^{(2)}(0) \) is shown for different magnon driving field intensities (\( \Omega_d/\gamma = 0.004, 0.01, 0.012 \)), with corresponding ratios \( \Omega_s/\Omega_d = 15, 6, 5 \), while the skyrmion driving field intensity is fixed at \( \Omega_s/\gamma = 0.06 \). Around \( \Delta/\gamma \approx 9.8 \), \( g^{(2)}(0) \) reaches \( 10^{-7.3} \), indicating a pronounced antibunching effect and a high-purity magnon blockade state. However, around \( \Delta/\gamma \approx -9.8 \), the antibunching effect becomes less sensitive to modulation, and \( g^{(2)}(0) \) reaches \( 10^{-3.3} \). In Fig.~\ref{fig:fig5}\hyperref[fig:fig5]{(b)}, \( g^{(2)}(0) \) is plotted as a function of the skyrmion driving field intensity (\( \Omega_s/\gamma = 0.001, 0.05, 0.09 \)), with corresponding ratios \( \Omega_s/\Omega_d = 0.1, 5, 9 \), while the magnon driving field intensity is fixed at \( \Omega_d/\gamma = 0.01 \). In this parameter regime, the MB effect is present but weaker, with \( g^{(2)}(0) \) reaching \( 10^{-2.7} \), \( 10^{-3.2} \), and \( 10^{-3.4} \), respectively. Notably, in both Fig.~\ref{fig:fig5}\hyperref[fig:fig5]{(a)} and Fig.~\ref{fig:fig5}\hyperref[fig:fig5]{(b)}, the numerical and analytical results agree almost perfectly, demonstrating the reliability of the model and theoretical calculations. Further analysis confirms that the optimal conditions are \( \Omega_s/\Omega_d \approx 6 \), $\Delta / \gamma \approx \pm \frac{1}{2} g_{ms} / \gamma$.
	
\begin{figure}
	\centering
	\includegraphics[width=0.9\linewidth]{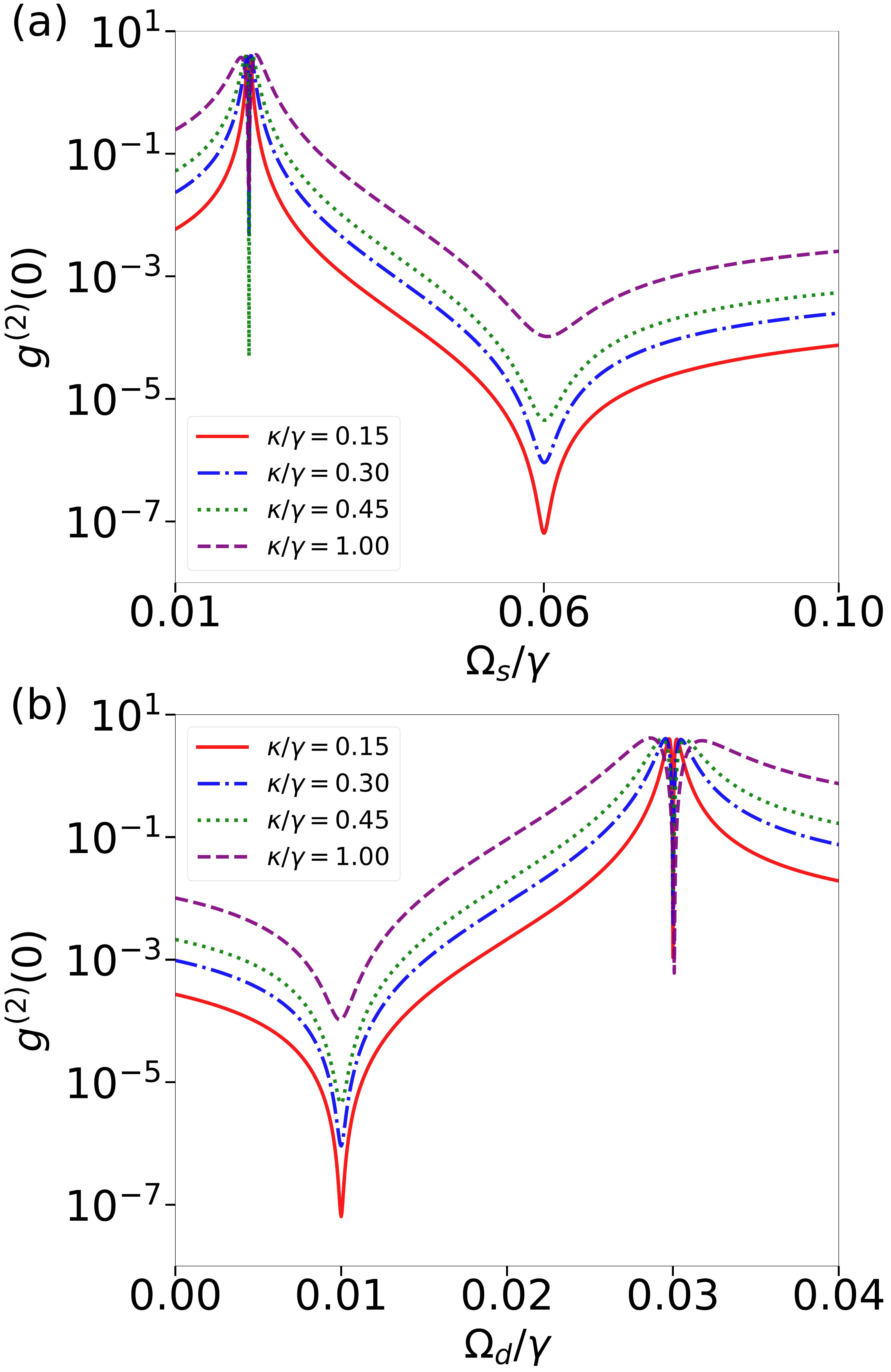}
	\caption{The dependence of the second-order correlation function $g^{(2)}(0)$ on both the skyrmion driving field $\Omega_s / \gamma$ and the magnon driving field $\Omega_d / \gamma$. In (a), $g^{(2)}(0)$ is plotted as a function of $\Omega_s / \gamma$ for detuning values $\kappa / \gamma = 0.15,0.30,0.45,1$, and $\Omega_d/\gamma = 0.01$. In (b), the behavior of $g^{(2)}(0)$ is depicted as a function of $\Omega_d / \gamma$ under the same detuning conditions, $\Omega_s/\gamma = 0.06$, while the other parameters are the same as those in Fig.~\ref{fig:fig5}.}
	\label{fig:fig6}
\end{figure}

In Fig.~\ref{fig:fig6}, we illustrate the relationship between \( \Omega_d \), \( g_{ms} \), \( \Omega_s \), and \( \kappa \), showing that the system achieves the strongest antibunching effect near \( \Omega_s/\gamma \approx 0.06 \) and \( \Omega_d/\gamma \approx 0.01 \). Stronger antibunching effects are observed at lower dissipation rates, such as \( \kappa/\gamma = 0.15 \), which is evidenced by the significantly reduced values of \( g^{(2)}(0) \). As the dissipation rate increases, the antibunching effect weakens; however, \( g^{(2)}(0) \) remains at a sufficiently low value even under higher dissipation conditions. For instance, when \( \kappa/\gamma = 1.00 \), \( g^{(2)}(0) \) can reach \( 10^{-4} \).

\begin{figure}
	\centering
	\includegraphics[width=0.9\linewidth]{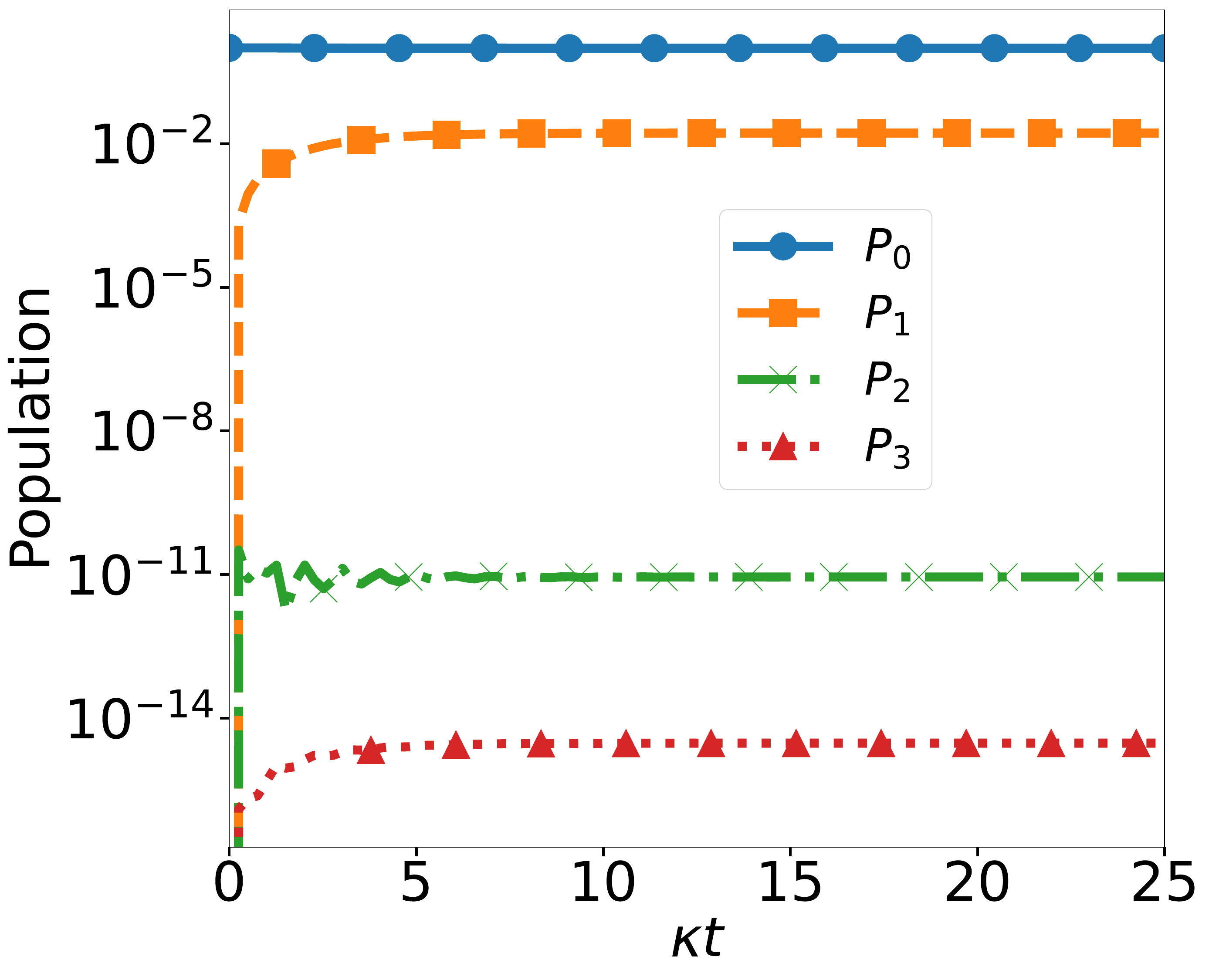}
	\caption{The population dynamics of the lowest few Fock states of the magnon mode under the optimal conditions \( \Delta/\gamma \approx \pm \frac{1}{2} g_\text{ms}/\gamma \) and \( \Omega_s = 6\Omega_d \). The parameters are \( g_\text{ms}/\gamma = 19.6 \), \( \Omega_d/\gamma = 0.01 \), and \( \kappa/\gamma = 0.15 \). The initial state is \( \rho(0) = |g'0\rangle \langle g'0| \).}
	\label{fig:fig7}
\end{figure}

Our approach to the MB effect aims to drive the magnon system into a quantum state where the first excited level is highly populated, while higher excited levels remain largely unoccupied, thereby suppressing multi-excitation events. This mechanism minimizes the second-order correlation function \( g^{(2)}(0) \), a key indicator of the MB effect. It is expressed as \( g^{(2)}(0) \approx 2P_2 / P_1^2 \), which aligns with Eqs.~\eqref{eq:8} and \eqref{eq:17}. Fig.~\ref{fig:fig7} shows the population dynamics of the lowest Fock states of the magnon mode under optimal conditions. The population of the single-excitation level $P_1$ is approximately $10^{-2}$, while the population of the double-excitation state $P_2$ is around  $10^{-11}$. With the direct coupling between magnons and qubits, a nearly perfect magnon blockade can be achieved \cite{PhysRevA.110.012459}, characterized by $P_1 \sim 0.40$ and $g^{(2)}(0) \sim 10^{-5}$. These results provide strong evidence of MB and highlight its potential for precise quantum control in hybrid quantum systems.

\section{discussions}
In the previous analysis, we investigated the case where the skyrmion qubit is solely modulated by an electric field ($E_z \approx 9.8 \, \text{GHz}$ and $K_0 \approx 0$) and observed a clear MB effect. This analysis provided valuable insights into the interaction mechanisms under such conditions. Moving forward, we consider another scenario where $E_z \approx 0$ and $K_0 \approx 9.8 \, \text{GHz}$. In this configuration, the system is designed to exclusively probe the magnon mode, with the external driving on the skyrmion qubit removed. The Hamiltonian is written as
\begin{equation}
		\begin{aligned}
			\hat{H}_{\text{total}}' = & \, w_m \hat{m}^\dagger \hat{m} + \frac{K_0}{2} \hat{\sigma}_z + \frac{\tilde{g}_{ms}}{2} (\hat{m} + \hat{m}^\dagger) \hat{\sigma}_x  \\
			& + \frac{g_{ms}}{2} (\hat{m} + \hat{m}^\dagger) \hat{\sigma}_z + \Omega_d (\hat{m}^\dagger e^{-i\omega_d t} + \hat{m} e^{i\omega_d t}).
		\end{aligned}
		\tag{18}
		\label{eq:total_hamiltonian11}
\end{equation}

Following the approach used in the previous analysis for handling the Hamiltonian, The effective Hamiltonian can then be written as
\[
\begin{aligned}
	\hat{H}_{\text{eff}}' = & \Delta_m \hat{m}^\dagger \hat{m} + \frac{\Delta_s}{2} \sigma_z+ \frac{\tilde{g}_{ms}}{2} \left( \hat{m} \sigma_+ + \hat{m}^\dagger \sigma_- \right) \\
	& + \Omega_d \left( \hat{m}^\dagger + \hat{m} \right),
\end{aligned}
\tag{19}
\]

where $\Delta_m \equiv \omega_m - \omega_d$ ($\Delta_s \equiv K_0 - \omega_d$) represents the detuning between the magnon mode and the driving field frequency, while $\Delta_s$ represents the detuning between the skyrmion qubit frequency and the driving frequency. Simultaneously, $\Delta = \Delta_s = \Delta_m$.
The statistical properties of the magnon are investigated by numerically solving the master equation\cite{Carmichael1999}
\begin{equation}
	\begin{aligned}
		\frac{\partial \rho}{\partial t} = &-i[\hat{H}_{\text{eff}}', \rho] + \frac{\kappa_m}{2} \mathcal{L}_m[\rho] + \frac{\kappa_s}{2} \mathcal{L}_{\sigma}[\rho],
		\label{eq:33}
	\end{aligned} \tag{20}
\end{equation}
the dissipator is defined as the Lindblad superoperators \( \mathcal{L}[\rho] = 2o \rho o^\dagger - o^\dagger o \rho - \rho o^\dagger o \), with \( o = \hat{m}, \sigma_{-} \) indicating, respectively.

\begin{figure}
	\centering
	\includegraphics[width=0.9\linewidth]{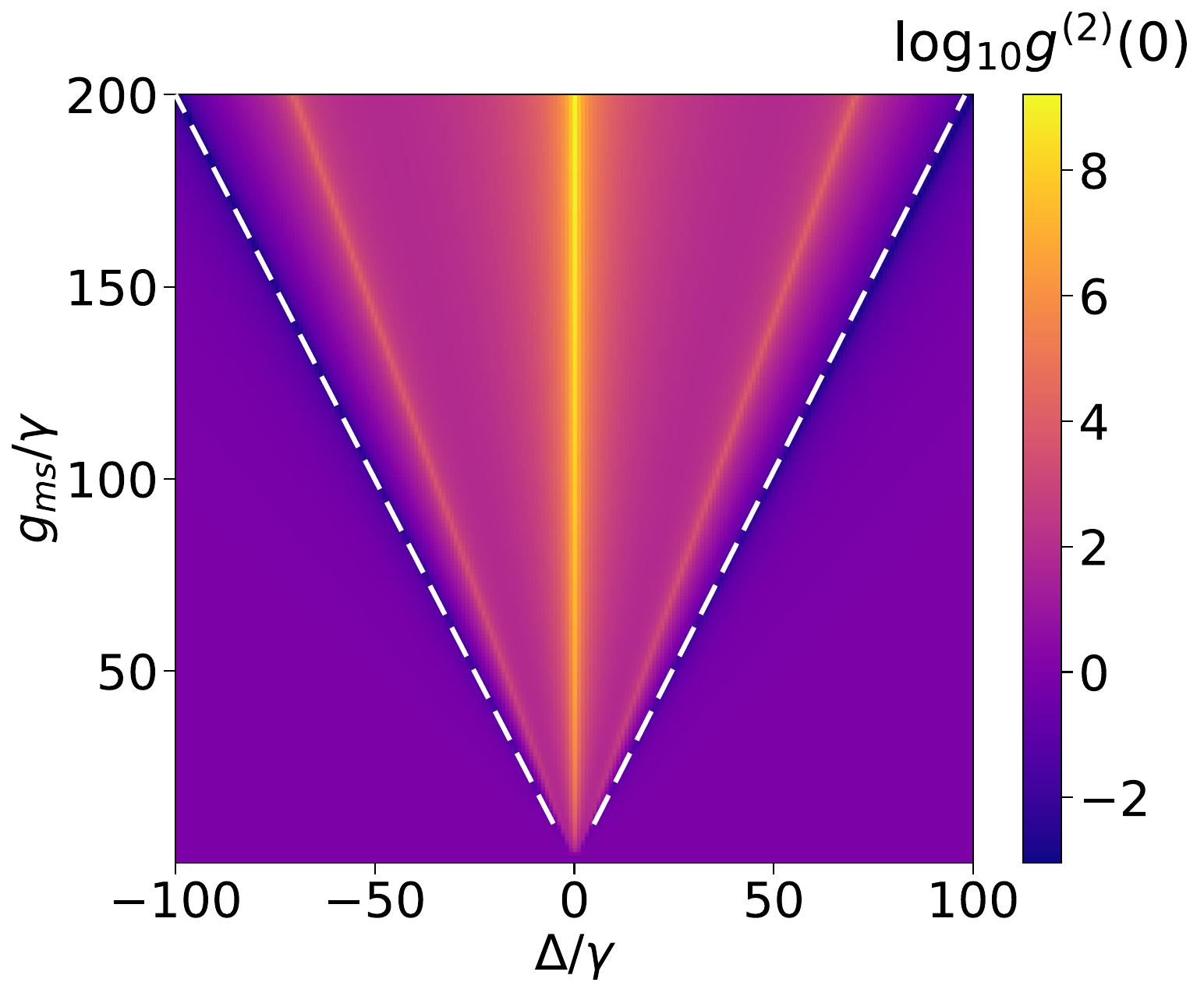}
	\caption{The variation of the steady-state equal-time second-order correlation function $\log_{10} g^{(2)}(0)$ is shown as a function of the driving detuning $\Delta/\gamma$ and the skyrmion qubit-magnon coupling strength $\tilde{g}_{ms}/\gamma$. The probe field intensity of the magnons is fixed at $\Omega_d/\gamma = 0.01$, while other system parameters are set to $\kappa_m/\gamma = \kappa_q/\gamma = 1$.}
	\label{fig:fig8}
\end{figure}
The Fig.~\ref{fig:fig8} shows the variation of the steady-state equal-time second-order correlation function $\log_{10} g^{(2)}(0)$ as a function of the driving detuning $\Delta/\gamma$ and the skyrmion qubit-magnon coupling strength $g_{ms}/\gamma$. It can also be observed that under the optimal conditions where $\Delta/\gamma \approx \pm \frac{1}{2} \tilde{g}_{ms}/\gamma$, $\log_{10} g^{(2)}(0)$ reaches -2, indicating the occurrence of a strong MB effect. 
\begin{figure}
	\centering
	\includegraphics[width=0.9\linewidth]{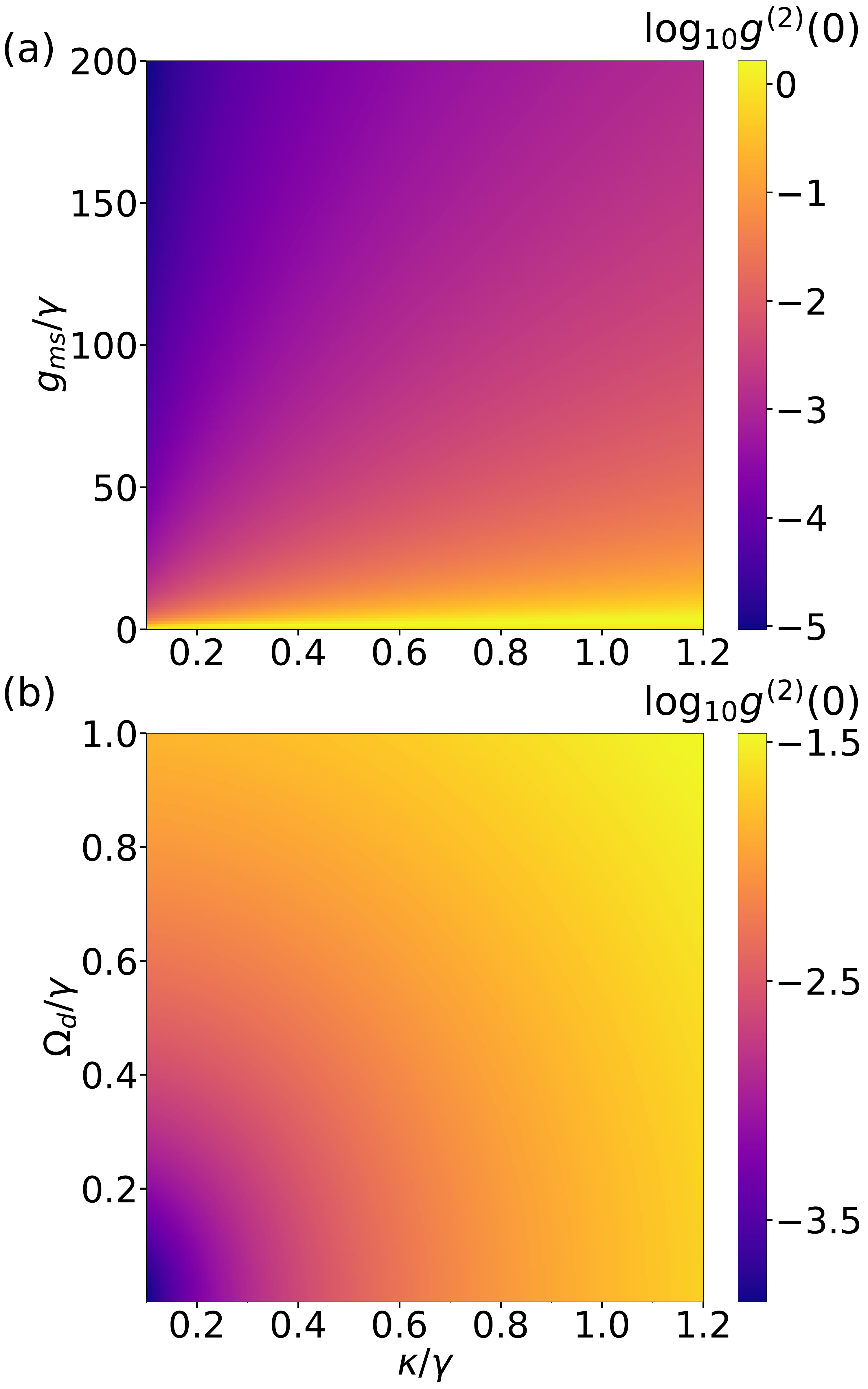}
	\caption{The steady-state $\log_{10} g^{(2)}(0)$ is presented as a function of (a) the skyrmion-qubit coupling strength $g_{ms}$ and (b) the control field intensity $\Omega_s/\gamma$, along with the dissipation rate $\kappa$, under the condition $\Delta/\gamma \approx \pm \frac{1}{2} \tilde{g}_{ms}/\gamma$. In panel (a), $\Omega_d/\gamma = 0.01$ is set, while in panel (b), $g_{ms}/\gamma = 50.1$ is chosen, and $n_{th} = 0$.}
	\label{fig:fig9}
\end{figure}

In the analysis of Fig.~\ref{fig:fig9}\hyperref[fig:fig9]{(a)}, the effects of the dissipation rate \(k\) and the skyrmion qubit-magnon coupling strength \(\tilde{g}_{ms}\) on the system’s behavior are explored. The result indicates that, under high coupling strength and low dissipation rate \(k\), \(\log_{10} g^{(2)}(0)\) remains at a low value of approximately -5.
In Fig.~\ref{fig:fig9}\hyperref[fig:fig9]{(b)}, the influence of the driving intensity \(\Omega_d\) on the system is further investigated. When the dissipation rate \(k\) is fixed, small variations in the driving intensity (e.g., \(0.001 \leq \Omega_d / \gamma \leq 0.01\)) have little effect on the system, with \(\log_{10} g^{(2)}(0)\) remaining stable at around -3.5. However, with a larger variation in the driving intensity, the system exhibits pronounced fluctuations, leading to noticeable changes in the second-order correlation function. For instance, as \(\Omega_d / \gamma\) increases from 0.1 to 0.6, \(\log_{10} g^{(2)}(0)\) rises from -3.5 to -1.5. Notably, even in the absence of an external drive applied to the skyrmion qubit, applying a drive solely to the magnons effectively induces a MB effect.
\section{conclusions}
In conclusion, we propose a scheme to realize a high-purity single-magnon state in a skyrmion qubit-magnon system. We consider two scenarios: one where the magnon is probed and the skyrmion qubit is driven, with the skyrmion qubit modulated by an electric field (\( E_z \approx 9.8 \, \text{GHz} \) and \( K_0 \approx 0 \)); and another where only the magnon is probed, with the skyrmion qubit modulated by a magnetic field (\( E_z \approx 0 \) and \( K_0 \approx 9.8 \, \text{GHz} \)). In the first scenario, we derive the analytical expression for the second-order correlation function and the optimal conditions through theoretical analysis and numerical simulations. Specifically, the detuning is one-half of the transverse coupling strength of the skyrmion qubit, and the driving field strength applied to the skyrmion qubit is six times the probing field strength applied to the magnon. Under these optimal conditions, and with the thermal magnon occupation number $n_{\text{th}} = 0$, a strong MB effect is observed, with the second-order correlation function $g^{(2)}(0) \sim 10^{-7}$. This demonstrates that detuning, driving field strength, and probing field strength significantly influence the enhancement of the MB effect. Moreover, we analyze the physical mechanism responsible for the magnon blockade and confirm the coexistence of CMB and UMB. In the second scenario, by appropriately tuning the coupling strength \( g_{\text{ms}} \) and the probing field \( \Omega_d \), the MB effect is also observed. Further analysis of the system's dissipation, driving fields shows that, under suitable parameter settings, the MB effect is still observed, with \( g^{(2)}(0) \) reaching approximately \( 10^{-2} \). These findings establish a theoretical framework for studying and advancing the potential realization of single-magnon emitters in hybrid quantum systems.

Additionally, the versatility of this platform allows for further exploration of its applicability in broader quantum technologies. By leveraging the tunability of coupling parameters and system configurations, this scheme may inspire the design of scalable quantum architectures and novel quantum devices. These advancements could contribute to enriching the functionality of hybrid quantum systems and unlocking new possibilities in magnon-based quantum technologies.
\section*{Acknowledgments}
	
We thank Feng-Yang Zhang, Mingsong Ding, and Ye-Ting Yan for constructive discussions. This work was supported by the National Natural Science Foundation of China (Grants No. 12274053).
	
\bibliographystyle{apsrevlong}
\bibliography{ref2}

\begin{thebibliography}{118}%
\makeatletter
\providecommand \@ifxundefined [1]{%
 \@ifx{#1\undefined}
}%
\providecommand \@ifnum [1]{%
 \ifnum #1\expandafter \@firstoftwo
 \else \expandafter \@secondoftwo
 \fi
}%
\providecommand \@ifx [1]{%
 \ifx #1\expandafter \@firstoftwo
 \else \expandafter \@secondoftwo
 \fi
}%
\providecommand \natexlab [1]{#1}%
\providecommand \enquote  [1]{``#1''}%
\providecommand \bibnamefont  [1]{#1}%
\providecommand \bibfnamefont [1]{#1}%
\providecommand \citenamefont [1]{#1}%
\providecommand \href@noop [0]{\@secondoftwo}%
\providecommand \href [0]{\begingroup \@sanitize@url \@href}%
\providecommand \@href[1]{\@@startlink{#1}\@@href}%
\providecommand \@@href[1]{\endgroup#1\@@endlink}%
\providecommand \@sanitize@url [0]{\catcode `\\12\catcode `\$12\catcode
  `\&12\catcode `\#12\catcode `\^12\catcode `\_12\catcode `\%12\relax}%
\providecommand \@@startlink[1]{}%
\providecommand \@@endlink[0]{}%
\providecommand \url  [0]{\begingroup\@sanitize@url \@url }%
\providecommand \@url [1]{\endgroup\@href {#1}{\urlprefix }}%
\providecommand \urlprefix  [0]{URL }%
\providecommand \Eprint [0]{\href }%
\providecommand \doibase [0]{http://dx.doi.org/}%
\providecommand \selectlanguage [0]{\@gobble}%
\providecommand \bibinfo  [0]{\@secondoftwo}%
\providecommand \bibfield  [0]{\@secondoftwo}%
\providecommand \translation [1]{[#1]}%
\providecommand \BibitemOpen [0]{}%
\providecommand \bibitemStop [0]{}%
\providecommand \bibitemNoStop [0]{.\EOS\space}%
\providecommand \EOS [0]{\spacefactor3000\relax}%
\providecommand \BibitemShut  [1]{\csname bibitem#1\endcsname}%
\let\auto@bib@innerbib\@empty
\bibitem [{\citenamefont {Wallquist}\ \emph {et~al.}(2009)\citenamefont
  {Wallquist}, \citenamefont {Hammerer}, \citenamefont {Rabl}, \citenamefont
  {Lukin},\ and\ \citenamefont
  {Zoller}}]{doi:10.1088/0031-8949/2009/t137/014001}%
  \BibitemOpen
  \bibfield  {author} {\bibinfo {author} {\bibfnamefont {M.}~\bibnamefont
  {Wallquist}}, \bibinfo {author} {\bibfnamefont {K.}~\bibnamefont {Hammerer}},
  \bibinfo {author} {\bibfnamefont {P.}~\bibnamefont {Rabl}}, \bibinfo {author}
  {\bibfnamefont {M.}~\bibnamefont {Lukin}}, \ and\ \bibinfo {author}
  {\bibfnamefont {P.}~\bibnamefont {Zoller}},\ }\bibfield  {title} {\emph
  {\bibinfo {title} {\textnormal{Hybrid quantum devices and quantum
  engineering}},\ }}\href {\doibase 10.1088/0031-8949/2009/t137/014001}
  {\bibfield  {journal} {\bibinfo  {journal} {Phys. Scr.}\ }\textbf {\bibinfo
  {volume} {T137}},\ \bibinfo {pages} {014001} (\bibinfo {year}
  {2009})}\BibitemShut {NoStop}%
\bibitem [{\citenamefont {Kimble}(2008)}]{doi:10.1038/nature07127}%
  \BibitemOpen
  \bibfield  {author} {\bibinfo {author} {\bibfnamefont {H.~J.}\ \bibnamefont
  {Kimble}},\ }\bibfield  {title} {\emph {\bibinfo {title} {\textnormal{The
  quantum internet}},\ }}\href {\doibase 10.1038/nature07127} {\bibfield
  {journal} {\bibinfo  {journal} {Nature}\ }\textbf {\bibinfo {volume} {453}},\
  \bibinfo {pages} {1023} (\bibinfo {year} {2008})}\BibitemShut {NoStop}%
\bibitem [{\citenamefont {Xiang}\ \emph {et~al.}(2013)\citenamefont {Xiang},
  \citenamefont {Ashhab}, \citenamefont {You},\ and\ \citenamefont
  {Nori}}]{RevModPhys.85.623}%
  \BibitemOpen
  \bibfield  {author} {\bibinfo {author} {\bibfnamefont {Z.-L.}\ \bibnamefont
  {Xiang}}, \bibinfo {author} {\bibfnamefont {S.}~\bibnamefont {Ashhab}},
  \bibinfo {author} {\bibfnamefont {J.~Q.}\ \bibnamefont {You}}, \ and\
  \bibinfo {author} {\bibfnamefont {F.}~\bibnamefont {Nori}},\ }\bibfield
  {title} {\emph {\bibinfo {title} {\textnormal{Hybrid quantum circuits:
  Superconducting circuits interacting with other quantum systems}},\ }}\href
  {\doibase 10.1103/RevModPhys.85.623} {\bibfield  {journal} {\bibinfo
  {journal} {Rev. Mod. Phys.}\ }\textbf {\bibinfo {volume} {85}},\ \bibinfo
  {pages} {623} (\bibinfo {year} {2013})}\BibitemShut {NoStop}%
\bibitem [{\citenamefont {Kurizki}\ \emph {et~al.}(2015)\citenamefont
  {Kurizki}, \citenamefont {Bertet}, \citenamefont {Kubo}, \citenamefont
  {Mølmer}, \citenamefont {Petrosyan}, \citenamefont {Rabl},\ and\
  \citenamefont {Schmiedmayer}}]{doi:10.1073/pnas.1419326112}%
  \BibitemOpen
  \bibfield  {author} {\bibinfo {author} {\bibfnamefont {G.}~\bibnamefont
  {Kurizki}}, \bibinfo {author} {\bibfnamefont {P.}~\bibnamefont {Bertet}},
  \bibinfo {author} {\bibfnamefont {Y.}~\bibnamefont {Kubo}}, \bibinfo {author}
  {\bibfnamefont {K.}~\bibnamefont {Mølmer}}, \bibinfo {author} {\bibfnamefont
  {D.}~\bibnamefont {Petrosyan}}, \bibinfo {author} {\bibfnamefont
  {P.}~\bibnamefont {Rabl}}, \ and\ \bibinfo {author} {\bibfnamefont
  {J.}~\bibnamefont {Schmiedmayer}},\ }\bibfield  {title} {\emph {\bibinfo
  {title} {\textnormal{Quantum technologies with hybrid systems}},\ }}\href
  {https://www.pnas.org/doi/abs/10.1073/pnas.1419326112} {\bibfield  {journal}
  {\bibinfo  {journal} {Proc. Natl. Acad. Sci. U.S.A.}\ }\textbf {\bibinfo
  {volume} {112}},\ \bibinfo {pages} {3866} (\bibinfo {year}
  {2015})}\BibitemShut {NoStop}%
\bibitem [{\citenamefont {Chen}\ \emph {et~al.}(2023)\citenamefont {Chen},
  \citenamefont {Fan}, \citenamefont {Xiong}, \citenamefont {Wang},\ and\
  \citenamefont {Ye}}]{PhysRevB.108.024105}%
  \BibitemOpen
  \bibfield  {author} {\bibinfo {author} {\bibfnamefont {J.}~\bibnamefont
  {Chen}}, \bibinfo {author} {\bibfnamefont {X.-G.}\ \bibnamefont {Fan}},
  \bibinfo {author} {\bibfnamefont {W.}~\bibnamefont {Xiong}}, \bibinfo
  {author} {\bibfnamefont {D.}~\bibnamefont {Wang}}, \ and\ \bibinfo {author}
  {\bibfnamefont {L.}~\bibnamefont {Ye}},\ }\bibfield  {title} {\emph {\bibinfo
  {title} {\textnormal{Nonreciprocal entanglement in cavity-magnon
  optomechanics}},\ }}\href {\doibase 10.1103/PhysRevB.108.024105} {\bibfield
  {journal} {\bibinfo  {journal} {Phys. Rev. B}\ }\textbf {\bibinfo {volume}
  {108}},\ \bibinfo {pages} {024105} (\bibinfo {year} {2023})}\BibitemShut
  {NoStop}%
\bibitem [{\citenamefont {Xiong}\ \emph {et~al.}(2023)\citenamefont {Xiong},
  \citenamefont {Wang}, \citenamefont {Zhang},\ and\ \citenamefont
  {Chen}}]{PhysRevA.107.033516}%
  \BibitemOpen
  \bibfield  {author} {\bibinfo {author} {\bibfnamefont {W.}~\bibnamefont
  {Xiong}}, \bibinfo {author} {\bibfnamefont {M.}~\bibnamefont {Wang}},
  \bibinfo {author} {\bibfnamefont {G.-Q.}\ \bibnamefont {Zhang}}, \ and\
  \bibinfo {author} {\bibfnamefont {J.}~\bibnamefont {Chen}},\ }\bibfield
  {title} {\emph {\bibinfo {title}
  {\textnormal{Optomechanical-interface-induced strong spin-magnon coupling}},\
  }}\href {\doibase 10.1103/PhysRevA.107.033516} {\bibfield  {journal}
  {\bibinfo  {journal} {Phys. Rev. A}\ }\textbf {\bibinfo {volume} {107}},\
  \bibinfo {pages} {033516} (\bibinfo {year} {2023})}\BibitemShut {NoStop}%
\bibitem [{\citenamefont {Zuo}\ \emph {et~al.}(2024)\citenamefont {Zuo},
  \citenamefont {Fan}, \citenamefont {Qian}, \citenamefont {Ding},
  \citenamefont {Tan}, \citenamefont {Xiong},\ and\ \citenamefont
  {Li}}]{doi:10.1088/1367-2630/ad327c}%
  \BibitemOpen
  \bibfield  {author} {\bibinfo {author} {\bibfnamefont {X.}~\bibnamefont
  {Zuo}}, \bibinfo {author} {\bibfnamefont {Z.-Y.}\ \bibnamefont {Fan}},
  \bibinfo {author} {\bibfnamefont {H.}~\bibnamefont {Qian}}, \bibinfo {author}
  {\bibfnamefont {M.-S.}\ \bibnamefont {Ding}}, \bibinfo {author}
  {\bibfnamefont {H.}~\bibnamefont {Tan}}, \bibinfo {author} {\bibfnamefont
  {H.}~\bibnamefont {Xiong}}, \ and\ \bibinfo {author} {\bibfnamefont
  {J.}~\bibnamefont {Li}},\ }\bibfield  {title} {\emph {\bibinfo {title}
  {\textnormal{Cavity magnomechanics: from classical to quantum}},\ }}\href
  {\doibase 10.1088/1367-2630/ad327c} {\bibfield  {journal} {\bibinfo
  {journal} {New Journal of Physics}\ }\textbf {\bibinfo {volume} {26}},\
  \bibinfo {pages} {031201} (\bibinfo {year} {2024})}\BibitemShut {NoStop}%
\bibitem [{\citenamefont {Osada}\ \emph {et~al.}(2016)\citenamefont {Osada},
  \citenamefont {Hisatomi}, \citenamefont {Noguchi}, \citenamefont {Tabuchi},
  \citenamefont {Yamazaki}, \citenamefont {Usami}, \citenamefont {Sadgrove},
  \citenamefont {Yalla}, \citenamefont {Nomura},\ and\ \citenamefont
  {Nakamura}}]{PhysRevLett.116.223601}%
  \BibitemOpen
  \bibfield  {author} {\bibinfo {author} {\bibfnamefont {A.}~\bibnamefont
  {Osada}}, \bibinfo {author} {\bibfnamefont {R.}~\bibnamefont {Hisatomi}},
  \bibinfo {author} {\bibfnamefont {A.}~\bibnamefont {Noguchi}}, \bibinfo
  {author} {\bibfnamefont {Y.}~\bibnamefont {Tabuchi}}, \bibinfo {author}
  {\bibfnamefont {R.}~\bibnamefont {Yamazaki}}, \bibinfo {author}
  {\bibfnamefont {K.}~\bibnamefont {Usami}}, \bibinfo {author} {\bibfnamefont
  {M.}~\bibnamefont {Sadgrove}}, \bibinfo {author} {\bibfnamefont
  {R.}~\bibnamefont {Yalla}}, \bibinfo {author} {\bibfnamefont
  {M.}~\bibnamefont {Nomura}}, \ and\ \bibinfo {author} {\bibfnamefont
  {Y.}~\bibnamefont {Nakamura}},\ }\bibfield  {title} {\emph {\bibinfo {title}
  {\textnormal{Cavity Optomagnonics with Spin-Orbit Coupled Photons}},\ }}\href
  {\doibase 10.1103/PhysRevLett.116.223601} {\bibfield  {journal} {\bibinfo
  {journal} {Phys. Rev. Lett.}\ }\textbf {\bibinfo {volume} {116}},\ \bibinfo
  {pages} {223601} (\bibinfo {year} {2016})}\BibitemShut {NoStop}%
\bibitem [{\citenamefont {Wolski}\ \emph {et~al.}(2020)\citenamefont {Wolski},
  \citenamefont {Lachance-Quirion}, \citenamefont {Tabuchi}, \citenamefont
  {Kono}, \citenamefont {Noguchi}, \citenamefont {Usami},\ and\ \citenamefont
  {Nakamura}}]{PhysRevLett.125.117701}%
  \BibitemOpen
  \bibfield  {author} {\bibinfo {author} {\bibfnamefont {S.~P.}\ \bibnamefont
  {Wolski}}, \bibinfo {author} {\bibfnamefont {D.}~\bibnamefont
  {Lachance-Quirion}}, \bibinfo {author} {\bibfnamefont {Y.}~\bibnamefont
  {Tabuchi}}, \bibinfo {author} {\bibfnamefont {S.}~\bibnamefont {Kono}},
  \bibinfo {author} {\bibfnamefont {A.}~\bibnamefont {Noguchi}}, \bibinfo
  {author} {\bibfnamefont {K.}~\bibnamefont {Usami}}, \ and\ \bibinfo {author}
  {\bibfnamefont {Y.}~\bibnamefont {Nakamura}},\ }\bibfield  {title} {\emph
  {\bibinfo {title} {\textnormal{Dissipation-Based Quantum Sensing of Magnons
  with a Superconducting Qubit}},\ }}\href {\doibase
  10.1103/PhysRevLett.125.117701} {\bibfield  {journal} {\bibinfo  {journal}
  {Phys. Rev. Lett.}\ }\textbf {\bibinfo {volume} {125}},\ \bibinfo {pages}
  {117701} (\bibinfo {year} {2020})}\BibitemShut {NoStop}%
\bibitem [{\citenamefont {Trifunovic}\ \emph {et~al.}(2013)\citenamefont
  {Trifunovic}, \citenamefont {Pedrocchi},\ and\ \citenamefont
  {Loss}}]{PhysRevX.3.041023}%
  \BibitemOpen
  \bibfield  {author} {\bibinfo {author} {\bibfnamefont {L.}~\bibnamefont
  {Trifunovic}}, \bibinfo {author} {\bibfnamefont {F.~L.}\ \bibnamefont
  {Pedrocchi}}, \ and\ \bibinfo {author} {\bibfnamefont {D.}~\bibnamefont
  {Loss}},\ }\bibfield  {title} {\emph {\bibinfo {title}
  {\textnormal{Long-Distance Entanglement of Spin Qubits via Ferromagnet}},\
  }}\href {\doibase 10.1103/PhysRevX.3.041023} {\bibfield  {journal} {\bibinfo
  {journal} {Phys. Rev. X}\ }\textbf {\bibinfo {volume} {3}},\ \bibinfo {pages}
  {041023} (\bibinfo {year} {2013})}\BibitemShut {NoStop}%
\bibitem [{\citenamefont {Fukami}\ \emph {et~al.}(2021)\citenamefont {Fukami},
  \citenamefont {Candido}, \citenamefont {Awschalom},\ and\ \citenamefont
  {Flatt\'e}}]{PRXQuantum.2.040314}%
  \BibitemOpen
  \bibfield  {author} {\bibinfo {author} {\bibfnamefont {M.}~\bibnamefont
  {Fukami}}, \bibinfo {author} {\bibfnamefont {D.~R.}\ \bibnamefont {Candido}},
  \bibinfo {author} {\bibfnamefont {D.~D.}\ \bibnamefont {Awschalom}}, \ and\
  \bibinfo {author} {\bibfnamefont {M.~E.}\ \bibnamefont {Flatt\'e}},\
  }\bibfield  {title} {\emph {\bibinfo {title} {\textnormal{Opportunities for
  Long-Range Magnon-Mediated Entanglement of Spin Qubits via On- and
  Off-Resonant Coupling}},\ }}\href {\doibase 10.1103/PRXQuantum.2.040314}
  {\bibfield  {journal} {\bibinfo  {journal} {Phys. Rev. X Quantum}\ }\textbf
  {\bibinfo {volume} {2}},\ \bibinfo {pages} {040314} (\bibinfo {year}
  {2021})}\BibitemShut {NoStop}%
\bibitem [{\citenamefont {Kippenberg}\ and\ \citenamefont
  {Vahala}(2008)}]{10.1126/science.1156032}%
  \BibitemOpen
  \bibfield  {author} {\bibinfo {author} {\bibfnamefont {T.~J.}\ \bibnamefont
  {Kippenberg}}\ and\ \bibinfo {author} {\bibfnamefont {K.~J.}\ \bibnamefont
  {Vahala}},\ }\bibfield  {title} {\emph {\bibinfo {title} {\textnormal {Cavity
  Optomechanics: Back-Action at the Mesoscale}},\ }}\href {\doibase
  10.1126/science.1156032} {\bibfield  {journal} {\bibinfo  {journal}
  {Science}\ }\textbf {\bibinfo {volume} {321}},\ \bibinfo {pages} {1172}
  (\bibinfo {year} {2008})}\BibitemShut {NoStop}%
\bibitem [{\citenamefont {Meystre}(2013)}]{10.1002/andp.201200226}%
  \BibitemOpen
  \bibfield  {author} {\bibinfo {author} {\bibfnamefont {P.}~\bibnamefont
  {Meystre}},\ }\bibfield  {title} {\emph {\bibinfo {title} {\textnormal {A
  Short Walk Through Quantum Optomechanics}},\ }}\href {\doibase
  https://doi.org/10.1002/andp.201200226} {\bibfield  {journal} {\bibinfo
  {journal} {Ann. Phys.}\ }\textbf {\bibinfo {volume} {525}},\ \bibinfo {pages}
  {215} (\bibinfo {year} {2013})}\BibitemShut {NoStop}%
\bibitem [{\citenamefont {Aspelmeyer}\ \emph {et~al.}(2014)\citenamefont
  {Aspelmeyer}, \citenamefont {Kippenberg},\ and\ \citenamefont
  {Marquardt}}]{10.1103/RevModPhys.86.1391}%
  \BibitemOpen
  \bibfield  {author} {\bibinfo {author} {\bibfnamefont {M.}~\bibnamefont
  {Aspelmeyer}}, \bibinfo {author} {\bibfnamefont {T.~J.}\ \bibnamefont
  {Kippenberg}}, \ and\ \bibinfo {author} {\bibfnamefont {F.}~\bibnamefont
  {Marquardt}},\ }\bibfield  {title} {\emph {\bibinfo {title} {\textnormal
  {Cavity Optomechanics}},\ }}\href {\doibase 10.1103/RevModPhys.86.1391}
  {\bibfield  {journal} {\bibinfo  {journal} {Rev. Mod. Phys.}\ }\textbf
  {\bibinfo {volume} {86}},\ \bibinfo {pages} {1391} (\bibinfo {year}
  {2014})}\BibitemShut {NoStop}%
\bibitem [{\citenamefont {Bowen}\ and\ \citenamefont
  {Milburn}(2015)}]{Bowen2015}%
  \BibitemOpen
  \bibfield  {author} {\bibinfo {author} {\bibfnamefont {W.~P.}\ \bibnamefont
  {Bowen}}\ and\ \bibinfo {author} {\bibfnamefont {G.~J.}\ \bibnamefont
  {Milburn}},\ }\href@noop {} {\emph {\bibinfo {title} {\textnormal {Quantum
  Optomechanics}}}}\ (\bibinfo  {publisher} {CRC Press},\ \bibinfo {address}
  {Boca Raton},\ \bibinfo {year} {2015})\BibitemShut {NoStop}%
\bibitem [{\citenamefont {Lai}\ \emph {et~al.}(2022)\citenamefont {Lai},
  \citenamefont {Liao}, \citenamefont {Miranowicz},\ and\ \citenamefont
  {Nori}}]{PhysRevLett.129.063602}%
  \BibitemOpen
  \bibfield  {author} {\bibinfo {author} {\bibfnamefont {D.-G.}\ \bibnamefont
  {Lai}}, \bibinfo {author} {\bibfnamefont {J.-Q.}\ \bibnamefont {Liao}},
  \bibinfo {author} {\bibfnamefont {A.}~\bibnamefont {Miranowicz}}, \ and\
  \bibinfo {author} {\bibfnamefont {F.}~\bibnamefont {Nori}},\ }\bibfield
  {title} {\emph {\bibinfo {title} {\textnormal {Noise-Tolerant Optomechanical
  Entanglement via Synthetic Magnetism}},\ }}\href {\doibase
  10.1103/PhysRevLett.129.063602} {\bibfield  {journal} {\bibinfo  {journal}
  {Phys. Rev. Lett.}\ }\textbf {\bibinfo {volume} {129}},\ \bibinfo {pages}
  {063602} (\bibinfo {year} {2022})}\BibitemShut {NoStop}%
\bibitem [{\citenamefont {Zeng}\ \emph {et~al.}(2023)\citenamefont {Zeng},
  \citenamefont {Zhou}, \citenamefont {Rinaldi}, \citenamefont {Gneiting},\
  and\ \citenamefont {Nori}}]{10.1103/PhysRevLett.131.050601}%
  \BibitemOpen
  \bibfield  {author} {\bibinfo {author} {\bibfnamefont {Y.}~\bibnamefont
  {Zeng}}, \bibinfo {author} {\bibfnamefont {Z.-Y.}\ \bibnamefont {Zhou}},
  \bibinfo {author} {\bibfnamefont {E.}~\bibnamefont {Rinaldi}}, \bibinfo
  {author} {\bibfnamefont {C.}~\bibnamefont {Gneiting}}, \ and\ \bibinfo
  {author} {\bibfnamefont {F.}~\bibnamefont {Nori}},\ }\bibfield  {title}
  {\emph {\bibinfo {title} {\textnormal {Approximate Autonomous Quantum Error
  Correction with Reinforcement Learning}},\ }}\href {\doibase
  10.1103/PhysRevLett.131.050601} {\bibfield  {journal} {\bibinfo  {journal}
  {Phys. Rev. Lett.}\ }\textbf {\bibinfo {volume} {131}},\ \bibinfo {pages}
  {050601} (\bibinfo {year} {2023})}\BibitemShut {NoStop}%
\bibitem [{\citenamefont {Serga}\ \emph {et~al.}(2010)\citenamefont {Serga},
  \citenamefont {Chumak},\ and\ \citenamefont
  {Hillebrands}}]{doi:10.1088/0022-3727/43/26/264002}%
  \BibitemOpen
  \bibfield  {author} {\bibinfo {author} {\bibfnamefont {A.~A.}\ \bibnamefont
  {Serga}}, \bibinfo {author} {\bibfnamefont {A.~V.}\ \bibnamefont {Chumak}}, \
  and\ \bibinfo {author} {\bibfnamefont {B.}~\bibnamefont {Hillebrands}},\
  }\bibfield  {title} {\emph {\bibinfo {title} {\textnormal{YIG magnonics}},\
  }}\href {\doibase 10.1088/0022-3727/43/26/264002} {\bibfield  {journal}
  {\bibinfo  {journal} {J. Phys. D: Appl. Phys.}\ }\textbf {\bibinfo {volume}
  {43}},\ \bibinfo {pages} {264002} (\bibinfo {year} {2010})}\BibitemShut
  {NoStop}%
\bibitem [{\citenamefont {Chumak}\ \emph {et~al.}(2015)\citenamefont {Chumak},
  \citenamefont {Vasyuchka},\ and\ \citenamefont
  {Serga}}]{doi:10.1038/nphys3347}%
  \BibitemOpen
  \bibfield  {author} {\bibinfo {author} {\bibfnamefont {A.~V.}\ \bibnamefont
  {Chumak}}, \bibinfo {author} {\bibfnamefont {V.~I.}\ \bibnamefont
  {Vasyuchka}}, \ and\ \bibinfo {author} {\bibfnamefont {B.}~\bibnamefont
  {Serga}, \bibfnamefont {A.~A.and~Hillebrands}},\ }\bibfield  {title} {\emph
  {\bibinfo {title} {\textnormal{Magnon spintronics}},\ }}\href {\doibase
  10.1038/nphys3347} {\bibfield  {journal} {\bibinfo  {journal} {Nat. Phys.}\
  }\textbf {\bibinfo {volume} {11}},\ \bibinfo {pages} {453} (\bibinfo {year}
  {2015})}\BibitemShut {NoStop}%
\bibitem [{\citenamefont {Zheng}\ \emph {et~al.}(2023)\citenamefont {Zheng},
  \citenamefont {Wang}, \citenamefont {Wang}, \citenamefont {Sun},
  \citenamefont {He}, \citenamefont {Yan},\ and\ \citenamefont
  {Yuan}}]{doi:10.1063/5.0152543}%
  \BibitemOpen
  \bibfield  {author} {\bibinfo {author} {\bibfnamefont {S.}~\bibnamefont
  {Zheng}}, \bibinfo {author} {\bibfnamefont {Z.}~\bibnamefont {Wang}},
  \bibinfo {author} {\bibfnamefont {Y.}~\bibnamefont {Wang}}, \bibinfo {author}
  {\bibfnamefont {F.}~\bibnamefont {Sun}}, \bibinfo {author} {\bibfnamefont
  {Q.}~\bibnamefont {He}}, \bibinfo {author} {\bibfnamefont {P.}~\bibnamefont
  {Yan}}, \ and\ \bibinfo {author} {\bibfnamefont {H.~Y.}\ \bibnamefont
  {Yuan}},\ }\bibfield  {title} {\emph {\bibinfo {title} {\textnormal{Tutorial:
  Nonlinear magnonics}},\ }}\href {\doibase 10.1063/5.0152543} {\bibfield
  {journal} {\bibinfo  {journal} {J. Appl. Phys.}\ }\textbf {\bibinfo {volume}
  {134}},\ \bibinfo {pages} {151101} (\bibinfo {year} {2023})}\BibitemShut
  {NoStop}%
\bibitem [{\citenamefont {Yuan}\ \emph {et~al.}(2022)\citenamefont {Yuan},
  \citenamefont {Cao}, \citenamefont {Kamra}, \citenamefont {Duine},\ and\
  \citenamefont {Yan}}]{YUAN20221}%
  \BibitemOpen
  \bibfield  {author} {\bibinfo {author} {\bibfnamefont {H.~Y.}\ \bibnamefont
  {Yuan}}, \bibinfo {author} {\bibfnamefont {Y.}~\bibnamefont {Cao}}, \bibinfo
  {author} {\bibfnamefont {A.}~\bibnamefont {Kamra}}, \bibinfo {author}
  {\bibfnamefont {R.~A.}\ \bibnamefont {Duine}}, \ and\ \bibinfo {author}
  {\bibfnamefont {P.}~\bibnamefont {Yan}},\ }\bibfield  {title} {\emph
  {\bibinfo {title} {\textnormal{Quantum magnonics: When magnon spintronics
  meets quantum information science}},\ }}\href {\doibase
  10.1016/j.physrep.2022.03.002} {\bibfield  {journal} {\bibinfo  {journal}
  {Phys. Rep.}\ }\textbf {\bibinfo {volume} {965}},\ \bibinfo {pages} {1}
  (\bibinfo {year} {2022})}\BibitemShut {NoStop}%
\bibitem [{\citenamefont {Rameshti}\ \emph {et~al.}(2022)\citenamefont
  {Rameshti}, \citenamefont {Kusminskiy}, \citenamefont {Haigh}, \citenamefont
  {Usami}, \citenamefont {Lachance-Quirion}, \citenamefont {Nakamura},
  \citenamefont {Hu}, \citenamefont {Tang}, \citenamefont {Bauer},\ and\
  \citenamefont {Blanter}}]{ZARERAMESHTI20221}%
  \BibitemOpen
  \bibfield  {author} {\bibinfo {author} {\bibfnamefont {B.~Z.}\ \bibnamefont
  {Rameshti}}, \bibinfo {author} {\bibfnamefont {S.~V.}\ \bibnamefont
  {Kusminskiy}}, \bibinfo {author} {\bibfnamefont {J.~A.}\ \bibnamefont
  {Haigh}}, \bibinfo {author} {\bibfnamefont {K.}~\bibnamefont {Usami}},
  \bibinfo {author} {\bibfnamefont {D.}~\bibnamefont {Lachance-Quirion}},
  \bibinfo {author} {\bibfnamefont {Y.}~\bibnamefont {Nakamura}}, \bibinfo
  {author} {\bibfnamefont {C.-M.}\ \bibnamefont {Hu}}, \bibinfo {author}
  {\bibfnamefont {H.~X.}\ \bibnamefont {Tang}}, \bibinfo {author}
  {\bibfnamefont {G.~E.~W.}\ \bibnamefont {Bauer}}, \ and\ \bibinfo {author}
  {\bibfnamefont {Y.~M.}\ \bibnamefont {Blanter}},\ }\bibfield  {title} {\emph
  {\bibinfo {title} {\textnormal {Cavity Magnonics}},\ }}\href {\doibase
  10.1016/j.physrep.2022.06.001} {\bibfield  {journal} {\bibinfo  {journal}
  {Phys. Rep.}\ }\textbf {\bibinfo {volume} {979}},\ \bibinfo {pages} {1}
  (\bibinfo {year} {2022})}\BibitemShut {NoStop}%
\bibitem [{\citenamefont {Huebl}\ \emph {et~al.}(2013)\citenamefont {Huebl},
  \citenamefont {Zollitsch}, \citenamefont {Lotze}, \citenamefont {Hocke},
  \citenamefont {Greifenstein}, \citenamefont {Marx}, \citenamefont {Gross},\
  and\ \citenamefont {Goennenwein}}]{PhysRevLett.111.127003}%
  \BibitemOpen
  \bibfield  {author} {\bibinfo {author} {\bibfnamefont {H.}~\bibnamefont
  {Huebl}}, \bibinfo {author} {\bibfnamefont {C.~W.}\ \bibnamefont
  {Zollitsch}}, \bibinfo {author} {\bibfnamefont {J.}~\bibnamefont {Lotze}},
  \bibinfo {author} {\bibfnamefont {F.}~\bibnamefont {Hocke}}, \bibinfo
  {author} {\bibfnamefont {M.}~\bibnamefont {Greifenstein}}, \bibinfo {author}
  {\bibfnamefont {A.}~\bibnamefont {Marx}}, \bibinfo {author} {\bibfnamefont
  {R.}~\bibnamefont {Gross}}, \ and\ \bibinfo {author} {\bibfnamefont
  {S.~T.~B.}\ \bibnamefont {Goennenwein}},\ }\bibfield  {title} {\emph
  {\bibinfo {title} {\textnormal{High Cooperativity in Coupled Microwave
  Resonator Ferrimagnetic Insulator Hybrids}},\ }}\href {\doibase
  10.1103/PhysRevLett.111.127003} {\bibfield  {journal} {\bibinfo  {journal}
  {Phys. Rev. Lett.}\ }\textbf {\bibinfo {volume} {111}},\ \bibinfo {pages}
  {127003} (\bibinfo {year} {2013})}\BibitemShut {NoStop}%
\bibitem [{\citenamefont {Tabuchi}\ \emph {et~al.}(2014)\citenamefont
  {Tabuchi}, \citenamefont {Ishino}, \citenamefont {Ishikawa}, \citenamefont
  {Yamazaki}, \citenamefont {Usami},\ and\ \citenamefont
  {Nakamura}}]{PhysRevLett.113.083603}%
  \BibitemOpen
  \bibfield  {author} {\bibinfo {author} {\bibfnamefont {Y.}~\bibnamefont
  {Tabuchi}}, \bibinfo {author} {\bibfnamefont {S.}~\bibnamefont {Ishino}},
  \bibinfo {author} {\bibfnamefont {T.}~\bibnamefont {Ishikawa}}, \bibinfo
  {author} {\bibfnamefont {R.}~\bibnamefont {Yamazaki}}, \bibinfo {author}
  {\bibfnamefont {K.}~\bibnamefont {Usami}}, \ and\ \bibinfo {author}
  {\bibfnamefont {Y.}~\bibnamefont {Nakamura}},\ }\bibfield  {title} {\emph
  {\bibinfo {title} {\textnormal{Hybridizing Ferromagnetic Magnons and
  Microwave Photons in the Quantum Limit}},\ }}\href {\doibase
  10.1103/PhysRevLett.113.083603} {\bibfield  {journal} {\bibinfo  {journal}
  {Phys. Rev. Lett.}\ }\textbf {\bibinfo {volume} {113}},\ \bibinfo {pages}
  {083603} (\bibinfo {year} {2014})}\BibitemShut {NoStop}%
\bibitem [{\citenamefont {Li}\ \emph {et~al.}(2019)\citenamefont {Li},
  \citenamefont {Polakovic}, \citenamefont {Wang}, \citenamefont {Xu},
  \citenamefont {Lendinez}, \citenamefont {Zhang}, \citenamefont {Ding},
  \citenamefont {Khaire}, \citenamefont {Saglam}, \citenamefont {Divan},
  \citenamefont {Pearson}, \citenamefont {Kwok}, \citenamefont {Xiao},
  \citenamefont {Novosad}, \citenamefont {Hoffmann},\ and\ \citenamefont
  {Zhang}}]{PhysRevLett.123.107701}%
  \BibitemOpen
  \bibfield  {author} {\bibinfo {author} {\bibfnamefont {Y.}~\bibnamefont
  {Li}}, \bibinfo {author} {\bibfnamefont {T.}~\bibnamefont {Polakovic}},
  \bibinfo {author} {\bibfnamefont {Y.-L.}\ \bibnamefont {Wang}}, \bibinfo
  {author} {\bibfnamefont {J.}~\bibnamefont {Xu}}, \bibinfo {author}
  {\bibfnamefont {S.}~\bibnamefont {Lendinez}}, \bibinfo {author}
  {\bibfnamefont {Z.}~\bibnamefont {Zhang}}, \bibinfo {author} {\bibfnamefont
  {J.}~\bibnamefont {Ding}}, \bibinfo {author} {\bibfnamefont {T.}~\bibnamefont
  {Khaire}}, \bibinfo {author} {\bibfnamefont {H.}~\bibnamefont {Saglam}},
  \bibinfo {author} {\bibfnamefont {R.}~\bibnamefont {Divan}}, \bibinfo
  {author} {\bibfnamefont {J.}~\bibnamefont {Pearson}}, \bibinfo {author}
  {\bibfnamefont {W.-K.}\ \bibnamefont {Kwok}}, \bibinfo {author}
  {\bibfnamefont {Z.}~\bibnamefont {Xiao}}, \bibinfo {author} {\bibfnamefont
  {V.}~\bibnamefont {Novosad}}, \bibinfo {author} {\bibfnamefont
  {A.}~\bibnamefont {Hoffmann}}, \ and\ \bibinfo {author} {\bibfnamefont
  {W.}~\bibnamefont {Zhang}},\ }\bibfield  {title} {\emph {\bibinfo {title}
  {\textnormal{Strong Coupling between Magnons and Microwave Photons in On-Chip
  Ferromagnet-Superconductor Thin-Film Devices}},\ }}\href {\doibase
  10.1103/PhysRevLett.123.107701} {\bibfield  {journal} {\bibinfo  {journal}
  {Phys. Rev. Lett.}\ }\textbf {\bibinfo {volume} {123}},\ \bibinfo {pages}
  {107701} (\bibinfo {year} {2019})}\BibitemShut {NoStop}%
\bibitem [{\citenamefont {Wang}\ \emph {et~al.}(2016)\citenamefont {Wang},
  \citenamefont {Zhang}, \citenamefont {Zhang}, \citenamefont {Luo},
  \citenamefont {Xiong}, \citenamefont {Wang}, \citenamefont {Li},
  \citenamefont {Hu},\ and\ \citenamefont {You}}]{PhysRevB.94.224410}%
  \BibitemOpen
  \bibfield  {author} {\bibinfo {author} {\bibfnamefont {Y.-P.}\ \bibnamefont
  {Wang}}, \bibinfo {author} {\bibfnamefont {G.-Q.}\ \bibnamefont {Zhang}},
  \bibinfo {author} {\bibfnamefont {D.}~\bibnamefont {Zhang}}, \bibinfo
  {author} {\bibfnamefont {X.-Q.}\ \bibnamefont {Luo}}, \bibinfo {author}
  {\bibfnamefont {W.}~\bibnamefont {Xiong}}, \bibinfo {author} {\bibfnamefont
  {S.-P.}\ \bibnamefont {Wang}}, \bibinfo {author} {\bibfnamefont {T.-F.}\
  \bibnamefont {Li}}, \bibinfo {author} {\bibfnamefont {C.-M.}\ \bibnamefont
  {Hu}}, \ and\ \bibinfo {author} {\bibfnamefont {J.~Q.}\ \bibnamefont {You}},\
  }\bibfield  {title} {\emph {\bibinfo {title} {\textnormal{Magnon Kerr effect
  in a strongly coupled cavity-magnon system}},\ }}\href {\doibase
  10.1103/PhysRevB.94.224410} {\bibfield  {journal} {\bibinfo  {journal} {Phys.
  Rev. B}\ }\textbf {\bibinfo {volume} {94}},\ \bibinfo {pages} {224410}
  (\bibinfo {year} {2016})}\BibitemShut {NoStop}%
\bibitem [{\citenamefont {Zhang}\ \emph {et~al.}(2022)\citenamefont {Zhang},
  \citenamefont {Wu},\ and\ \citenamefont
  {Yang}}]{10.1103/PhysRevA.106.012609}%
  \BibitemOpen
  \bibfield  {author} {\bibinfo {author} {\bibfnamefont {F.-Y.}\ \bibnamefont
  {Zhang}}, \bibinfo {author} {\bibfnamefont {Q.-C.}\ \bibnamefont {Wu}}, \
  and\ \bibinfo {author} {\bibfnamefont {C.-P.}\ \bibnamefont {Yang}},\
  }\bibfield  {title} {\emph {\bibinfo {title} {\textnormal {Non-Hermitian
  Shortcut to Adiabaticity in Floquet Cavity Electromagnonics}},\ }}\href
  {\doibase 10.1103/PhysRevA.106.012609} {\bibfield  {journal} {\bibinfo
  {journal} {Phys. Rev. A}\ }\textbf {\bibinfo {volume} {106}},\ \bibinfo
  {pages} {012609} (\bibinfo {year} {2022})}\BibitemShut {NoStop}%
\bibitem [{\citenamefont {Zhang}\ \emph {et~al.}(2023)\citenamefont {Zhang},
  \citenamefont {Zeng}, \citenamefont {Wu},\ and\ \citenamefont
  {Yang}}]{10.1063/5.0138391}%
  \BibitemOpen
  \bibfield  {author} {\bibinfo {author} {\bibfnamefont {F.-Y.}\ \bibnamefont
  {Zhang}}, \bibinfo {author} {\bibfnamefont {Y.-X.}\ \bibnamefont {Zeng}},
  \bibinfo {author} {\bibfnamefont {Q.-C.}\ \bibnamefont {Wu}}, \ and\ \bibinfo
  {author} {\bibfnamefont {C.-P.}\ \bibnamefont {Yang}},\ }\bibfield  {title}
  {\emph {\bibinfo {title} {\textnormal {Cat-State Encoding of a Quantum
  Information Processor Module with a Cavity–Magnon System}},\ }}\href
  {\doibase 10.1063/5.0138391} {\bibfield  {journal} {\bibinfo  {journal}
  {Appl. Phys. Lett.}\ }\textbf {\bibinfo {volume} {122}},\ \bibinfo {pages}
  {084001} (\bibinfo {year} {2023})}\BibitemShut {NoStop}%
\bibitem [{\citenamefont {Osada}\ \emph {et~al.}(2018)\citenamefont {Osada},
  \citenamefont {Gloppe}, \citenamefont {Hisatomi}, \citenamefont {Noguchi},
  \citenamefont {Yamazaki}, \citenamefont {Nomura}, \citenamefont {Nakamura},\
  and\ \citenamefont {Usami}}]{PhysRevLett.120.133602}%
  \BibitemOpen
  \bibfield  {author} {\bibinfo {author} {\bibfnamefont {A.}~\bibnamefont
  {Osada}}, \bibinfo {author} {\bibfnamefont {A.}~\bibnamefont {Gloppe}},
  \bibinfo {author} {\bibfnamefont {R.}~\bibnamefont {Hisatomi}}, \bibinfo
  {author} {\bibfnamefont {A.}~\bibnamefont {Noguchi}}, \bibinfo {author}
  {\bibfnamefont {R.}~\bibnamefont {Yamazaki}}, \bibinfo {author}
  {\bibfnamefont {M.}~\bibnamefont {Nomura}}, \bibinfo {author} {\bibfnamefont
  {Y.}~\bibnamefont {Nakamura}}, \ and\ \bibinfo {author} {\bibfnamefont
  {K.}~\bibnamefont {Usami}},\ }\bibfield  {title} {\emph {\bibinfo {title}
  {\textnormal{Brillouin Light Scattering by Magnetic Quasivortices in Cavity
  Optomagnonics}},\ }}\href {\doibase 10.1103/PhysRevLett.120.133602}
  {\bibfield  {journal} {\bibinfo  {journal} {Phys. Rev. Lett.}\ }\textbf
  {\bibinfo {volume} {120}},\ \bibinfo {pages} {133602} (\bibinfo {year}
  {2018})}\BibitemShut {NoStop}%
\bibitem [{\citenamefont {Zhang}\ \emph {et~al.}(2016)\citenamefont {Zhang},
  \citenamefont {Zou}, \citenamefont {Jiang},\ and\ \citenamefont
  {Tang}}]{doi:10.1126/sciadv.1501286}%
  \BibitemOpen
  \bibfield  {author} {\bibinfo {author} {\bibfnamefont {X.}~\bibnamefont
  {Zhang}}, \bibinfo {author} {\bibfnamefont {C.-L.}\ \bibnamefont {Zou}},
  \bibinfo {author} {\bibfnamefont {L.}~\bibnamefont {Jiang}}, \ and\ \bibinfo
  {author} {\bibfnamefont {H.~X.}\ \bibnamefont {Tang}},\ }\bibfield  {title}
  {\emph {\bibinfo {title} {\textnormal{Cavity magnomechanics}},\ }}\href
  {\doibase 10.1126/sciadv.1501286} {\bibfield  {journal} {\bibinfo  {journal}
  {Sci. Adv}\ }\textbf {\bibinfo {volume} {2}},\ \bibinfo {pages} {e1501286}
  (\bibinfo {year} {2016})}\BibitemShut {NoStop}%
\bibitem [{\citenamefont {Shen}\ \emph {et~al.}(2022)\citenamefont {Shen},
  \citenamefont {Xu}, \citenamefont {Zhang}, \citenamefont {Zhang},
  \citenamefont {Wang}, \citenamefont {Chai}, \citenamefont {Zou},
  \citenamefont {Guo},\ and\ \citenamefont {Dong}}]{PhysRevLett.129.243601}%
  \BibitemOpen
  \bibfield  {author} {\bibinfo {author} {\bibfnamefont {Z.}~\bibnamefont
  {Shen}}, \bibinfo {author} {\bibfnamefont {G.-T.}\ \bibnamefont {Xu}},
  \bibinfo {author} {\bibfnamefont {M.}~\bibnamefont {Zhang}}, \bibinfo
  {author} {\bibfnamefont {Y.-L.}\ \bibnamefont {Zhang}}, \bibinfo {author}
  {\bibfnamefont {Y.}~\bibnamefont {Wang}}, \bibinfo {author} {\bibfnamefont
  {C.-Z.}\ \bibnamefont {Chai}}, \bibinfo {author} {\bibfnamefont {C.-L.}\
  \bibnamefont {Zou}}, \bibinfo {author} {\bibfnamefont {G.-C.}\ \bibnamefont
  {Guo}}, \ and\ \bibinfo {author} {\bibfnamefont {C.-H.}\ \bibnamefont
  {Dong}},\ }\bibfield  {title} {\emph {\bibinfo {title} {\textnormal{Coherent
  Coupling between Phonons, Magnons, and Photons}},\ }}\href {\doibase
  10.1103/PhysRevLett.129.243601} {\bibfield  {journal} {\bibinfo  {journal}
  {Phys. Rev. Lett.}\ }\textbf {\bibinfo {volume} {129}},\ \bibinfo {pages}
  {243601} (\bibinfo {year} {2022})}\BibitemShut {NoStop}%
\bibitem [{\citenamefont {Mergenthaler}\ \emph {et~al.}(2017)\citenamefont
  {Mergenthaler}, \citenamefont {Liu}, \citenamefont {Le~Roy}, \citenamefont
  {Ares}, \citenamefont {Thompson}, \citenamefont {Bogani}, \citenamefont
  {Luis}, \citenamefont {Blundell}, \citenamefont {Lancaster}, \citenamefont
  {Ardavan}, \citenamefont {Briggs}, \citenamefont {Leek},\ and\ \citenamefont
  {Laird}}]{PhysRevLett.119.147701}%
  \BibitemOpen
  \bibfield  {author} {\bibinfo {author} {\bibfnamefont {M.}~\bibnamefont
  {Mergenthaler}}, \bibinfo {author} {\bibfnamefont {J.}~\bibnamefont {Liu}},
  \bibinfo {author} {\bibfnamefont {J.~J.}\ \bibnamefont {Le~Roy}}, \bibinfo
  {author} {\bibfnamefont {N.}~\bibnamefont {Ares}}, \bibinfo {author}
  {\bibfnamefont {A.~L.}\ \bibnamefont {Thompson}}, \bibinfo {author}
  {\bibfnamefont {L.}~\bibnamefont {Bogani}}, \bibinfo {author} {\bibfnamefont
  {F.}~\bibnamefont {Luis}}, \bibinfo {author} {\bibfnamefont {S.~J.}\
  \bibnamefont {Blundell}}, \bibinfo {author} {\bibfnamefont {T.}~\bibnamefont
  {Lancaster}}, \bibinfo {author} {\bibfnamefont {A.}~\bibnamefont {Ardavan}},
  \bibinfo {author} {\bibfnamefont {G.~A.~D.}\ \bibnamefont {Briggs}}, \bibinfo
  {author} {\bibfnamefont {P.~J.}\ \bibnamefont {Leek}}, \ and\ \bibinfo
  {author} {\bibfnamefont {E.~A.}\ \bibnamefont {Laird}},\ }\bibfield  {title}
  {\emph {\bibinfo {title} {\textnormal{Strong Coupling of Microwave Photons to
  Antiferromagnetic Fluctuations in an Organic Magnet}},\ }}\href {\doibase
  10.1103/PhysRevLett.119.147701} {\bibfield  {journal} {\bibinfo  {journal}
  {Phys. Rev. Lett.}\ }\textbf {\bibinfo {volume} {119}},\ \bibinfo {pages}
  {147701} (\bibinfo {year} {2017})}\BibitemShut {NoStop}%
\bibitem [{\citenamefont {Parvini}\ \emph {et~al.}(2020)\citenamefont
  {Parvini}, \citenamefont {Bittencourt},\ and\ \citenamefont
  {Kusminskiy}}]{PhysRevResearch.2.022027}%
  \BibitemOpen
  \bibfield  {author} {\bibinfo {author} {\bibfnamefont {T.~S.}\ \bibnamefont
  {Parvini}}, \bibinfo {author} {\bibfnamefont {V.~A. S.~V.}\ \bibnamefont
  {Bittencourt}}, \ and\ \bibinfo {author} {\bibfnamefont {S.~V.}\ \bibnamefont
  {Kusminskiy}},\ }\bibfield  {title} {\emph {\bibinfo {title}
  {\textnormal{Antiferromagnetic cavity optomagnonics}},\ }}\href {\doibase
  10.1103/PhysRevResearch.2.022027} {\bibfield  {journal} {\bibinfo  {journal}
  {Phys. Rev. Res.}\ }\textbf {\bibinfo {volume} {2}},\ \bibinfo {pages}
  {022027} (\bibinfo {year} {2020})}\BibitemShut {NoStop}%
\bibitem [{\citenamefont {Kounalakis}\ \emph {et~al.}(2022)\citenamefont
  {Kounalakis}, \citenamefont {Bauer},\ and\ \citenamefont
  {Blanter}}]{PhysRevLett.129.037205}%
  \BibitemOpen
  \bibfield  {author} {\bibinfo {author} {\bibfnamefont {M.}~\bibnamefont
  {Kounalakis}}, \bibinfo {author} {\bibfnamefont {G.~E.~W.}\ \bibnamefont
  {Bauer}}, \ and\ \bibinfo {author} {\bibfnamefont {Y.~M.}\ \bibnamefont
  {Blanter}},\ }\bibfield  {title} {\emph {\bibinfo {title} {\textnormal{Analog
  Quantum Control of Magnonic Cat States on a Chip by a Superconducting
  Qubit}},\ }}\href {\doibase 10.1103/PhysRevLett.129.037205} {\bibfield
  {journal} {\bibinfo  {journal} {Phys. Rev. Lett.}\ }\textbf {\bibinfo
  {volume} {129}},\ \bibinfo {pages} {037205} (\bibinfo {year}
  {2022})}\BibitemShut {NoStop}%
\bibitem [{\citenamefont {Liu}\ \emph {et~al.}(2019)\citenamefont {Liu},
  \citenamefont {Xiong},\ and\ \citenamefont {Wu}}]{PhysRevB.100.134421}%
  \BibitemOpen
  \bibfield  {author} {\bibinfo {author} {\bibfnamefont {Z.-X.}\ \bibnamefont
  {Liu}}, \bibinfo {author} {\bibfnamefont {H.}~\bibnamefont {Xiong}}, \ and\
  \bibinfo {author} {\bibfnamefont {Y.}~\bibnamefont {Wu}},\ }\bibfield
  {title} {\emph {\bibinfo {title} {\textnormal{Magnon blockade in a hybrid
  ferromagnet-superconductor quantum system}},\ }}\href {\doibase
  10.1103/PhysRevB.100.134421} {\bibfield  {journal} {\bibinfo  {journal}
  {Phys. Rev. B}\ }\textbf {\bibinfo {volume} {100}},\ \bibinfo {pages}
  {134421} (\bibinfo {year} {2019})}\BibitemShut {NoStop}%
\bibitem [{\citenamefont {Xie}\ \emph {et~al.}(2020)\citenamefont {Xie},
  \citenamefont {Ma},\ and\ \citenamefont {Li}}]{PhysRevA.101.042331}%
  \BibitemOpen
  \bibfield  {author} {\bibinfo {author} {\bibfnamefont {J.-k.}\ \bibnamefont
  {Xie}}, \bibinfo {author} {\bibfnamefont {S.-l.}\ \bibnamefont {Ma}}, \ and\
  \bibinfo {author} {\bibfnamefont {F.-l.}\ \bibnamefont {Li}},\ }\bibfield
  {title} {\emph {\bibinfo {title} {\textnormal{Quantum-interference-enhanced
  magnon blockade in an yttrium-iron-garnet sphere coupled to superconducting
  circuits}},\ }}\href {\doibase 10.1103/PhysRevA.101.042331} {\bibfield
  {journal} {\bibinfo  {journal} {Phys. Rev. A}\ }\textbf {\bibinfo {volume}
  {101}},\ \bibinfo {pages} {042331} (\bibinfo {year} {2020})}\BibitemShut
  {NoStop}%
\bibitem [{\citenamefont {Xu}\ \emph {et~al.}(2023)\citenamefont {Xu},
  \citenamefont {Gu}, \citenamefont {Li}, \citenamefont {Weng}, \citenamefont
  {Wang}, \citenamefont {Li}, \citenamefont {Wang}, \citenamefont {Zhu},\ and\
  \citenamefont {You}}]{PhysRevLett.130.193603}%
  \BibitemOpen
  \bibfield  {author} {\bibinfo {author} {\bibfnamefont {D.}~\bibnamefont
  {Xu}}, \bibinfo {author} {\bibfnamefont {X.-K.}\ \bibnamefont {Gu}}, \bibinfo
  {author} {\bibfnamefont {H.-K.}\ \bibnamefont {Li}}, \bibinfo {author}
  {\bibfnamefont {Y.-C.}\ \bibnamefont {Weng}}, \bibinfo {author}
  {\bibfnamefont {Y.-P.}\ \bibnamefont {Wang}}, \bibinfo {author}
  {\bibfnamefont {J.}~\bibnamefont {Li}}, \bibinfo {author} {\bibfnamefont
  {H.}~\bibnamefont {Wang}}, \bibinfo {author} {\bibfnamefont {S.-Y.}\
  \bibnamefont {Zhu}}, \ and\ \bibinfo {author} {\bibfnamefont {J.~Q.}\
  \bibnamefont {You}},\ }\bibfield  {title} {\emph {\bibinfo {title}
  {\textnormal{Quantum Control of a Single Magnon in a Macroscopic Spin
  System}},\ }}\href {\doibase 10.1103/PhysRevLett.130.193603} {\bibfield
  {journal} {\bibinfo  {journal} {Phys. Rev. Lett.}\ }\textbf {\bibinfo
  {volume} {130}},\ \bibinfo {pages} {193603} (\bibinfo {year}
  {2023})}\BibitemShut {NoStop}%
\bibitem [{\citenamefont {He}\ \emph {et~al.}(2023{\natexlab{a}})\citenamefont
  {He}, \citenamefont {Xin}, \citenamefont {Zhang},\ and\ \citenamefont
  {Li}}]{PhysRevA.107.023709}%
  \BibitemOpen
  \bibfield  {author} {\bibinfo {author} {\bibfnamefont {S.}~\bibnamefont
  {He}}, \bibinfo {author} {\bibfnamefont {X.}~\bibnamefont {Xin}}, \bibinfo
  {author} {\bibfnamefont {F.-Y.}\ \bibnamefont {Zhang}}, \ and\ \bibinfo
  {author} {\bibfnamefont {C.}~\bibnamefont {Li}},\ }\bibfield  {title} {\emph
  {\bibinfo {title} {\textnormal {Generation of a Schr\"odinger cat state in a
  hybrid ferromagnet-superconductor system}},\ }}\href {\doibase
  10.1103/PhysRevA.107.023709} {\bibfield  {journal} {\bibinfo  {journal}
  {Phys. Rev. A}\ }\textbf {\bibinfo {volume} {107}},\ \bibinfo {pages}
  {023709} (\bibinfo {year} {2023}{\natexlab{a}})}\BibitemShut {NoStop}%
\bibitem [{\citenamefont {He}\ \emph {et~al.}(2024)\citenamefont {He},
  \citenamefont {Xin}, \citenamefont {Wang}, \citenamefont {Zhang},\ and\
  \citenamefont {Li}}]{10.1103/PhysRevA.110.053710}%
  \BibitemOpen
  \bibfield  {author} {\bibinfo {author} {\bibfnamefont {S.}~\bibnamefont
  {He}}, \bibinfo {author} {\bibfnamefont {X.}~\bibnamefont {Xin}}, \bibinfo
  {author} {\bibfnamefont {Z.}~\bibnamefont {Wang}}, \bibinfo {author}
  {\bibfnamefont {F.-Y.}\ \bibnamefont {Zhang}}, \ and\ \bibinfo {author}
  {\bibfnamefont {C.}~\bibnamefont {Li}},\ }\bibfield  {title} {\emph {\bibinfo
  {title} {\textnormal {Generation of a Squeezed Schr\"odinger Cat State in an
  Anisotropic Ferromagnet-Superconductor Coupled System}},\ }}\href {\doibase
  10.1103/PhysRevA.110.053710} {\bibfield  {journal} {\bibinfo  {journal}
  {Phys. Rev. A}\ }\textbf {\bibinfo {volume} {110}},\ \bibinfo {pages}
  {053710} (\bibinfo {year} {2024})}\BibitemShut {NoStop}%
\bibitem [{\citenamefont {Zhang}\ \emph
  {et~al.}(2024{\natexlab{a}})\citenamefont {Zhang}, \citenamefont {He},
  \citenamefont {Liu}, \citenamefont {Wu},\ and\ \citenamefont
  {Yang}}]{10.1103/PhysRevA.109.022442}%
  \BibitemOpen
  \bibfield  {author} {\bibinfo {author} {\bibfnamefont {F.-Y.}\ \bibnamefont
  {Zhang}}, \bibinfo {author} {\bibfnamefont {S.-W.}\ \bibnamefont {He}},
  \bibinfo {author} {\bibfnamefont {T.}~\bibnamefont {Liu}}, \bibinfo {author}
  {\bibfnamefont {Q.-C.}\ \bibnamefont {Wu}}, \ and\ \bibinfo {author}
  {\bibfnamefont {C.-P.}\ \bibnamefont {Yang}},\ }\bibfield  {title} {\emph
  {\bibinfo {title} {\textnormal {Tunable Strong Coupling Between a Transmon
  and a Magnon}},\ }}\href {\doibase 10.1103/PhysRevA.109.022442} {\bibfield
  {journal} {\bibinfo  {journal} {Phys. Rev. A}\ }\textbf {\bibinfo {volume}
  {109}},\ \bibinfo {pages} {022442} (\bibinfo {year}
  {2024}{\natexlab{a}})}\BibitemShut {NoStop}%
\bibitem [{\citenamefont {Lu}\ \emph {et~al.}(2021)\citenamefont {Lu},
  \citenamefont {Zhang}, \citenamefont {Zhang},\ and\ \citenamefont
  {Jing}}]{PhysRevA.103.063708}%
  \BibitemOpen
  \bibfield  {author} {\bibinfo {author} {\bibfnamefont {T.-X.}\ \bibnamefont
  {Lu}}, \bibinfo {author} {\bibfnamefont {H.}~\bibnamefont {Zhang}}, \bibinfo
  {author} {\bibfnamefont {Q.}~\bibnamefont {Zhang}}, \ and\ \bibinfo {author}
  {\bibfnamefont {H.}~\bibnamefont {Jing}},\ }\bibfield  {title} {\emph
  {\bibinfo {title} {\textnormal{Exceptional-point-engineered cavity
  magnomechanics}},\ }}\href {\doibase 10.1103/PhysRevA.103.063708} {\bibfield
  {journal} {\bibinfo  {journal} {Phys. Rev. A}\ }\textbf {\bibinfo {volume}
  {103}},\ \bibinfo {pages} {063708} (\bibinfo {year} {2021})}\BibitemShut
  {NoStop}%
\bibitem [{\citenamefont {Xiong}(2023)}]{XIONG20238}%
  \BibitemOpen
  \bibfield  {author} {\bibinfo {author} {\bibfnamefont {H.}~\bibnamefont
  {Xiong}},\ }\bibfield  {title} {\emph {\bibinfo {title} {\textnormal{Magnonic
  frequency combs based on the resonantly enhanced magnetostrictive effect}},\
  }}\href {\doibase https://doi.org/10.1016/j.fmre.2022.08.017} {\bibfield
  {journal} {\bibinfo  {journal} {Fundam. Res}\ }\textbf {\bibinfo {volume}
  {3}},\ \bibinfo {pages} {8} (\bibinfo {year} {2023})}\BibitemShut {NoStop}%
\bibitem [{\citenamefont {Mukhopadhyay}\ \emph {et~al.}(2022)\citenamefont
  {Mukhopadhyay}, \citenamefont {Nair},\ and\ \citenamefont
  {Agarwal}}]{PhysRevB.106.184426}%
  \BibitemOpen
  \bibfield  {author} {\bibinfo {author} {\bibfnamefont {D.}~\bibnamefont
  {Mukhopadhyay}}, \bibinfo {author} {\bibfnamefont {J.~M.~P.}\ \bibnamefont
  {Nair}}, \ and\ \bibinfo {author} {\bibfnamefont {G.~S.}\ \bibnamefont
  {Agarwal}},\ }\bibfield  {title} {\emph {\bibinfo {title}
  {\textnormal{Quantum amplification of spin currents in cavity magnonics by a
  parametric drive induced long-lived mode}},\ }}\href {\doibase
  10.1103/PhysRevB.106.184426} {\bibfield  {journal} {\bibinfo  {journal}
  {Phys. Rev. B}\ }\textbf {\bibinfo {volume} {106}},\ \bibinfo {pages}
  {184426} (\bibinfo {year} {2022})}\BibitemShut {NoStop}%
\bibitem [{\citenamefont {Liu}\ \emph {et~al.}(2023)\citenamefont {Liu},
  \citenamefont {Bergman}, \citenamefont {Bagrov}, \citenamefont {Delin},
  \citenamefont {Thonig}, \citenamefont {Pereiro}, \citenamefont {Eriksson},
  \citenamefont {Streib}, \citenamefont {Sjöqvist},\ and\ \citenamefont
  {Azimi-Mousolou}}]{doi:10.1088/1367-2630/ad0b20}%
  \BibitemOpen
  \bibfield  {author} {\bibinfo {author} {\bibfnamefont {Y.}~\bibnamefont
  {Liu}}, \bibinfo {author} {\bibfnamefont {A.}~\bibnamefont {Bergman}},
  \bibinfo {author} {\bibfnamefont {A.}~\bibnamefont {Bagrov}}, \bibinfo
  {author} {\bibfnamefont {A.}~\bibnamefont {Delin}}, \bibinfo {author}
  {\bibfnamefont {D.}~\bibnamefont {Thonig}}, \bibinfo {author} {\bibfnamefont
  {M.}~\bibnamefont {Pereiro}}, \bibinfo {author} {\bibfnamefont
  {O.}~\bibnamefont {Eriksson}}, \bibinfo {author} {\bibfnamefont
  {S.}~\bibnamefont {Streib}}, \bibinfo {author} {\bibfnamefont
  {E.}~\bibnamefont {Sjöqvist}}, \ and\ \bibinfo {author} {\bibfnamefont
  {V.}~\bibnamefont {Azimi-Mousolou}},\ }\bibfield  {title} {\emph {\bibinfo
  {title} {\textnormal{Tunable phonon-driven magnon–magnon entanglement at
  room temperature}},\ }}\href {\doibase 10.1088/1367-2630/ad0b20} {\bibfield
  {journal} {\bibinfo  {journal} {New J. Phys}\ }\textbf {\bibinfo {volume}
  {25}},\ \bibinfo {pages} {113032} (\bibinfo {year} {2023})}\BibitemShut
  {NoStop}%
\bibitem [{\citenamefont {Ren}\ \emph {et~al.}(2022)\citenamefont {Ren},
  \citenamefont {Xie}, \citenamefont {Li}, \citenamefont {Ma},\ and\
  \citenamefont {Li}}]{PhysRevB.105.094422}%
  \BibitemOpen
  \bibfield  {author} {\bibinfo {author} {\bibfnamefont {Y.-l.}\ \bibnamefont
  {Ren}}, \bibinfo {author} {\bibfnamefont {J.-k.}\ \bibnamefont {Xie}},
  \bibinfo {author} {\bibfnamefont {X.-k.}\ \bibnamefont {Li}}, \bibinfo
  {author} {\bibfnamefont {S.-l.}\ \bibnamefont {Ma}}, \ and\ \bibinfo {author}
  {\bibfnamefont {F.-l.}\ \bibnamefont {Li}},\ }\bibfield  {title} {\emph
  {\bibinfo {title} {\textnormal{Long-range generation of a magnon-magnon
  entangled state}},\ }}\href {\doibase 10.1103/PhysRevB.105.094422} {\bibfield
   {journal} {\bibinfo  {journal} {Phys. Rev. B}\ }\textbf {\bibinfo {volume}
  {105}},\ \bibinfo {pages} {094422} (\bibinfo {year} {2022})}\BibitemShut
  {NoStop}%
\bibitem [{\citenamefont {Wang}\ \emph {et~al.}(2019)\citenamefont {Wang},
  \citenamefont {Rao}, \citenamefont {Yang}, \citenamefont {Xu}, \citenamefont
  {Gui}, \citenamefont {Yao}, \citenamefont {You},\ and\ \citenamefont
  {Hu}}]{PhysRevLett.123.127202}%
  \BibitemOpen
  \bibfield  {author} {\bibinfo {author} {\bibfnamefont {Y.-P.}\ \bibnamefont
  {Wang}}, \bibinfo {author} {\bibfnamefont {J.~W.}\ \bibnamefont {Rao}},
  \bibinfo {author} {\bibfnamefont {Y.}~\bibnamefont {Yang}}, \bibinfo {author}
  {\bibfnamefont {P.-C.}\ \bibnamefont {Xu}}, \bibinfo {author} {\bibfnamefont
  {Y.~S.}\ \bibnamefont {Gui}}, \bibinfo {author} {\bibfnamefont {B.~M.}\
  \bibnamefont {Yao}}, \bibinfo {author} {\bibfnamefont {J.~Q.}\ \bibnamefont
  {You}}, \ and\ \bibinfo {author} {\bibfnamefont {C.-M.}\ \bibnamefont {Hu}},\
  }\bibfield  {title} {\emph {\bibinfo {title} {\textnormal{Nonreciprocity and
  Unidirectional Invisibility in Cavity Magnonics}},\ }}\href {\doibase
  10.1103/PhysRevLett.123.127202} {\bibfield  {journal} {\bibinfo  {journal}
  {Phys. Rev. Lett.}\ }\textbf {\bibinfo {volume} {123}},\ \bibinfo {pages}
  {127202} (\bibinfo {year} {2019})}\BibitemShut {NoStop}%
\bibitem [{\citenamefont {Zhao}\ \emph
  {et~al.}(2020{\natexlab{a}})\citenamefont {Zhao}, \citenamefont {Liu},
  \citenamefont {Wu}, \citenamefont {Duan}, \citenamefont {Liu},\ and\
  \citenamefont {Du}}]{PhysRevApplied.13.014053}%
  \BibitemOpen
  \bibfield  {author} {\bibinfo {author} {\bibfnamefont {J.}~\bibnamefont
  {Zhao}}, \bibinfo {author} {\bibfnamefont {Y.}~\bibnamefont {Liu}}, \bibinfo
  {author} {\bibfnamefont {L.}~\bibnamefont {Wu}}, \bibinfo {author}
  {\bibfnamefont {C.-K.}\ \bibnamefont {Duan}}, \bibinfo {author}
  {\bibfnamefont {Y.-x.}\ \bibnamefont {Liu}}, \ and\ \bibinfo {author}
  {\bibfnamefont {J.}~\bibnamefont {Du}},\ }\bibfield  {title} {\emph {\bibinfo
  {title} {\textnormal{Observation of Anti-$\mathcal{P}\mathcal{T}$-Symmetry
  Phase Transition in the Magnon-Cavity-Magnon Coupled System}},\ }}\href
  {\doibase 10.1103/PhysRevApplied.13.014053} {\bibfield  {journal} {\bibinfo
  {journal} {Phys. Rev. Appl.}\ }\textbf {\bibinfo {volume} {13}},\ \bibinfo
  {pages} {014053} (\bibinfo {year} {2020}{\natexlab{a}})}\BibitemShut
  {NoStop}%
\bibitem [{\citenamefont {Hei}\ \emph {et~al.}(2021)\citenamefont {Hei},
  \citenamefont {Dong}, \citenamefont {Chen}, \citenamefont {Shen},
  \citenamefont {Qiao},\ and\ \citenamefont {Li}}]{PhysRevA.103.043706}%
  \BibitemOpen
  \bibfield  {author} {\bibinfo {author} {\bibfnamefont {X.-L.}\ \bibnamefont
  {Hei}}, \bibinfo {author} {\bibfnamefont {X.-L.}\ \bibnamefont {Dong}},
  \bibinfo {author} {\bibfnamefont {J.-Q.}\ \bibnamefont {Chen}}, \bibinfo
  {author} {\bibfnamefont {C.-P.}\ \bibnamefont {Shen}}, \bibinfo {author}
  {\bibfnamefont {Y.-F.}\ \bibnamefont {Qiao}}, \ and\ \bibinfo {author}
  {\bibfnamefont {P.-B.}\ \bibnamefont {Li}},\ }\bibfield  {title} {\emph
  {\bibinfo {title} {\textnormal{Enhancing spin-photon coupling with a
  micromagnet}},\ }}\href {\doibase 10.1103/PhysRevA.103.043706} {\bibfield
  {journal} {\bibinfo  {journal} {Phys. Rev. A}\ }\textbf {\bibinfo {volume}
  {103}},\ \bibinfo {pages} {043706} (\bibinfo {year} {2021})}\BibitemShut
  {NoStop}%
\bibitem [{\citenamefont {Yuan}\ \emph {et~al.}(2020)\citenamefont {Yuan},
  \citenamefont {Yan}, \citenamefont {Zheng}, \citenamefont {He}, \citenamefont
  {Xia},\ and\ \citenamefont {Yung}}]{PhysRevLett.124.053602}%
  \BibitemOpen
  \bibfield  {author} {\bibinfo {author} {\bibfnamefont {H.~Y.}\ \bibnamefont
  {Yuan}}, \bibinfo {author} {\bibfnamefont {P.}~\bibnamefont {Yan}}, \bibinfo
  {author} {\bibfnamefont {S.}~\bibnamefont {Zheng}}, \bibinfo {author}
  {\bibfnamefont {Q.~Y.}\ \bibnamefont {He}}, \bibinfo {author} {\bibfnamefont
  {K.}~\bibnamefont {Xia}}, \ and\ \bibinfo {author} {\bibfnamefont {M.-H.}\
  \bibnamefont {Yung}},\ }\bibfield  {title} {\emph {\bibinfo {title}
  {\textnormal{Steady Bell State Generation via Magnon-Photon Coupling}},\
  }}\href {\doibase 10.1103/PhysRevLett.124.053602} {\bibfield  {journal}
  {\bibinfo  {journal} {Phys. Rev. Lett.}\ }\textbf {\bibinfo {volume} {124}},\
  \bibinfo {pages} {053602} (\bibinfo {year} {2020})}\BibitemShut {NoStop}%
\bibitem [{\citenamefont {Qi}\ and\ \citenamefont
  {Jing}(2022)}]{PhysRevA.105.022624}%
  \BibitemOpen
  \bibfield  {author} {\bibinfo {author} {\bibfnamefont {S.-f.}\ \bibnamefont
  {Qi}}\ and\ \bibinfo {author} {\bibfnamefont {J.}~\bibnamefont {Jing}},\
  }\bibfield  {title} {\emph {\bibinfo {title} {\textnormal{Generation of Bell
  and Greenberger-Horne-Zeilinger states from a hybrid qubit-photon-magnon
  system}},\ }}\href {\doibase 10.1103/PhysRevA.105.022624} {\bibfield
  {journal} {\bibinfo  {journal} {Phys. Rev. A}\ }\textbf {\bibinfo {volume}
  {105}},\ \bibinfo {pages} {022624} (\bibinfo {year} {2022})}\BibitemShut
  {NoStop}%
\bibitem [{\citenamefont {Hisatomi}\ \emph {et~al.}(2016)\citenamefont
  {Hisatomi}, \citenamefont {Osada}, \citenamefont {Tabuchi}, \citenamefont
  {Ishikawa}, \citenamefont {Noguchi}, \citenamefont {Yamazaki}, \citenamefont
  {Usami},\ and\ \citenamefont {Nakamura}}]{PhysRevB.93.174427}%
  \BibitemOpen
  \bibfield  {author} {\bibinfo {author} {\bibfnamefont {R.}~\bibnamefont
  {Hisatomi}}, \bibinfo {author} {\bibfnamefont {A.}~\bibnamefont {Osada}},
  \bibinfo {author} {\bibfnamefont {Y.}~\bibnamefont {Tabuchi}}, \bibinfo
  {author} {\bibfnamefont {T.}~\bibnamefont {Ishikawa}}, \bibinfo {author}
  {\bibfnamefont {A.}~\bibnamefont {Noguchi}}, \bibinfo {author} {\bibfnamefont
  {R.}~\bibnamefont {Yamazaki}}, \bibinfo {author} {\bibfnamefont
  {K.}~\bibnamefont {Usami}}, \ and\ \bibinfo {author} {\bibfnamefont
  {Y.}~\bibnamefont {Nakamura}},\ }\bibfield  {title} {\emph {\bibinfo {title}
  {\textnormal{Bidirectional conversion between microwave and light via
  ferromagnetic magnons}},\ }}\href {\doibase 10.1103/PhysRevB.93.174427}
  {\bibfield  {journal} {\bibinfo  {journal} {Phys. Rev. B}\ }\textbf {\bibinfo
  {volume} {93}},\ \bibinfo {pages} {174427} (\bibinfo {year}
  {2016})}\BibitemShut {NoStop}%
\bibitem [{\citenamefont {Zhao}\ \emph
  {et~al.}(2020{\natexlab{b}})\citenamefont {Zhao}, \citenamefont {Li},
  \citenamefont {Chao}, \citenamefont {Peng}, \citenamefont {Li},\ and\
  \citenamefont {Zhou}}]{PhysRevA.101.063838}%
  \BibitemOpen
  \bibfield  {author} {\bibinfo {author} {\bibfnamefont {C.}~\bibnamefont
  {Zhao}}, \bibinfo {author} {\bibfnamefont {X.}~\bibnamefont {Li}}, \bibinfo
  {author} {\bibfnamefont {S.}~\bibnamefont {Chao}}, \bibinfo {author}
  {\bibfnamefont {R.}~\bibnamefont {Peng}}, \bibinfo {author} {\bibfnamefont
  {C.}~\bibnamefont {Li}}, \ and\ \bibinfo {author} {\bibfnamefont
  {L.}~\bibnamefont {Zhou}},\ }\bibfield  {title} {\emph {\bibinfo {title}
  {\textnormal{Simultaneous blockade of a photon, phonon, and magnon induced by
  a two-level atom}},\ }}\href {\doibase 10.1103/PhysRevA.101.063838}
  {\bibfield  {journal} {\bibinfo  {journal} {Phys. Rev. A}\ }\textbf {\bibinfo
  {volume} {101}},\ \bibinfo {pages} {063838} (\bibinfo {year}
  {2020}{\natexlab{b}})}\BibitemShut {NoStop}%
\bibitem [{\citenamefont {jun Xu}\ \emph {et~al.}(2021)\citenamefont {jun Xu},
  \citenamefont {le~Yang}, \citenamefont {Lin},\ and\ \citenamefont
  {Song}}]{Xu:21}%
  \BibitemOpen
  \bibfield  {author} {\bibinfo {author} {\bibfnamefont {Y.}~\bibnamefont {jun
  Xu}}, \bibinfo {author} {\bibfnamefont {T.}~\bibnamefont {le~Yang}}, \bibinfo
  {author} {\bibfnamefont {L.}~\bibnamefont {Lin}}, \ and\ \bibinfo {author}
  {\bibfnamefont {J.}~\bibnamefont {Song}},\ }\bibfield  {title} {\emph
  {\bibinfo {title} {\textnormal{Conventional and unconventional magnon
  blockades in a qubit-magnon hybrid quantum system}},\ }}\href {\doibase
  10.1364/JOSAB.414600} {\bibfield  {journal} {\bibinfo  {journal} {J. Opt.
  Soc. Am. B}\ }\textbf {\bibinfo {volume} {38}},\ \bibinfo {pages} {876}
  (\bibinfo {year} {2021})}\BibitemShut {NoStop}%
\bibitem [{\citenamefont {Wu}\ \emph {et~al.}(2021)\citenamefont {Wu},
  \citenamefont {Zhong}, \citenamefont {Cheng},\ and\ \citenamefont
  {Chen}}]{PhysRevA.103.052411}%
  \BibitemOpen
  \bibfield  {author} {\bibinfo {author} {\bibfnamefont {K.}~\bibnamefont
  {Wu}}, \bibinfo {author} {\bibfnamefont {W.-x.}\ \bibnamefont {Zhong}},
  \bibinfo {author} {\bibfnamefont {G.-l.}\ \bibnamefont {Cheng}}, \ and\
  \bibinfo {author} {\bibfnamefont {A.-x.}\ \bibnamefont {Chen}},\ }\bibfield
  {title} {\emph {\bibinfo {title} {\textnormal{Phase-controlled multimagnon
  blockade and magnon-induced tunneling in a hybrid superconducting system}},\
  }}\href {\doibase 10.1103/PhysRevA.103.052411} {\bibfield  {journal}
  {\bibinfo  {journal} {Phys. Rev. A}\ }\textbf {\bibinfo {volume} {103}},\
  \bibinfo {pages} {052411} (\bibinfo {year} {2021})}\BibitemShut {NoStop}%
\bibitem [{\citenamefont {Wang}\ \emph {et~al.}(2022)\citenamefont {Wang},
  \citenamefont {Gou}, \citenamefont {Xu},\ and\ \citenamefont
  {Gong}}]{PhysRevA.106.013705}%
  \BibitemOpen
  \bibfield  {author} {\bibinfo {author} {\bibfnamefont {F.}~\bibnamefont
  {Wang}}, \bibinfo {author} {\bibfnamefont {C.}~\bibnamefont {Gou}}, \bibinfo
  {author} {\bibfnamefont {J.}~\bibnamefont {Xu}}, \ and\ \bibinfo {author}
  {\bibfnamefont {C.}~\bibnamefont {Gong}},\ }\bibfield  {title} {\emph
  {\bibinfo {title} {\textnormal{Hybrid magnon-atom entanglement and magnon
  blockade via quantum interference}},\ }}\href {\doibase
  10.1103/PhysRevA.106.013705} {\bibfield  {journal} {\bibinfo  {journal}
  {Phys. Rev. A}\ }\textbf {\bibinfo {volume} {106}},\ \bibinfo {pages}
  {013705} (\bibinfo {year} {2022})}\BibitemShut {NoStop}%
\bibitem [{\citenamefont {Jin}\ and\ \citenamefont
  {Jing}(2023)}]{PhysRevA.108.053702}%
  \BibitemOpen
  \bibfield  {author} {\bibinfo {author} {\bibfnamefont {Z.-y.}\ \bibnamefont
  {Jin}}\ and\ \bibinfo {author} {\bibfnamefont {J.}~\bibnamefont {Jing}},\
  }\bibfield  {title} {\emph {\bibinfo {title} {\textnormal{Magnon blockade in
  magnon-qubit systems}},\ }}\href {\doibase 10.1103/PhysRevA.108.053702}
  {\bibfield  {journal} {\bibinfo  {journal} {Phys. Rev. A}\ }\textbf {\bibinfo
  {volume} {108}},\ \bibinfo {pages} {053702} (\bibinfo {year}
  {2023})}\BibitemShut {NoStop}%
\bibitem [{\citenamefont {Yuan}\ and\ \citenamefont
  {Duine}(2020)}]{PhysRevB.102.100402}%
  \BibitemOpen
  \bibfield  {author} {\bibinfo {author} {\bibfnamefont {H.~Y.}\ \bibnamefont
  {Yuan}}\ and\ \bibinfo {author} {\bibfnamefont {R.~A.}\ \bibnamefont
  {Duine}},\ }\bibfield  {title} {\emph {\bibinfo {title} {\textnormal{Magnon
  antibunching in a nanomagnet}},\ }}\href {\doibase
  10.1103/PhysRevB.102.100402} {\bibfield  {journal} {\bibinfo  {journal}
  {Phys. Rev. B}\ }\textbf {\bibinfo {volume} {102}},\ \bibinfo {pages}
  {100402} (\bibinfo {year} {2020})}\BibitemShut {NoStop}%
\bibitem [{\citenamefont {Yan}\ \emph {et~al.}(2024)\citenamefont {Yan},
  \citenamefont {Zhao}, \citenamefont {Wang}, \citenamefont {Yang},\ and\
  \citenamefont {Zhou}}]{PhysRevA.109.023710}%
  \BibitemOpen
  \bibfield  {author} {\bibinfo {author} {\bibfnamefont {Y.-T.}\ \bibnamefont
  {Yan}}, \bibinfo {author} {\bibfnamefont {C.}~\bibnamefont {Zhao}}, \bibinfo
  {author} {\bibfnamefont {D.-W.}\ \bibnamefont {Wang}}, \bibinfo {author}
  {\bibfnamefont {J.}~\bibnamefont {Yang}}, \ and\ \bibinfo {author}
  {\bibfnamefont {L.}~\bibnamefont {Zhou}},\ }\bibfield  {title} {\emph
  {\bibinfo {title} {\textnormal{Simultaneous blockade of two remote magnons
  induced by an atom}},\ }}\href {\doibase 10.1103/PhysRevA.109.023710}
  {\bibfield  {journal} {\bibinfo  {journal} {Phys. Rev. A}\ }\textbf {\bibinfo
  {volume} {109}},\ \bibinfo {pages} {023710} (\bibinfo {year}
  {2024})}\BibitemShut {NoStop}%
\bibitem [{\citenamefont {Jin}\ and\ \citenamefont
  {Jing}(2024)}]{PhysRevA.110.012459}%
  \BibitemOpen
  \bibfield  {author} {\bibinfo {author} {\bibfnamefont {Z.-y.}\ \bibnamefont
  {Jin}}\ and\ \bibinfo {author} {\bibfnamefont {J.}~\bibnamefont {Jing}},\
  }\bibfield  {title} {\emph {\bibinfo {title} {\textnormal{Stabilizing a
  single-magnon state by optimizing magnon blockade}},\ }}\href {\doibase
  10.1103/PhysRevA.110.012459} {\bibfield  {journal} {\bibinfo  {journal}
  {Phys. Rev. A}\ }\textbf {\bibinfo {volume} {110}},\ \bibinfo {pages}
  {012459} (\bibinfo {year} {2024})}\BibitemShut {NoStop}%
\bibitem [{\citenamefont {Wang}\ \emph {et~al.}(2020)\citenamefont {Wang},
  \citenamefont {Bai}, \citenamefont {Han}, \citenamefont {Liu}, \citenamefont
  {Zhang},\ and\ \citenamefont {Wang}}]{Wang:20}%
  \BibitemOpen
  \bibfield  {author} {\bibinfo {author} {\bibfnamefont {D.-Y.}\ \bibnamefont
  {Wang}}, \bibinfo {author} {\bibfnamefont {C.-H.}\ \bibnamefont {Bai}},
  \bibinfo {author} {\bibfnamefont {X.}~\bibnamefont {Han}}, \bibinfo {author}
  {\bibfnamefont {S.}~\bibnamefont {Liu}}, \bibinfo {author} {\bibfnamefont
  {S.}~\bibnamefont {Zhang}}, \ and\ \bibinfo {author} {\bibfnamefont {H.-F.}\
  \bibnamefont {Wang}},\ }\bibfield  {title} {\emph {\bibinfo {title}
  {\textnormal{Enhanced photon blockade in an optomechanical system with
  parametric amplification}},\ }}\href {\doibase 10.1364/OL.392514} {\bibfield
  {journal} {\bibinfo  {journal} {Opt. Lett.}\ }\textbf {\bibinfo {volume}
  {45}},\ \bibinfo {pages} {2604} (\bibinfo {year} {2020})}\BibitemShut
  {NoStop}%
\bibitem [{\citenamefont {Zhu}\ \emph {et~al.}(2021)\citenamefont {Zhu},
  \citenamefont {Hou}, \citenamefont {Yang},\ and\ \citenamefont
  {Deng}}]{Zhu:21}%
  \BibitemOpen
  \bibfield  {author} {\bibinfo {author} {\bibfnamefont {C.~J.}\ \bibnamefont
  {Zhu}}, \bibinfo {author} {\bibfnamefont {K.}~\bibnamefont {Hou}}, \bibinfo
  {author} {\bibfnamefont {Y.~P.}\ \bibnamefont {Yang}}, \ and\ \bibinfo
  {author} {\bibfnamefont {L.}~\bibnamefont {Deng}},\ }\bibfield  {title}
  {\emph {\bibinfo {title} {\textnormal{Hybrid level anharmonicity and
  interference-induced photon blockade in a two-qubit cavity QED system with
  dipole–dipole interaction}},\ }}\href {\doibase 10.1364/PRJ.421234}
  {\bibfield  {journal} {\bibinfo  {journal} {Photon. Res.}\ }\textbf {\bibinfo
  {volume} {9}},\ \bibinfo {pages} {1264} (\bibinfo {year} {2021})}\BibitemShut
  {NoStop}%
\bibitem [{\citenamefont {Lu}\ \emph {et~al.}(2023)\citenamefont {Lu},
  \citenamefont {Shang}, \citenamefont {Wu},\ and\ \citenamefont
  {L\"u}}]{PhysRevA.108.053703}%
  \BibitemOpen
  \bibfield  {author} {\bibinfo {author} {\bibfnamefont {Z.-G.}\ \bibnamefont
  {Lu}}, \bibinfo {author} {\bibfnamefont {C.}~\bibnamefont {Shang}}, \bibinfo
  {author} {\bibfnamefont {Y.}~\bibnamefont {Wu}}, \ and\ \bibinfo {author}
  {\bibfnamefont {X.-Y.}\ \bibnamefont {L\"u}},\ }\bibfield  {title} {\emph
  {\bibinfo {title} {\textnormal{Analytical approach to higher-order
  correlation functions in U(1) symmetric systems}},\ }}\href {\doibase
  10.1103/PhysRevA.108.053703} {\bibfield  {journal} {\bibinfo  {journal}
  {Phys. Rev. A}\ }\textbf {\bibinfo {volume} {108}},\ \bibinfo {pages}
  {053703} (\bibinfo {year} {2023})}\BibitemShut {NoStop}%
\bibitem [{\citenamefont {Wu}\ \emph {et~al.}(2024)\citenamefont {Wu},
  \citenamefont {Li},\ and\ \citenamefont {Wu}}]{PhysRevA.109.033709}%
  \BibitemOpen
  \bibfield  {author} {\bibinfo {author} {\bibfnamefont {Z.}~\bibnamefont
  {Wu}}, \bibinfo {author} {\bibfnamefont {J.}~\bibnamefont {Li}}, \ and\
  \bibinfo {author} {\bibfnamefont {Y.}~\bibnamefont {Wu}},\ }\bibfield
  {title} {\emph {\bibinfo {title} {\textnormal{Phase-engineered photon
  correlations in weakly coupled nanofiber cavity QED}},\ }}\href {\doibase
  10.1103/PhysRevA.109.033709} {\bibfield  {journal} {\bibinfo  {journal}
  {Phys. Rev. A}\ }\textbf {\bibinfo {volume} {109}},\ \bibinfo {pages}
  {033709} (\bibinfo {year} {2024})}\BibitemShut {NoStop}%
\bibitem [{\citenamefont {Couteau}\ \emph {et~al.}(2023)\citenamefont
  {Couteau}, \citenamefont {Barz}, \citenamefont {Durt}, \citenamefont
  {Gerrits}, \citenamefont {Huwer}, \citenamefont {Prevedel}, \citenamefont
  {Rarity}, \citenamefont {Shields},\ and\ \citenamefont
  {Weihs}}]{doi:10.1038/s42254-023-00583-2}%
  \BibitemOpen
  \bibfield  {author} {\bibinfo {author} {\bibfnamefont {C.}~\bibnamefont
  {Couteau}}, \bibinfo {author} {\bibfnamefont {S.}~\bibnamefont {Barz}},
  \bibinfo {author} {\bibfnamefont {T.}~\bibnamefont {Durt}}, \bibinfo {author}
  {\bibfnamefont {T.}~\bibnamefont {Gerrits}}, \bibinfo {author} {\bibfnamefont
  {J.}~\bibnamefont {Huwer}}, \bibinfo {author} {\bibfnamefont
  {R.}~\bibnamefont {Prevedel}}, \bibinfo {author} {\bibfnamefont
  {J.}~\bibnamefont {Rarity}}, \bibinfo {author} {\bibfnamefont
  {A.}~\bibnamefont {Shields}}, \ and\ \bibinfo {author} {\bibfnamefont
  {G.}~\bibnamefont {Weihs}},\ }\bibfield  {title} {\emph {\bibinfo {title}
  {\textnormal{Applications of single photons to quantum communication and
  computing}},\ }}\href {\doibase 10.1038/s42254-023-00583-2} {\bibfield
  {journal} {\bibinfo  {journal} {Nat. Rev. Phys}\ }\textbf {\bibinfo {volume}
  {5}},\ \bibinfo {pages} {326} (\bibinfo {year} {2023})}\BibitemShut {NoStop}%
\bibitem [{\citenamefont {Wang}\ \emph {et~al.}(2024)\citenamefont {Wang},
  \citenamefont {Huang},\ and\ \citenamefont {Xiong}}]{PhysRevA.110.033702}%
  \BibitemOpen
  \bibfield  {author} {\bibinfo {author} {\bibfnamefont {X.}~\bibnamefont
  {Wang}}, \bibinfo {author} {\bibfnamefont {K.-W.}\ \bibnamefont {Huang}}, \
  and\ \bibinfo {author} {\bibfnamefont {H.}~\bibnamefont {Xiong}},\ }\bibfield
   {title} {\emph {\bibinfo {title} {\textnormal{Magnon blockade in a QED
  system with a giant spin ensemble and a giant atom coupled to a waveguide}},\
  }}\href {\doibase 10.1103/PhysRevA.110.033702} {\bibfield  {journal}
  {\bibinfo  {journal} {Phys. Rev. A}\ }\textbf {\bibinfo {volume} {110}},\
  \bibinfo {pages} {033702} (\bibinfo {year} {2024})}\BibitemShut {NoStop}%
\bibitem [{\citenamefont {He}\ \emph {et~al.}(2023{\natexlab{b}})\citenamefont
  {He}, \citenamefont {Xin}, \citenamefont {Zhang},\ and\ \citenamefont
  {Li}}]{doi:10.1103/PhysRevA.107.023709}%
  \BibitemOpen
  \bibfield  {author} {\bibinfo {author} {\bibfnamefont {S.}~\bibnamefont
  {He}}, \bibinfo {author} {\bibfnamefont {X.}~\bibnamefont {Xin}}, \bibinfo
  {author} {\bibfnamefont {F.-Y.}\ \bibnamefont {Zhang}}, \ and\ \bibinfo
  {author} {\bibfnamefont {C.}~\bibnamefont {Li}},\ }\bibfield  {title} {\emph
  {\bibinfo {title} {\textnormal{Generation of a Schr\"odinger cat state in a
  hybrid ferromagnet-superconductor system}},\ }}\href
  {https://link.aps.org/doi/10.1103/PhysRevA.107.023709} {\bibfield  {journal}
  {\bibinfo  {journal} {Phys. Rev. A}\ }\textbf {\bibinfo {volume} {107}},\
  \bibinfo {pages} {023709} (\bibinfo {year} {2023}{\natexlab{b}})}\BibitemShut
  {NoStop}%
\bibitem [{\citenamefont {Reichhardt}\ \emph {et~al.}(2022)\citenamefont
  {Reichhardt}, \citenamefont {Reichhardt},\ and\ \citenamefont {Milo\ifmmode
  \check{s}\else \v{s}\fi{}evi\ifmmode~\acute{c}\else
  \'{c}\fi{}}}]{RevModPhys.94.035005}%
  \BibitemOpen
  \bibfield  {author} {\bibinfo {author} {\bibfnamefont {C.}~\bibnamefont
  {Reichhardt}}, \bibinfo {author} {\bibfnamefont {C.~J.~O.}\ \bibnamefont
  {Reichhardt}}, \ and\ \bibinfo {author} {\bibfnamefont {M.~V.}\ \bibnamefont
  {Milo\ifmmode \check{s}\else \v{s}\fi{}evi\ifmmode~\acute{c}\else
  \'{c}\fi{}}},\ }\bibfield  {title} {\emph {\bibinfo {title}
  {\textnormal{Statics and dynamics of skyrmions interacting with disorder and
  nanostructures}},\ }}\href {\doibase 10.1103/RevModPhys.94.035005} {\bibfield
   {journal} {\bibinfo  {journal} {Rev. Mod. Phys.}\ }\textbf {\bibinfo
  {volume} {94}},\ \bibinfo {pages} {035005} (\bibinfo {year}
  {2022})}\BibitemShut {NoStop}%
\bibitem [{\citenamefont {Everschor-Sitte}\ \emph {et~al.}(2018)\citenamefont
  {Everschor-Sitte}, \citenamefont {Masell}, \citenamefont {Reeve},\ and\
  \citenamefont {Kläui}}]{doi:10.1063/1.5048972}%
  \BibitemOpen
  \bibfield  {author} {\bibinfo {author} {\bibfnamefont {K.}~\bibnamefont
  {Everschor-Sitte}}, \bibinfo {author} {\bibfnamefont {J.}~\bibnamefont
  {Masell}}, \bibinfo {author} {\bibfnamefont {R.~M.}\ \bibnamefont {Reeve}}, \
  and\ \bibinfo {author} {\bibfnamefont {M.}~\bibnamefont {Kläui}},\
  }\bibfield  {title} {\emph {\bibinfo {title} {\textnormal{Perspective:
  Magnetic Skyrmions—Overview of Recent Progress in an Active Research
  Field}},\ }}\href@noop {} {\bibfield  {journal} {\bibinfo  {journal}
  {\href{https://doi.org/10.1063/1.5048972}{J. Appl. Phys.}}\ }\textbf
  {\bibinfo {volume} {\href{https://doi.org/10.1063/1.5048972}{124}}} (\bibinfo
  {year} {\href{https://doi.org/10.1063/1.5048972}{2018}})}\BibitemShut
  {NoStop}%
\bibitem [{\citenamefont {Liensberger}\ \emph {et~al.}(2021)\citenamefont
  {Liensberger}, \citenamefont {Haslbeck}, \citenamefont {Bauer}, \citenamefont
  {Berger}, \citenamefont {Gross}, \citenamefont {Huebl}, \citenamefont
  {Pfleiderer},\ and\ \citenamefont {Weiler}}]{PhysRevB.104.L100415}%
  \BibitemOpen
  \bibfield  {author} {\bibinfo {author} {\bibfnamefont {L.}~\bibnamefont
  {Liensberger}}, \bibinfo {author} {\bibfnamefont {F.~X.}\ \bibnamefont
  {Haslbeck}}, \bibinfo {author} {\bibfnamefont {A.}~\bibnamefont {Bauer}},
  \bibinfo {author} {\bibfnamefont {H.}~\bibnamefont {Berger}}, \bibinfo
  {author} {\bibfnamefont {R.}~\bibnamefont {Gross}}, \bibinfo {author}
  {\bibfnamefont {H.}~\bibnamefont {Huebl}}, \bibinfo {author} {\bibfnamefont
  {C.}~\bibnamefont {Pfleiderer}}, \ and\ \bibinfo {author} {\bibfnamefont
  {M.}~\bibnamefont {Weiler}},\ }\bibfield  {title} {\emph {\bibinfo {title}
  {\textnormal{Tunable cooperativity in coupled spin-cavity systems}},\ }}\href
  {\doibase 10.1103/PhysRevB.104.L100415} {\bibfield  {journal} {\bibinfo
  {journal} {Phys. Rev. B}\ }\textbf {\bibinfo {volume} {104}},\ \bibinfo
  {pages} {L100415} (\bibinfo {year} {2021})}\BibitemShut {NoStop}%
\bibitem [{\citenamefont {Hirosawa}\ \emph {et~al.}(2022)\citenamefont
  {Hirosawa}, \citenamefont {Mook}, \citenamefont {Klinovaja},\ and\
  \citenamefont {Loss}}]{PRXQuantum.3.040321}%
  \BibitemOpen
  \bibfield  {author} {\bibinfo {author} {\bibfnamefont {T.}~\bibnamefont
  {Hirosawa}}, \bibinfo {author} {\bibfnamefont {A.}~\bibnamefont {Mook}},
  \bibinfo {author} {\bibfnamefont {J.}~\bibnamefont {Klinovaja}}, \ and\
  \bibinfo {author} {\bibfnamefont {D.}~\bibnamefont {Loss}},\ }\bibfield
  {title} {\emph {\bibinfo {title} {\textnormal{Magnetoelectric Cavity
  Magnonics in Skyrmion Crystals}},\ }}\href {\doibase
  10.1103/PRXQuantum.3.040321} {\bibfield  {journal} {\bibinfo  {journal}
  {Phys. Rev. X Quantum}\ }\textbf {\bibinfo {volume} {3}},\ \bibinfo {pages}
  {040321} (\bibinfo {year} {2022})}\BibitemShut {NoStop}%
\bibitem [{\citenamefont {Ochoa}\ and\ \citenamefont
  {Tserkovnyak}(2019)}]{doi:10.1142/s0217979219300056}%
  \BibitemOpen
  \bibfield  {author} {\bibinfo {author} {\bibfnamefont {H.}~\bibnamefont
  {Ochoa}}\ and\ \bibinfo {author} {\bibfnamefont {Y.}~\bibnamefont
  {Tserkovnyak}},\ }\bibfield  {title} {\emph {\bibinfo {title}
  {\textnormal{Quantum skyrmionics}},\ }}\href {\doibase
  10.1142/s0217979219300056} {\bibfield  {journal} {\bibinfo  {journal} {Int.
  J. Mod. Phys. B}\ }\textbf {\bibinfo {volume} {33}},\ \bibinfo {pages}
  {1930005} (\bibinfo {year} {2019})}\BibitemShut {NoStop}%
\bibitem [{\citenamefont {Back}\ \emph {et~al.}(2020)\citenamefont {Back},
  \citenamefont {Cros}, \citenamefont {Ebert}, \citenamefont {Everschor-Sitte},
  \citenamefont {Fert}, \citenamefont {Garst}, \citenamefont {Ma},
  \citenamefont {Mankovsky}, \citenamefont {Monchesky}, \citenamefont
  {Mostovoy}, \citenamefont {Nagaosa}, \citenamefont {Parkin}, \citenamefont
  {Pfleiderer}, \citenamefont {Reyren}, \citenamefont {Rosch}, \citenamefont
  {Taguchi}, \citenamefont {Tokura}, \citenamefont {von Bergmann},\ and\
  \citenamefont {Zang}}]{doi:10.1088/1361-6463/ab8418}%
  \BibitemOpen
  \bibfield  {author} {\bibinfo {author} {\bibfnamefont {C.}~\bibnamefont
  {Back}}, \bibinfo {author} {\bibfnamefont {V.}~\bibnamefont {Cros}}, \bibinfo
  {author} {\bibfnamefont {H.}~\bibnamefont {Ebert}}, \bibinfo {author}
  {\bibfnamefont {K.}~\bibnamefont {Everschor-Sitte}}, \bibinfo {author}
  {\bibfnamefont {A.}~\bibnamefont {Fert}}, \bibinfo {author} {\bibfnamefont
  {M.}~\bibnamefont {Garst}}, \bibinfo {author} {\bibfnamefont
  {T.}~\bibnamefont {Ma}}, \bibinfo {author} {\bibfnamefont {S.}~\bibnamefont
  {Mankovsky}}, \bibinfo {author} {\bibfnamefont {T.~L.}\ \bibnamefont
  {Monchesky}}, \bibinfo {author} {\bibfnamefont {M.}~\bibnamefont {Mostovoy}},
  \bibinfo {author} {\bibfnamefont {N.}~\bibnamefont {Nagaosa}}, \bibinfo
  {author} {\bibfnamefont {S.~S.~P.}\ \bibnamefont {Parkin}}, \bibinfo {author}
  {\bibfnamefont {C.}~\bibnamefont {Pfleiderer}}, \bibinfo {author}
  {\bibfnamefont {N.}~\bibnamefont {Reyren}}, \bibinfo {author} {\bibfnamefont
  {A.}~\bibnamefont {Rosch}}, \bibinfo {author} {\bibfnamefont
  {Y.}~\bibnamefont {Taguchi}}, \bibinfo {author} {\bibfnamefont
  {Y.}~\bibnamefont {Tokura}}, \bibinfo {author} {\bibfnamefont
  {K.}~\bibnamefont {von Bergmann}}, \ and\ \bibinfo {author} {\bibfnamefont
  {J.}~\bibnamefont {Zang}},\ }\bibfield  {title} {\emph {\bibinfo {title}
  {\textnormal{The 2020 skyrmionics roadmap}},\ }}\href {\doibase
  10.1088/1361-6463/ab8418} {\bibfield  {journal} {\bibinfo  {journal} {J.
  Phys. D: Appl. Phys.}\ }\textbf {\bibinfo {volume} {53}},\ \bibinfo {pages}
  {363001} (\bibinfo {year} {2020})}\BibitemShut {NoStop}%
\bibitem [{\citenamefont {Psaroudaki}\ and\ \citenamefont
  {Panagopoulos}(2021)}]{PhysRevLett.127.067201}%
  \BibitemOpen
  \bibfield  {author} {\bibinfo {author} {\bibfnamefont {C.}~\bibnamefont
  {Psaroudaki}}\ and\ \bibinfo {author} {\bibfnamefont {C.}~\bibnamefont
  {Panagopoulos}},\ }\bibfield  {title} {\emph {\bibinfo {title}
  {\textnormal{Skyrmion Qubits: A New Class of Quantum Logic Elements Based on
  Nanoscale Magnetization}},\ }}\href {\doibase 10.1103/PhysRevLett.127.067201}
  {\bibfield  {journal} {\bibinfo  {journal} {Phys. Rev. Lett.}\ }\textbf
  {\bibinfo {volume} {127}},\ \bibinfo {pages} {067201} (\bibinfo {year}
  {2021})}\BibitemShut {NoStop}%
\bibitem [{\citenamefont {Pan}\ \emph {et~al.}(2024{\natexlab{a}})\citenamefont
  {Pan}, \citenamefont {Li}, \citenamefont {Hei}, \citenamefont {Zhang},
  \citenamefont {Mochizuki}, \citenamefont {Li},\ and\ \citenamefont
  {Nori}}]{PhysRevLett.132.193601}%
  \BibitemOpen
  \bibfield  {author} {\bibinfo {author} {\bibfnamefont {X.-F.}\ \bibnamefont
  {Pan}}, \bibinfo {author} {\bibfnamefont {P.-B.}\ \bibnamefont {Li}},
  \bibinfo {author} {\bibfnamefont {X.-L.}\ \bibnamefont {Hei}}, \bibinfo
  {author} {\bibfnamefont {X.}~\bibnamefont {Zhang}}, \bibinfo {author}
  {\bibfnamefont {M.}~\bibnamefont {Mochizuki}}, \bibinfo {author}
  {\bibfnamefont {F.-L.}\ \bibnamefont {Li}}, \ and\ \bibinfo {author}
  {\bibfnamefont {F.}~\bibnamefont {Nori}},\ }\bibfield  {title} {\emph
  {\bibinfo {title} {\textnormal{Magnon-Skyrmion Hybrid Quantum Systems:
  Tailoring Interactions via Magnons}},\ }}\href {\doibase
  10.1103/PhysRevLett.132.193601} {\bibfield  {journal} {\bibinfo  {journal}
  {Phys. Rev. Lett.}\ }\textbf {\bibinfo {volume} {132}},\ \bibinfo {pages}
  {193601} (\bibinfo {year} {2024}{\natexlab{a}})}\BibitemShut {NoStop}%
\bibitem [{\citenamefont {Hsu}\ \emph {et~al.}(2016)\citenamefont {Hsu},
  \citenamefont {Kubetzka}, \citenamefont {Finco}, \citenamefont {Romming},
  \citenamefont {von Bergmann},\ and\ \citenamefont
  {Wiesendanger}}]{doi:10.1038/nnano.2016.234}%
  \BibitemOpen
  \bibfield  {author} {\bibinfo {author} {\bibfnamefont {P.-J.}\ \bibnamefont
  {Hsu}}, \bibinfo {author} {\bibfnamefont {A.}~\bibnamefont {Kubetzka}},
  \bibinfo {author} {\bibfnamefont {A.}~\bibnamefont {Finco}}, \bibinfo
  {author} {\bibfnamefont {N.}~\bibnamefont {Romming}}, \bibinfo {author}
  {\bibfnamefont {K.}~\bibnamefont {von Bergmann}}, \ and\ \bibinfo {author}
  {\bibfnamefont {R.}~\bibnamefont {Wiesendanger}},\ }\bibfield  {title} {\emph
  {\bibinfo {title} {\textnormal{Electric-field-driven switching of individual
  magnetic skyrmions}},\ }}\href
  {https://www.nature.com/articles/nnano.2016.234} {\bibfield  {journal}
  {\bibinfo  {journal} {Nat. Nanotechnol.}\ }\textbf {\bibinfo {volume} {12}},\
  \bibinfo {pages} {123} (\bibinfo {year} {2016})}\BibitemShut {NoStop}%
\bibitem [{\citenamefont {Birch}\ \emph {et~al.}(2024)\citenamefont {Birch},
  \citenamefont {Belopolski}, \citenamefont {Fujishiro}, \citenamefont
  {Kawamura}, \citenamefont {Kikkawa}, \citenamefont {Taguchi}, \citenamefont
  {Hirschberger}, \citenamefont {Nagaosa},\ and\ \citenamefont
  {Tokura}}]{doi:10.1038/s41586-024-07859-2}%
  \BibitemOpen
  \bibfield  {author} {\bibinfo {author} {\bibfnamefont {M.~T.}\ \bibnamefont
  {Birch}}, \bibinfo {author} {\bibfnamefont {I.}~\bibnamefont {Belopolski}},
  \bibinfo {author} {\bibfnamefont {Y.}~\bibnamefont {Fujishiro}}, \bibinfo
  {author} {\bibfnamefont {M.}~\bibnamefont {Kawamura}}, \bibinfo {author}
  {\bibfnamefont {A.}~\bibnamefont {Kikkawa}}, \bibinfo {author} {\bibfnamefont
  {Y.}~\bibnamefont {Taguchi}}, \bibinfo {author} {\bibfnamefont
  {M.}~\bibnamefont {Hirschberger}}, \bibinfo {author} {\bibfnamefont
  {N.}~\bibnamefont {Nagaosa}}, \ and\ \bibinfo {author} {\bibfnamefont
  {Y.}~\bibnamefont {Tokura}},\ }\bibfield  {title} {\emph {\bibinfo {title}
  {\textnormal{Dynamic transition and Galilean relativity of current-driven
  skyrmions}},\ }}\href {https://www.nature.com/articles/s41586-024-07859-2}
  {\bibfield  {journal} {\bibinfo  {journal} {Nature}\ }\textbf {\bibinfo
  {volume} {633}},\ \bibinfo {pages} {554} (\bibinfo {year}
  {2024})}\BibitemShut {NoStop}%
\bibitem [{\citenamefont {Lonsky}\ and\ \citenamefont
  {Hoffmann}(2020)}]{doi:10.1063/5.0027042}%
  \BibitemOpen
  \bibfield  {author} {\bibinfo {author} {\bibfnamefont {M.}~\bibnamefont
  {Lonsky}}\ and\ \bibinfo {author} {\bibfnamefont {A.}~\bibnamefont
  {Hoffmann}},\ }\bibfield  {title} {\emph {\bibinfo {title}
  {\textnormal{Dynamic Excitations of Chiral Magnetic Textures}},\ }}\href@noop
  {} {\bibfield  {journal} {\bibinfo  {journal}
  {\href{https://doi.org/10.1063/5.0027042}{APL Materials}}\ }\textbf {\bibinfo
  {volume} {\href{https://doi.org/10.1063/5.0027042}{8}}} (\bibinfo {year}
  {\href{https://doi.org/10.1063/5.0027042}{2020}})}\BibitemShut {NoStop}%
\bibitem [{\citenamefont {Bogdanov}\ and\ \citenamefont
  {Hubert}(1994)}]{BOGDANOV1994255}%
  \BibitemOpen
  \bibfield  {author} {\bibinfo {author} {\bibfnamefont {A.}~\bibnamefont
  {Bogdanov}}\ and\ \bibinfo {author} {\bibfnamefont {A.}~\bibnamefont
  {Hubert}},\ }\bibfield  {title} {\emph {\bibinfo {title}
  {\textnormal{Thermodynamically stable magnetic vortex states in magnetic
  crystals}},\ }}\href {\doibase https://doi.org/10.1016/0304-8853(94)90046-9}
  {\bibfield  {journal} {\bibinfo  {journal} {J. Magn. Magn. Mater.}\ }\textbf
  {\bibinfo {volume} {138}},\ \bibinfo {pages} {255} (\bibinfo {year}
  {1994})}\BibitemShut {NoStop}%
\bibitem [{\citenamefont {Rößler}\ \emph {et~al.}(2006)\citenamefont
  {Rößler}, \citenamefont {Bogdanov},\ and\ \citenamefont
  {Pfleiderer}}]{doi:10.1038/nature05056}%
  \BibitemOpen
  \bibfield  {author} {\bibinfo {author} {\bibfnamefont {U.~K.}\ \bibnamefont
  {Rößler}}, \bibinfo {author} {\bibfnamefont {A.~N.}\ \bibnamefont
  {Bogdanov}}, \ and\ \bibinfo {author} {\bibfnamefont {C.}~\bibnamefont
  {Pfleiderer}},\ }\bibfield  {title} {\emph {\bibinfo {title}
  {\textnormal{Spontaneous skyrmion ground states in magnetic metals}},\
  }}\href {\doibase 10.1038/nature05056} {\bibfield  {journal} {\bibinfo
  {journal} {Nature}\ }\textbf {\bibinfo {volume} {442}},\ \bibinfo {pages}
  {797} (\bibinfo {year} {2006})}\BibitemShut {NoStop}%
\bibitem [{\citenamefont {Mühlbauer}\ \emph {et~al.}(2009)\citenamefont
  {Mühlbauer}, \citenamefont {Binz}, \citenamefont {Jonietz}, \citenamefont
  {Pfleiderer}, \citenamefont {Rosch}, \citenamefont {Neubauer}, \citenamefont
  {Georgii},\ and\ \citenamefont {Böni}}]{doi:10.1126/science.1166767}%
  \BibitemOpen
  \bibfield  {author} {\bibinfo {author} {\bibfnamefont {S.}~\bibnamefont
  {Mühlbauer}}, \bibinfo {author} {\bibfnamefont {B.}~\bibnamefont {Binz}},
  \bibinfo {author} {\bibfnamefont {F.}~\bibnamefont {Jonietz}}, \bibinfo
  {author} {\bibfnamefont {C.}~\bibnamefont {Pfleiderer}}, \bibinfo {author}
  {\bibfnamefont {A.}~\bibnamefont {Rosch}}, \bibinfo {author} {\bibfnamefont
  {A.}~\bibnamefont {Neubauer}}, \bibinfo {author} {\bibfnamefont
  {R.}~\bibnamefont {Georgii}}, \ and\ \bibinfo {author} {\bibfnamefont
  {P.}~\bibnamefont {Böni}},\ }\bibfield  {title} {\emph {\bibinfo {title}
  {\textnormal{Skyrmion Lattice in a Chiral Magnet}},\ }}\href {\doibase
  10.1126/science.1166767} {\bibfield  {journal} {\bibinfo  {journal}
  {Science}\ }\textbf {\bibinfo {volume} {323}},\ \bibinfo {pages} {915}
  (\bibinfo {year} {2009})}\BibitemShut {NoStop}%
\bibitem [{\citenamefont {Yu}\ \emph {et~al.}(2010)\citenamefont {Yu},
  \citenamefont {Onose}, \citenamefont {Kanazawa}, \citenamefont {Park},
  \citenamefont {Han}, \citenamefont {Matsui}, \citenamefont {Nagaosa},\ and\
  \citenamefont {Tokura}}]{doi:10.1038/nature09124}%
  \BibitemOpen
  \bibfield  {author} {\bibinfo {author} {\bibfnamefont {X.~Z.}\ \bibnamefont
  {Yu}}, \bibinfo {author} {\bibfnamefont {Y.}~\bibnamefont {Onose}}, \bibinfo
  {author} {\bibfnamefont {N.}~\bibnamefont {Kanazawa}}, \bibinfo {author}
  {\bibfnamefont {J.~H.}\ \bibnamefont {Park}}, \bibinfo {author}
  {\bibfnamefont {J.~H.}\ \bibnamefont {Han}}, \bibinfo {author} {\bibfnamefont
  {Y.}~\bibnamefont {Matsui}}, \bibinfo {author} {\bibfnamefont
  {N.}~\bibnamefont {Nagaosa}}, \ and\ \bibinfo {author} {\bibfnamefont
  {Y.}~\bibnamefont {Tokura}},\ }\bibfield  {title} {\emph {\bibinfo {title}
  {\textnormal{Real-space observation of a two-dimensional skyrmion crystal}},\
  }}\href {\doibase 10.1038/nature09124} {\bibfield  {journal} {\bibinfo
  {journal} {Nature(London)}\ }\textbf {\bibinfo {volume} {465}},\ \bibinfo
  {pages} {901} (\bibinfo {year} {2010})}\BibitemShut {NoStop}%
\bibitem [{\citenamefont {Jonietz}\ \emph {et~al.}(2010)\citenamefont
  {Jonietz}, \citenamefont {Mühlbauer}, \citenamefont {Pfleiderer},
  \citenamefont {Neubauer}, \citenamefont {Münzer}, \citenamefont {Bauer},
  \citenamefont {Adams}, \citenamefont {Georgii}, \citenamefont {Böni},
  \citenamefont {Duine}, \citenamefont {Everschor}, \citenamefont {Garst},\
  and\ \citenamefont {Rosch}}]{doi:10.1126/science.1195709}%
  \BibitemOpen
  \bibfield  {author} {\bibinfo {author} {\bibfnamefont {F.}~\bibnamefont
  {Jonietz}}, \bibinfo {author} {\bibfnamefont {S.}~\bibnamefont {Mühlbauer}},
  \bibinfo {author} {\bibfnamefont {C.}~\bibnamefont {Pfleiderer}}, \bibinfo
  {author} {\bibfnamefont {A.}~\bibnamefont {Neubauer}}, \bibinfo {author}
  {\bibfnamefont {W.}~\bibnamefont {Münzer}}, \bibinfo {author} {\bibfnamefont
  {A.}~\bibnamefont {Bauer}}, \bibinfo {author} {\bibfnamefont
  {T.}~\bibnamefont {Adams}}, \bibinfo {author} {\bibfnamefont
  {R.}~\bibnamefont {Georgii}}, \bibinfo {author} {\bibfnamefont
  {P.}~\bibnamefont {Böni}}, \bibinfo {author} {\bibfnamefont {R.~A.}\
  \bibnamefont {Duine}}, \bibinfo {author} {\bibfnamefont {K.}~\bibnamefont
  {Everschor}}, \bibinfo {author} {\bibfnamefont {M.}~\bibnamefont {Garst}}, \
  and\ \bibinfo {author} {\bibfnamefont {A.}~\bibnamefont {Rosch}},\ }\bibfield
   {title} {\emph {\bibinfo {title} {\textnormal{Spin Transfer Torques in MnSi
  at Ultralow Current Densities}},\ }}\href {\doibase 10.1126/science.1195709}
  {\bibfield  {journal} {\bibinfo  {journal} {Science}\ }\textbf {\bibinfo
  {volume} {330}},\ \bibinfo {pages} {1648} (\bibinfo {year}
  {2010})}\BibitemShut {NoStop}%
\bibitem [{\citenamefont {Schulz}\ \emph {et~al.}(2012)\citenamefont {Schulz},
  \citenamefont {Ritz}, \citenamefont {Bauer}, \citenamefont {Halder},
  \citenamefont {Wagner}, \citenamefont {Franz}, \citenamefont {Pfleiderer},
  \citenamefont {Everschor}, \citenamefont {Garst},\ and\ \citenamefont
  {Rosch}}]{doi:10.1038/nphys2231}%
  \BibitemOpen
  \bibfield  {author} {\bibinfo {author} {\bibfnamefont {T.}~\bibnamefont
  {Schulz}}, \bibinfo {author} {\bibfnamefont {R.}~\bibnamefont {Ritz}},
  \bibinfo {author} {\bibfnamefont {A.}~\bibnamefont {Bauer}}, \bibinfo
  {author} {\bibfnamefont {M.}~\bibnamefont {Halder}}, \bibinfo {author}
  {\bibfnamefont {M.}~\bibnamefont {Wagner}}, \bibinfo {author} {\bibfnamefont
  {C.}~\bibnamefont {Franz}}, \bibinfo {author} {\bibfnamefont
  {C.}~\bibnamefont {Pfleiderer}}, \bibinfo {author} {\bibfnamefont
  {K.}~\bibnamefont {Everschor}}, \bibinfo {author} {\bibfnamefont
  {M.}~\bibnamefont {Garst}}, \ and\ \bibinfo {author} {\bibfnamefont
  {A.}~\bibnamefont {Rosch}},\ }\bibfield  {title} {\emph {\bibinfo {title}
  {\textnormal{Emergent electrodynamics of skyrmions in a chiral magnet}},\
  }}\href {\doibase 10.1038/nphys2231} {\bibfield  {journal} {\bibinfo
  {journal} {Nat. Phys.}\ }\textbf {\bibinfo {volume} {8}},\ \bibinfo {pages}
  {301} (\bibinfo {year} {2012})}\BibitemShut {NoStop}%
\bibitem [{\citenamefont {Huang}\ and\ \citenamefont
  {Chien}(2012)}]{PhysRevLett.108.267201}%
  \BibitemOpen
  \bibfield  {author} {\bibinfo {author} {\bibfnamefont {S.~X.}\ \bibnamefont
  {Huang}}\ and\ \bibinfo {author} {\bibfnamefont {C.~L.}\ \bibnamefont
  {Chien}},\ }\bibfield  {title} {\emph {\bibinfo {title} {\textnormal{Extended
  Skyrmion Phase in Epitaxial $\mathrm{FeGe}(111)$ Thin Films}},\ }}\href
  {\doibase 10.1103/PhysRevLett.108.267201} {\bibfield  {journal} {\bibinfo
  {journal} {Phys. Rev. Lett.}\ }\textbf {\bibinfo {volume} {108}},\ \bibinfo
  {pages} {267201} (\bibinfo {year} {2012})}\BibitemShut {NoStop}%
\bibitem [{\citenamefont {Nagaosa}\ and\ \citenamefont
  {Tokura}(2013)}]{doi:10.1038/nnano.2013.243}%
  \BibitemOpen
  \bibfield  {author} {\bibinfo {author} {\bibfnamefont {N.}~\bibnamefont
  {Nagaosa}}\ and\ \bibinfo {author} {\bibfnamefont {Y.}~\bibnamefont
  {Tokura}},\ }\bibfield  {title} {\emph {\bibinfo {title}
  {\textnormal{Topological Properties and Dynamics of Magnetic Skyrmions}},\
  }}\href@noop {} {\bibfield  {journal} {\bibinfo  {journal}
  {\href{https://doi.org/10.1038/nnano.2013.243}{Nat. Nanotechnol.}}\ }\textbf
  {\bibinfo {volume} {\href{https://doi.org/10.1038/nnano.2013.243}{8}}}
  (\bibinfo {year}
  {\href{https://doi.org/10.1038/nnano.2013.243}{2013}})}\BibitemShut {NoStop}%
\bibitem [{\citenamefont {Fert}\ \emph {et~al.}(2013)\citenamefont {Fert},
  \citenamefont {Cros},\ and\ \citenamefont
  {Sampaio}}]{doi:10.1038/nnano.2013.29}%
  \BibitemOpen
  \bibfield  {author} {\bibinfo {author} {\bibfnamefont {A.}~\bibnamefont
  {Fert}}, \bibinfo {author} {\bibfnamefont {V.}~\bibnamefont {Cros}}, \ and\
  \bibinfo {author} {\bibfnamefont {J.}~\bibnamefont {Sampaio}},\ }\bibfield
  {title} {\emph {\bibinfo {title} {\textnormal{Skyrmions on the track}},\
  }}\href {\doibase 10.1038/nnano.2013.29} {\bibfield  {journal} {\bibinfo
  {journal} {Nat. Nanotechnol.}\ }\textbf {\bibinfo {volume} {8}},\ \bibinfo
  {pages} {152} (\bibinfo {year} {2013})}\BibitemShut {NoStop}%
\bibitem [{\citenamefont {Fert}\ \emph {et~al.}(2017)\citenamefont {Fert},
  \citenamefont {Reyren},\ and\ \citenamefont
  {Cros}}]{doi:10.1038/natrevmats.2017.31}%
  \BibitemOpen
  \bibfield  {author} {\bibinfo {author} {\bibfnamefont {A.}~\bibnamefont
  {Fert}}, \bibinfo {author} {\bibfnamefont {N.}~\bibnamefont {Reyren}}, \ and\
  \bibinfo {author} {\bibfnamefont {V.}~\bibnamefont {Cros}},\ }\bibfield
  {title} {\emph {\bibinfo {title} {\textnormal{Magnetic Skyrmions: Advances in
  Physics and Potential Applications}},\ }}\href@noop {} {\bibfield  {journal}
  {\bibinfo  {journal} {\href{https://doi.org/10.1038/natrevmats.2017.31}{Nat.
  Rev. Mater.}}\ }\textbf {\bibinfo {volume}
  {\href{https://doi.org/10.1038/natrevmats.2017.31}{2}}} (\bibinfo {year}
  {\href{https://doi.org/10.1038/natrevmats.2017.31}{2017}})}\BibitemShut
  {NoStop}%
\bibitem [{\citenamefont {Bhowal}\ and\ \citenamefont
  {Spaldin}(2022)}]{PhysRevLett.128.227204}%
  \BibitemOpen
  \bibfield  {author} {\bibinfo {author} {\bibfnamefont {S.}~\bibnamefont
  {Bhowal}}\ and\ \bibinfo {author} {\bibfnamefont {N.~A.}\ \bibnamefont
  {Spaldin}},\ }\bibfield  {title} {\emph {\bibinfo {title}
  {\textnormal{Magnetoelectric Classification of Skyrmions}},\ }}\href
  {\doibase 10.1103/PhysRevLett.128.227204} {\bibfield  {journal} {\bibinfo
  {journal} {Phys. Rev. Lett.}\ }\textbf {\bibinfo {volume} {128}},\ \bibinfo
  {pages} {227204} (\bibinfo {year} {2022})}\BibitemShut {NoStop}%
\bibitem [{\citenamefont {Kurumaji}\ \emph {et~al.}(2019)\citenamefont
  {Kurumaji}, \citenamefont {Nakajima}, \citenamefont {Hirschberger},
  \citenamefont {Kikkawa}, \citenamefont {Yamasaki}, \citenamefont {Sagayama},
  \citenamefont {Nakao}, \citenamefont {Taguchi}, \citenamefont {hisa Arima},\
  and\ \citenamefont {Tokura}}]{doi:10.1126/science.aau0968}%
  \BibitemOpen
  \bibfield  {author} {\bibinfo {author} {\bibfnamefont {T.}~\bibnamefont
  {Kurumaji}}, \bibinfo {author} {\bibfnamefont {T.}~\bibnamefont {Nakajima}},
  \bibinfo {author} {\bibfnamefont {M.}~\bibnamefont {Hirschberger}}, \bibinfo
  {author} {\bibfnamefont {A.}~\bibnamefont {Kikkawa}}, \bibinfo {author}
  {\bibfnamefont {Y.}~\bibnamefont {Yamasaki}}, \bibinfo {author}
  {\bibfnamefont {H.}~\bibnamefont {Sagayama}}, \bibinfo {author}
  {\bibfnamefont {H.}~\bibnamefont {Nakao}}, \bibinfo {author} {\bibfnamefont
  {Y.}~\bibnamefont {Taguchi}}, \bibinfo {author} {\bibfnamefont
  {T.}~\bibnamefont {hisa Arima}}, \ and\ \bibinfo {author} {\bibfnamefont
  {Y.}~\bibnamefont {Tokura}},\ }\bibfield  {title} {\emph {\bibinfo {title}
  {\textnormal{Skyrmion lattice with a giant topological Hall effect in a
  frustrated triangular-lattice magnet}},\ }}\href {\doibase
  10.1126/science.aau0968} {\bibfield  {journal} {\bibinfo  {journal}
  {Science}\ }\textbf {\bibinfo {volume} {365}},\ \bibinfo {pages} {914}
  (\bibinfo {year} {2019})}\BibitemShut {NoStop}%
\bibitem [{\citenamefont {Leonov}\ and\ \citenamefont
  {Mostovoy}(2015)}]{doi:10.1038/ncomms9275}%
  \BibitemOpen
  \bibfield  {author} {\bibinfo {author} {\bibfnamefont {A.~O.}\ \bibnamefont
  {Leonov}}\ and\ \bibinfo {author} {\bibfnamefont {M.}~\bibnamefont
  {Mostovoy}},\ }\bibfield  {title} {\emph {\bibinfo {title}
  {\textnormal{Multiply Periodic States and Isolated Skyrmions in an
  Anisotropic Frustrated Magnet}},\ }}\href@noop {} {\bibfield  {journal}
  {\bibinfo  {journal} {\href{https://doi.org/10.1038/ncomms9275}{Nat.
  Commun.}}\ }\textbf {\bibinfo {volume}
  {\href{https://doi.org/10.1038/ncomms9275}{6}}} (\bibinfo {year}
  {\href{https://doi.org/10.1038/ncomms9275}{2015}})}\BibitemShut {NoStop}%
\bibitem [{\citenamefont {Lin}\ and\ \citenamefont
  {Hayami}(2016)}]{PhysRevB.93.064430}%
  \BibitemOpen
  \bibfield  {author} {\bibinfo {author} {\bibfnamefont {S.-Z.}\ \bibnamefont
  {Lin}}\ and\ \bibinfo {author} {\bibfnamefont {S.}~\bibnamefont {Hayami}},\
  }\bibfield  {title} {\emph {\bibinfo {title} {\textnormal{Ginzburg-Landau
  theory for skyrmions in inversion-symmetric magnets with competing
  interactions}},\ }}\href {\doibase 10.1103/PhysRevB.93.064430} {\bibfield
  {journal} {\bibinfo  {journal} {Phys. Rev. B}\ }\textbf {\bibinfo {volume}
  {93}},\ \bibinfo {pages} {064430} (\bibinfo {year} {2016})}\BibitemShut
  {NoStop}%
\bibitem [{\citenamefont {Zhang}\ \emph {et~al.}(2017)\citenamefont {Zhang},
  \citenamefont {Xia}, \citenamefont {Zhou}, \citenamefont {Liu}, \citenamefont
  {Zhang},\ and\ \citenamefont {Ezawa}}]{pmid:29167418}%
  \BibitemOpen
  \bibfield  {author} {\bibinfo {author} {\bibfnamefont {X.}~\bibnamefont
  {Zhang}}, \bibinfo {author} {\bibfnamefont {J.}~\bibnamefont {Xia}}, \bibinfo
  {author} {\bibfnamefont {Y.}~\bibnamefont {Zhou}}, \bibinfo {author}
  {\bibfnamefont {X.}~\bibnamefont {Liu}}, \bibinfo {author} {\bibfnamefont
  {H.}~\bibnamefont {Zhang}}, \ and\ \bibinfo {author} {\bibfnamefont
  {M.}~\bibnamefont {Ezawa}},\ }\bibfield  {title} {\emph {\bibinfo {title}
  {\textnormal{Skyrmion dynamics in a frustrated ferromagnetic film and
  current-induced helicity locking-unlocking transition}},\ }}\href {\doibase
  10.1038/s41467-017-01785-w} {\bibfield  {journal} {\bibinfo  {journal} {Nat.
  Commun.}\ }\textbf {\bibinfo {volume} {8}},\ \bibinfo {pages} {1717}
  (\bibinfo {year} {2017})}\BibitemShut {NoStop}%
\bibitem [{\citenamefont {Yao}\ \emph {et~al.}(2020)\citenamefont {Yao},
  \citenamefont {Chen},\ and\ \citenamefont
  {Dong}}]{doi:10.1088/1367-2630/aba1b3}%
  \BibitemOpen
  \bibfield  {author} {\bibinfo {author} {\bibfnamefont {X.}~\bibnamefont
  {Yao}}, \bibinfo {author} {\bibfnamefont {J.}~\bibnamefont {Chen}}, \ and\
  \bibinfo {author} {\bibfnamefont {S.}~\bibnamefont {Dong}},\ }\bibfield
  {title} {\emph {\bibinfo {title} {\textnormal{Controlling the helicity of
  magnetic skyrmions by electrical field in frustrated magnets}},\ }}\href
  {\doibase 10.1088/1367-2630/aba1b3} {\bibfield  {journal} {\bibinfo
  {journal} {New J. Phys}\ }\textbf {\bibinfo {volume} {22}},\ \bibinfo {pages}
  {083032} (\bibinfo {year} {2020})}\BibitemShut {NoStop}%
\bibitem [{\citenamefont {Xia}\ \emph {et~al.}(2023)\citenamefont {Xia},
  \citenamefont {Zhang}, \citenamefont {Liu}, \citenamefont {Zhou},\ and\
  \citenamefont {Ezawa}}]{PhysRevLett.130.106701}%
  \BibitemOpen
  \bibfield  {author} {\bibinfo {author} {\bibfnamefont {J.}~\bibnamefont
  {Xia}}, \bibinfo {author} {\bibfnamefont {X.}~\bibnamefont {Zhang}}, \bibinfo
  {author} {\bibfnamefont {X.}~\bibnamefont {Liu}}, \bibinfo {author}
  {\bibfnamefont {Y.}~\bibnamefont {Zhou}}, \ and\ \bibinfo {author}
  {\bibfnamefont {M.}~\bibnamefont {Ezawa}},\ }\bibfield  {title} {\emph
  {\bibinfo {title} {\textnormal{Universal Quantum Computation Based on
  Nanoscale Skyrmion Helicity Qubits in Frustrated Magnets}},\ }}\href
  {\doibase 10.1103/PhysRevLett.130.106701} {\bibfield  {journal} {\bibinfo
  {journal} {Phys. Rev. Lett.}\ }\textbf {\bibinfo {volume} {130}},\ \bibinfo
  {pages} {106701} (\bibinfo {year} {2023})}\BibitemShut {NoStop}%
\bibitem [{\citenamefont {Li}\ \emph {et~al.}(2021)\citenamefont {Li},
  \citenamefont {Wang}, \citenamefont {Wu}, \citenamefont {Yang},\ and\
  \citenamefont {Chen}}]{PhysRevB.104.224434}%
  \BibitemOpen
  \bibfield  {author} {\bibinfo {author} {\bibfnamefont {X.}~\bibnamefont
  {Li}}, \bibinfo {author} {\bibfnamefont {X.}~\bibnamefont {Wang}}, \bibinfo
  {author} {\bibfnamefont {Z.}~\bibnamefont {Wu}}, \bibinfo {author}
  {\bibfnamefont {W.-X.}\ \bibnamefont {Yang}}, \ and\ \bibinfo {author}
  {\bibfnamefont {A.}~\bibnamefont {Chen}},\ }\bibfield  {title} {\emph
  {\bibinfo {title} {\textnormal{Tunable magnon antibunching in a hybrid
  ferromagnet-superconductor system with two qubits}},\ }}\href {\doibase
  10.1103/PhysRevB.104.224434} {\bibfield  {journal} {\bibinfo  {journal}
  {Phys. Rev. B}\ }\textbf {\bibinfo {volume} {104}},\ \bibinfo {pages}
  {224434} (\bibinfo {year} {2021})}\BibitemShut {NoStop}%
\bibitem [{\citenamefont {Zhang}\ \emph
  {et~al.}(2024{\natexlab{b}})\citenamefont {Zhang}, \citenamefont {Liu},
  \citenamefont {Zhang},\ and\ \citenamefont {Wang}}]{PhysRevA.109.043712}%
  \BibitemOpen
  \bibfield  {author} {\bibinfo {author} {\bibfnamefont {W.}~\bibnamefont
  {Zhang}}, \bibinfo {author} {\bibfnamefont {S.}~\bibnamefont {Liu}}, \bibinfo
  {author} {\bibfnamefont {S.}~\bibnamefont {Zhang}}, \ and\ \bibinfo {author}
  {\bibfnamefont {H.-F.}\ \bibnamefont {Wang}},\ }\bibfield  {title} {\emph
  {\bibinfo {title} {\textnormal{Magnon blockade induced by parametric
  amplification}},\ }}\href {\doibase 10.1103/PhysRevA.109.043712} {\bibfield
  {journal} {\bibinfo  {journal} {Phys. Rev. A}\ }\textbf {\bibinfo {volume}
  {109}},\ \bibinfo {pages} {043712} (\bibinfo {year}
  {2024}{\natexlab{b}})}\BibitemShut {NoStop}%
\bibitem [{\citenamefont {Rabl}(2011)}]{PhysRevLett.107.063601}%
  \BibitemOpen
  \bibfield  {author} {\bibinfo {author} {\bibfnamefont {P.}~\bibnamefont
  {Rabl}},\ }\bibfield  {title} {\emph {\bibinfo {title} {\textnormal{Photon
  Blockade Effect in Optomechanical Systems}},\ }}\href {\doibase
  10.1103/PhysRevLett.107.063601} {\bibfield  {journal} {\bibinfo  {journal}
  {Phys. Rev. Lett.}\ }\textbf {\bibinfo {volume} {107}},\ \bibinfo {pages}
  {063601} (\bibinfo {year} {2011})}\BibitemShut {NoStop}%
\bibitem [{\citenamefont {Liu}\ \emph {et~al.}(2010)\citenamefont {Liu},
  \citenamefont {Miranowicz}, \citenamefont {Gao}, \citenamefont {Bajer},
  \citenamefont {Sun},\ and\ \citenamefont {Nori}}]{PhysRevA.82.032101}%
  \BibitemOpen
  \bibfield  {author} {\bibinfo {author} {\bibfnamefont {Y.-x.}\ \bibnamefont
  {Liu}}, \bibinfo {author} {\bibfnamefont {A.}~\bibnamefont {Miranowicz}},
  \bibinfo {author} {\bibfnamefont {Y.~B.}\ \bibnamefont {Gao}}, \bibinfo
  {author} {\bibfnamefont {J.~c.~v.}\ \bibnamefont {Bajer}}, \bibinfo {author}
  {\bibfnamefont {C.~P.}\ \bibnamefont {Sun}}, \ and\ \bibinfo {author}
  {\bibfnamefont {F.}~\bibnamefont {Nori}},\ }\bibfield  {title} {\emph
  {\bibinfo {title} {\textnormal{Qubit-induced phonon blockade as a signature
  of quantum behavior in nanomechanical resonators}},\ }}\href {\doibase
  10.1103/PhysRevA.82.032101} {\bibfield  {journal} {\bibinfo  {journal} {Phys.
  Rev. A}\ }\textbf {\bibinfo {volume} {82}},\ \bibinfo {pages} {032101}
  (\bibinfo {year} {2010})}\BibitemShut {NoStop}%
\bibitem [{\citenamefont {Zheng}\ \emph {et~al.}(2019)\citenamefont {Zheng},
  \citenamefont {Yin}, \citenamefont {Bin}, \citenamefont {L\"u},\ and\
  \citenamefont {Wu}}]{PhysRevA.99.013804}%
  \BibitemOpen
  \bibfield  {author} {\bibinfo {author} {\bibfnamefont {L.-L.}\ \bibnamefont
  {Zheng}}, \bibinfo {author} {\bibfnamefont {T.-S.}\ \bibnamefont {Yin}},
  \bibinfo {author} {\bibfnamefont {Q.}~\bibnamefont {Bin}}, \bibinfo {author}
  {\bibfnamefont {X.-Y.}\ \bibnamefont {L\"u}}, \ and\ \bibinfo {author}
  {\bibfnamefont {Y.}~\bibnamefont {Wu}},\ }\bibfield  {title} {\emph {\bibinfo
  {title} {\textnormal{Single-photon-induced phonon blockade in a hybrid
  spin-optomechanical system}},\ }}\href {\doibase 10.1103/PhysRevA.99.013804}
  {\bibfield  {journal} {\bibinfo  {journal} {Phys. Rev. A}\ }\textbf {\bibinfo
  {volume} {99}},\ \bibinfo {pages} {013804} (\bibinfo {year}
  {2019})}\BibitemShut {NoStop}%
\bibitem [{\citenamefont {Lu}\ \emph {et~al.}(2022)\citenamefont {Lu},
  \citenamefont {Liu}, \citenamefont {Li}, \citenamefont {Wu}, \citenamefont
  {Tan},\ and\ \citenamefont {Li}}]{doi:10.1088/1367-2630/ac6a46}%
  \BibitemOpen
  \bibfield  {author} {\bibinfo {author} {\bibfnamefont {Y.-W.}\ \bibnamefont
  {Lu}}, \bibinfo {author} {\bibfnamefont {J.-F.}\ \bibnamefont {Liu}},
  \bibinfo {author} {\bibfnamefont {R.}~\bibnamefont {Li}}, \bibinfo {author}
  {\bibfnamefont {Y.}~\bibnamefont {Wu}}, \bibinfo {author} {\bibfnamefont
  {H.}~\bibnamefont {Tan}}, \ and\ \bibinfo {author} {\bibfnamefont
  {Y.}~\bibnamefont {Li}},\ }\bibfield  {title} {\emph {\bibinfo {title}
  {\textnormal{Single-photon blockade in quasichiral atom–photon interaction:
  simultaneous high purity and high efficiency}},\ }}\href {\doibase
  10.1088/1367-2630/ac6a46} {\bibfield  {journal} {\bibinfo  {journal} {New J.
  Phys}\ }\textbf {\bibinfo {volume} {24}},\ \bibinfo {pages} {053029}
  (\bibinfo {year} {2022})}\BibitemShut {NoStop}%
\bibitem [{\citenamefont {Flayac}\ and\ \citenamefont
  {Savona}(2017)}]{PhysRevA.96.053810}%
  \BibitemOpen
  \bibfield  {author} {\bibinfo {author} {\bibfnamefont {H.}~\bibnamefont
  {Flayac}}\ and\ \bibinfo {author} {\bibfnamefont {V.}~\bibnamefont
  {Savona}},\ }\bibfield  {title} {\emph {\bibinfo {title}
  {\textnormal{Unconventional photon blockade}},\ }}\href {\doibase
  10.1103/PhysRevA.96.053810} {\bibfield  {journal} {\bibinfo  {journal} {Phys.
  Rev. A}\ }\textbf {\bibinfo {volume} {96}},\ \bibinfo {pages} {053810}
  (\bibinfo {year} {2017})}\BibitemShut {NoStop}%
\bibitem [{\citenamefont {Huang}\ \emph {et~al.}(2018)\citenamefont {Huang},
  \citenamefont {Miranowicz}, \citenamefont {Liao}, \citenamefont {Nori},\ and\
  \citenamefont {Jing}}]{PhysRevLett.121.153601}%
  \BibitemOpen
  \bibfield  {author} {\bibinfo {author} {\bibfnamefont {R.}~\bibnamefont
  {Huang}}, \bibinfo {author} {\bibfnamefont {A.}~\bibnamefont {Miranowicz}},
  \bibinfo {author} {\bibfnamefont {J.-Q.}\ \bibnamefont {Liao}}, \bibinfo
  {author} {\bibfnamefont {F.}~\bibnamefont {Nori}}, \ and\ \bibinfo {author}
  {\bibfnamefont {H.}~\bibnamefont {Jing}},\ }\bibfield  {title} {\emph
  {\bibinfo {title} {\textnormal{Nonreciprocal Photon Blockade}},\ }}\href
  {\doibase 10.1103/PhysRevLett.121.153601} {\bibfield  {journal} {\bibinfo
  {journal} {Phys. Rev. Lett.}\ }\textbf {\bibinfo {volume} {121}},\ \bibinfo
  {pages} {153601} (\bibinfo {year} {2018})}\BibitemShut {NoStop}%
\bibitem [{\citenamefont {Xie}\ \emph {et~al.}(2017)\citenamefont {Xie},
  \citenamefont {Liao}, \citenamefont {Shang}, \citenamefont {Ye},\ and\
  \citenamefont {Lin}}]{PhysRevA.96.013861}%
  \BibitemOpen
  \bibfield  {author} {\bibinfo {author} {\bibfnamefont {H.}~\bibnamefont
  {Xie}}, \bibinfo {author} {\bibfnamefont {C.-G.}\ \bibnamefont {Liao}},
  \bibinfo {author} {\bibfnamefont {X.}~\bibnamefont {Shang}}, \bibinfo
  {author} {\bibfnamefont {M.-Y.}\ \bibnamefont {Ye}}, \ and\ \bibinfo {author}
  {\bibfnamefont {X.-M.}\ \bibnamefont {Lin}},\ }\bibfield  {title} {\emph
  {\bibinfo {title} {\textnormal{Phonon blockade in a quadratically coupled
  optomechanical system}},\ }}\href {\doibase 10.1103/PhysRevA.96.013861}
  {\bibfield  {journal} {\bibinfo  {journal} {Phys. Rev. A}\ }\textbf {\bibinfo
  {volume} {96}},\ \bibinfo {pages} {013861} (\bibinfo {year}
  {2017})}\BibitemShut {NoStop}%
\bibitem [{\citenamefont {Xu}\ \emph {et~al.}(2016)\citenamefont {Xu},
  \citenamefont {Chen},\ and\ \citenamefont {Liu}}]{PhysRevA.94.063853}%
  \BibitemOpen
  \bibfield  {author} {\bibinfo {author} {\bibfnamefont {X.-W.}\ \bibnamefont
  {Xu}}, \bibinfo {author} {\bibfnamefont {A.-X.}\ \bibnamefont {Chen}}, \ and\
  \bibinfo {author} {\bibfnamefont {Y.-x.}\ \bibnamefont {Liu}},\ }\bibfield
  {title} {\emph {\bibinfo {title} {\textnormal{Phonon blockade in a
  nanomechanical resonator resonantly coupled to a qubit}},\ }}\href {\doibase
  10.1103/PhysRevA.94.063853} {\bibfield  {journal} {\bibinfo  {journal} {Phys.
  Rev. A}\ }\textbf {\bibinfo {volume} {94}},\ \bibinfo {pages} {063853}
  (\bibinfo {year} {2016})}\BibitemShut {NoStop}%
\bibitem [{\citenamefont {Zeng}\ \emph {et~al.}(2021)\citenamefont {Zeng},
  \citenamefont {Gebremariam}, \citenamefont {Shen}, \citenamefont {Xiong},\
  and\ \citenamefont {Li}}]{APL2021}%
  \BibitemOpen
  \bibfield  {author} {\bibinfo {author} {\bibfnamefont {Y.-X.}\ \bibnamefont
  {Zeng}}, \bibinfo {author} {\bibfnamefont {T.}~\bibnamefont {Gebremariam}},
  \bibinfo {author} {\bibfnamefont {J.}~\bibnamefont {Shen}}, \bibinfo {author}
  {\bibfnamefont {B.}~\bibnamefont {Xiong}}, \ and\ \bibinfo {author}
  {\bibfnamefont {C.}~\bibnamefont {Li}},\ }\bibfield  {title} {\emph {\bibinfo
  {title} {\textnormal{Application of machine learning for predicting strong
  phonon blockade}},\ }}\href {\doibase 10.1063/5.0035498} {\bibfield
  {journal} {\bibinfo  {journal} {Appl. Phys. Lett.}\ }\textbf {\bibinfo
  {volume} {118}},\ \bibinfo {pages} {164003} (\bibinfo {year}
  {2021})}\BibitemShut {NoStop}%
\bibitem [{\citenamefont {Carmichael}(1999)}]{Carmichael1999}%
  \BibitemOpen
  \bibfield  {author} {\bibinfo {author} {\bibfnamefont {H.}~\bibnamefont
  {Carmichael}},\ }\href@noop {} {\emph {\bibinfo {title} {Statistical Methods
  in Quantum Optics}}}\ (\bibinfo  {publisher} {Springer},\ \bibinfo {address}
  {Berlin},\ \bibinfo {year} {1999})\BibitemShut {NoStop}%
\bibitem [{\citenamefont {Hou}\ and\ \citenamefont
  {Liu}(2019)}]{PhysRevLett.123.107702}%
  \BibitemOpen
  \bibfield  {author} {\bibinfo {author} {\bibfnamefont {J.~T.}\ \bibnamefont
  {Hou}}\ and\ \bibinfo {author} {\bibfnamefont {L.}~\bibnamefont {Liu}},\
  }\bibfield  {title} {\emph {\bibinfo {title} {\textnormal{Strong Coupling
  between Microwave Photons and Nanomagnet Magnons}},\ }}\href {\doibase
  10.1103/PhysRevLett.123.107702} {\bibfield  {journal} {\bibinfo  {journal}
  {Phys. Rev. Lett.}\ }\textbf {\bibinfo {volume} {123}},\ \bibinfo {pages}
  {107702} (\bibinfo {year} {2019})}\BibitemShut {NoStop}%
\bibitem [{\citenamefont {Gonzalez-Ballestero}\ \emph
  {et~al.}(2020{\natexlab{a}})\citenamefont {Gonzalez-Ballestero},
  \citenamefont {H\"ummer}, \citenamefont {Gieseler},\ and\ \citenamefont
  {Romero-Isart}}]{PhysRevB.101.125404}%
  \BibitemOpen
  \bibfield  {author} {\bibinfo {author} {\bibfnamefont {C.}~\bibnamefont
  {Gonzalez-Ballestero}}, \bibinfo {author} {\bibfnamefont {D.}~\bibnamefont
  {H\"ummer}}, \bibinfo {author} {\bibfnamefont {J.}~\bibnamefont {Gieseler}},
  \ and\ \bibinfo {author} {\bibfnamefont {O.}~\bibnamefont {Romero-Isart}},\
  }\bibfield  {title} {\emph {\bibinfo {title} {\textnormal{Theory of quantum
  acoustomagnonics and acoustomechanics with a micromagnet}},\ }}\href
  {\doibase 10.1103/PhysRevB.101.125404} {\bibfield  {journal} {\bibinfo
  {journal} {Phys. Rev. B}\ }\textbf {\bibinfo {volume} {101}},\ \bibinfo
  {pages} {125404} (\bibinfo {year} {2020}{\natexlab{a}})}\BibitemShut
  {NoStop}%
\bibitem [{\citenamefont {Gonzalez-Ballestero}\ \emph
  {et~al.}(2020{\natexlab{b}})\citenamefont {Gonzalez-Ballestero},
  \citenamefont {Gieseler},\ and\ \citenamefont
  {Romero-Isart}}]{PhysRevLett.124.093602}%
  \BibitemOpen
  \bibfield  {author} {\bibinfo {author} {\bibfnamefont {C.}~\bibnamefont
  {Gonzalez-Ballestero}}, \bibinfo {author} {\bibfnamefont {J.}~\bibnamefont
  {Gieseler}}, \ and\ \bibinfo {author} {\bibfnamefont {O.}~\bibnamefont
  {Romero-Isart}},\ }\bibfield  {title} {\emph {\bibinfo {title}
  {\textnormal{Quantum Acoustomechanics with a Micromagnet}},\ }}\href
  {\doibase 10.1103/PhysRevLett.124.093602} {\bibfield  {journal} {\bibinfo
  {journal} {Phys. Rev. Lett.}\ }\textbf {\bibinfo {volume} {124}},\ \bibinfo
  {pages} {093602} (\bibinfo {year} {2020}{\natexlab{b}})}\BibitemShut
  {NoStop}%
\bibitem [{\citenamefont {Psaroudaki}\ \emph {et~al.}(2023)\citenamefont
  {Psaroudaki}, \citenamefont {Peraticos},\ and\ \citenamefont
  {Panagopoulos}}]{doi:10.1063/5.0177864}%
  \BibitemOpen
  \bibfield  {author} {\bibinfo {author} {\bibfnamefont {C.}~\bibnamefont
  {Psaroudaki}}, \bibinfo {author} {\bibfnamefont {E.}~\bibnamefont
  {Peraticos}}, \ and\ \bibinfo {author} {\bibfnamefont {C.}~\bibnamefont
  {Panagopoulos}},\ }\bibfield  {title} {\emph {\bibinfo {title}
  {\textnormal{Skyrmion qubits: Challenges for future quantum computing
  applications}},\ }}\href {\doibase 10.1063/5.0177864} {\bibfield  {journal}
  {\bibinfo  {journal} {Appl. Phys. Lett.}\ }\textbf {\bibinfo {volume}
  {123}},\ \bibinfo {pages} {26} (\bibinfo {year} {2023})}\BibitemShut
  {NoStop}%
\bibitem [{\citenamefont {Pan}\ \emph {et~al.}(2024{\natexlab{b}})\citenamefont
  {Pan}, \citenamefont {Hei}, \citenamefont {Yao}, \citenamefont {Chen},
  \citenamefont {Ren}, \citenamefont {Dong}, \citenamefont {Qiao},\ and\
  \citenamefont {Li}}]{PhysRevResearch.6.023067}%
  \BibitemOpen
  \bibfield  {author} {\bibinfo {author} {\bibfnamefont {X.-F.}\ \bibnamefont
  {Pan}}, \bibinfo {author} {\bibfnamefont {X.-L.}\ \bibnamefont {Hei}},
  \bibinfo {author} {\bibfnamefont {X.-Y.}\ \bibnamefont {Yao}}, \bibinfo
  {author} {\bibfnamefont {J.-Q.}\ \bibnamefont {Chen}}, \bibinfo {author}
  {\bibfnamefont {Y.-M.}\ \bibnamefont {Ren}}, \bibinfo {author} {\bibfnamefont
  {X.-L.}\ \bibnamefont {Dong}}, \bibinfo {author} {\bibfnamefont {Y.-F.}\
  \bibnamefont {Qiao}}, \ and\ \bibinfo {author} {\bibfnamefont {P.-B.}\
  \bibnamefont {Li}},\ }\bibfield  {title} {\emph {\bibinfo {title}
  {\textnormal{Skyrmion-mechanical hybrid quantum systems: Manipulation of
  skyrmion qubits via phonons}},\ }}\href {\doibase
  10.1103/PhysRevResearch.6.023067} {\bibfield  {journal} {\bibinfo  {journal}
  {Phys. Rev. Res.}\ }\textbf {\bibinfo {volume} {6}},\ \bibinfo {pages}
  {023067} (\bibinfo {year} {2024}{\natexlab{b}})}\BibitemShut {NoStop}%
\bibitem [{\citenamefont {Fujita}\ \emph {et~al.}(2016)\citenamefont {Fujita},
  \citenamefont {Suzuki},\ and\ \citenamefont
  {Ho}}]{doi:10.1007/s10773-016-3180-y}%
  \BibitemOpen
  \bibfield  {author} {\bibinfo {author} {\bibfnamefont {S.}~\bibnamefont
  {Fujita}}, \bibinfo {author} {\bibfnamefont {A.}~\bibnamefont {Suzuki}}, \
  and\ \bibinfo {author} {\bibfnamefont {H.-C.}\ \bibnamefont {Ho}},\
  }\bibfield  {title} {\emph {\bibinfo {title} {\textnormal{Composite-Particles
  (Boson, Fermion) Theory of Fractional Quantum Hall Effect}},\ }}\href
  {\doibase 10.1007/s10773-016-3180-y} {\bibfield  {journal} {\bibinfo
  {journal} {Int. J. Theor. Phys.}\ }\textbf {\bibinfo {volume} {56}},\
  \bibinfo {pages} {396} (\bibinfo {year} {2016})}\BibitemShut {NoStop}%
\bibitem [{\citenamefont {K\'om\'ar}\ \emph {et~al.}(2013)\citenamefont
  {K\'om\'ar}, \citenamefont {Bennett}, \citenamefont {Stannigel},
  \citenamefont {Habraken}, \citenamefont {Rabl}, \citenamefont {Zoller},\ and\
  \citenamefont {Lukin}}]{PhysRevA.87.013839}%
  \BibitemOpen
  \bibfield  {author} {\bibinfo {author} {\bibfnamefont {P.}~\bibnamefont
  {K\'om\'ar}}, \bibinfo {author} {\bibfnamefont {S.~D.}\ \bibnamefont
  {Bennett}}, \bibinfo {author} {\bibfnamefont {K.}~\bibnamefont {Stannigel}},
  \bibinfo {author} {\bibfnamefont {S.~J.~M.}\ \bibnamefont {Habraken}},
  \bibinfo {author} {\bibfnamefont {P.}~\bibnamefont {Rabl}}, \bibinfo {author}
  {\bibfnamefont {P.}~\bibnamefont {Zoller}}, \ and\ \bibinfo {author}
  {\bibfnamefont {M.~D.}\ \bibnamefont {Lukin}},\ }\bibfield  {title} {\emph
  {\bibinfo {title} {\textnormal{Single-photon nonlinearities in two-mode
  optomechanics}},\ }}\href
  {https://link.aps.org/doi/10.1103/PhysRevA.87.013839} {\bibfield  {journal}
  {\bibinfo  {journal} {Phys. Rev. A}\ }\textbf {\bibinfo {volume} {87}},\
  \bibinfo {pages} {013839} (\bibinfo {year} {2013})}\BibitemShut {NoStop}%
\bibitem [{\citenamefont {Tang}\ \emph {et~al.}(2019)\citenamefont {Tang},
  \citenamefont {Deng},\ and\ \citenamefont {Lee}}]{PhysRevApplied.12.044065}%
  \BibitemOpen
  \bibfield  {author} {\bibinfo {author} {\bibfnamefont {J.}~\bibnamefont
  {Tang}}, \bibinfo {author} {\bibfnamefont {Y.}~\bibnamefont {Deng}}, \ and\
  \bibinfo {author} {\bibfnamefont {C.}~\bibnamefont {Lee}},\ }\bibfield
  {title} {\emph {\bibinfo {title} {\textnormal{Strong Photon Blockade Mediated
  by Optical Stark Shift in a Single-Atom--Cavity System}},\ }}\href
  {https://link.aps.org/doi/10.1103/PhysRevApplied.12.044065} {\bibfield
  {journal} {\bibinfo  {journal} {Phys. Rev. Appl.}\ }\textbf {\bibinfo
  {volume} {12}},\ \bibinfo {pages} {044065} (\bibinfo {year}
  {2019})}\BibitemShut {NoStop}%
\bibitem [{\citenamefont {del Valle}\ \emph {et~al.}(2009)\citenamefont {del
  Valle}, \citenamefont {Laussy},\ and\ \citenamefont
  {Tejedor}}]{PhysRevB2009}%
  \BibitemOpen
  \bibfield  {author} {\bibinfo {author} {\bibfnamefont {E.}~\bibnamefont {del
  Valle}}, \bibinfo {author} {\bibfnamefont {F.~P.}\ \bibnamefont {Laussy}}, \
  and\ \bibinfo {author} {\bibfnamefont {C.}~\bibnamefont {Tejedor}},\
  }\bibfield  {title} {\emph {\bibinfo {title} {\textnormal{Luminescence
  spectra of quantum dots in microcavities. II. Fermions}},\ }}\href {\doibase
  10.1103/PhysRevB.79.235326} {\bibfield  {journal} {\bibinfo  {journal} {Phys.
  Rev. B}\ }\textbf {\bibinfo {volume} {79}},\ \bibinfo {pages} {235326}
  (\bibinfo {year} {2009})}\BibitemShut {NoStop}%
\bibitem [{\citenamefont {Scully}\ and\ \citenamefont
  {Zubairy}(1997)}]{Scully1997}%
  \BibitemOpen
  \bibfield  {author} {\bibinfo {author} {\bibfnamefont {M.~O.}\ \bibnamefont
  {Scully}}\ and\ \bibinfo {author} {\bibfnamefont {M.~S.}\ \bibnamefont
  {Zubairy}},\ }\href@noop {} {\emph {\bibinfo {title} {Quantum Optics}}}\
  (\bibinfo  {publisher} {Cambridge University Press},\ \bibinfo {address}
  {Cambridge},\ \bibinfo {year} {1997})\BibitemShut {NoStop}%
\bibitem [{\citenamefont {Johansson}\ \emph {et~al.}(2013)\citenamefont
  {Johansson}, \citenamefont {Nation},\ and\ \citenamefont
  {Nori}}]{JOHANSSON20131234}%
  \BibitemOpen
  \bibfield  {author} {\bibinfo {author} {\bibfnamefont {J.~R.}\ \bibnamefont
  {Johansson}}, \bibinfo {author} {\bibfnamefont {P.~D.}\ \bibnamefont
  {Nation}}, \ and\ \bibinfo {author} {\bibfnamefont {F.}~\bibnamefont
  {Nori}},\ }\bibfield  {title} {\emph {\bibinfo {title} {\textnormal{QuTiP 2:
  A Python framework for the dynamics of open quantum systems}},\ }}\href
  {\doibase 10.1016/j.cpc.2012.11.019} {\bibfield  {journal} {\bibinfo
  {journal} {Comput. Phys. Commun.}\ }\textbf {\bibinfo {volume} {184}},\
  \bibinfo {pages} {1234} (\bibinfo {year} {2013})}\BibitemShut {NoStop}%
\bibitem [{\citenamefont {Johansson}\ \emph {et~al.}(2012)\citenamefont
  {Johansson}, \citenamefont {Nation},\ and\ \citenamefont
  {Nori}}]{JOHANSSON20121760}%
  \BibitemOpen
  \bibfield  {author} {\bibinfo {author} {\bibfnamefont {J.}~\bibnamefont
  {Johansson}}, \bibinfo {author} {\bibfnamefont {P.}~\bibnamefont {Nation}}, \
  and\ \bibinfo {author} {\bibfnamefont {F.}~\bibnamefont {Nori}},\ }\bibfield
  {title} {\emph {\bibinfo {title} {\textnormal{QuTiP: An open-source Python
  framework for the dynamics of open quantum systems}},\ }}\href {\doibase
  10.1016/j.cpc.2012.02.021} {\bibfield  {journal} {\bibinfo  {journal}
  {Computer Phys. Commun.}\ }\textbf {\bibinfo {volume} {183}},\ \bibinfo
  {pages} {1760} (\bibinfo {year} {2012})}\BibitemShut {NoStop}%
\end{thebibliography}%
	
\end{document}